\documentclass[prd,superscriptaddress,twocolumn,showpacs,nofootinbib,eqsecnum,floatfix,nofootinbib]{revtex4}

\usepackage[usenames]{color}
\usepackage{amsfonts}
\usepackage{amsmath}
\usepackage{amssymb}
\usepackage{bm}
\usepackage{dcolumn}
\usepackage{epsfig}
\usepackage{graphicx}
\usepackage{graphics}
\usepackage[latin1]{inputenc}
\usepackage{latexsym}
\usepackage{rotating}
\usepackage{hyperref}
\usepackage{multirow}

\newcommand{\<}{\begin{equation}}
\newcommand{\?}{\end{equation}}

\DeclareMathOperator{\sech}{sech}

\newcommand{\od}[2]{\left( #1 \right)_{ #2 }}
\newcommand{\ods}[2]{\bigl( #1 \bigr)_{ #2 }}

\newcommand{\sA}{\mathcal{A}}
\newcommand{\sB}{\mathcal{B}}
\newcommand{\sC}{\mathcal{C}}
\newcommand{\sE}{\mathcal{E}}
\newcommand{\sI}{\mathcal{I}}
\newcommand{\sJ}{\mathcal{J}}
\newcommand{\sM}{\mathcal{M}}
\newcommand{\sP}{\mathcal{P}}
\newcommand{\sR}{\mathcal{R}}
\newcommand{\sT}{\mathcal{T}}
\newcommand{\sX}{\mathcal{X}}

\newcommand{\bB}{\bar{\mathcal{B}}}
\newcommand{\bC}{\bar{\mathcal{C}}}
\newcommand{\bE}{\bar{\mathcal{E}}}
\newcommand{\bM}{\mathcal{K}} 

\newcommand{\tr}{\tilde{r}}

\newcommand{\tx}{\tilde{x}}

\newcommand{\R}{\mathbb{R}}

\newcommand{\rP}{\mathrm{P}}
\newcommand{\rNP}{\mathrm{NP}}
\newcommand{\rEW}{\mathrm{EW}}
\newcommand{\Schw}{\mathrm{(S)}}

\newcommand\be{\begin{equation}}
\newcommand\ba{\begin{eqnarray}}
\newcommand\ee{\end{equation}}
\newcommand\ea{\end{eqnarray}}

\begin{document}

\title{Conformally curved binary black hole initial data \\ including tidal deformations and outgoing radiation}

\author{Nathan~K.~Johnson-McDaniel}

\affiliation{Institute for Gravitation and the Cosmos,
  Center for Gravitational Wave Physics,
  Department of Physics, The Pennsylvania State University, University
  Park, PA 16802, USA}

\author{Nicol\'{a}s~Yunes}

\affiliation{Department of Physics, Princeton University,
             Princeton, NJ 08544, USA}

\author{Wolfgang~Tichy}

\affiliation{Department of Physics, Florida Atlantic University,
             Boca Raton, FL 33431, USA}

\author{Benjamin~J.~Owen}

\affiliation{Institute for Gravitation and the Cosmos,
  Center for Gravitational Wave Physics,
  Department of Physics, The Pennsylvania State University, University
  Park, PA 16802, USA}

\date{\today}


\begin{abstract}

By asymptotically matching a post-Newtonian (PN) metric to two perturbed Schwarzschild
metrics, we generate approximate initial data
(in the form of an approximate $4$-metric) for a nonspinning black
hole binary in a circular orbit. We carry out this matching through
$O(v^4)$ in the binary's orbital velocity $v$, and thus
the resulting data, like the $O(v^4)$ PN metric, are conformally
curved. The matching procedure also fixes the quadrupole and octupole tidal
deformations of the holes, including the $1$PN corrections to the quadrupole fields. Far
from the holes, we use the appropriate PN metric that accounts for
retardation, which we construct using the highest-order PN expressions available
to compute the binary's past history. The data set's uncontrolled remainders are thus
$O(v^5)$
throughout the timeslice; we also generate an extension to the data set that has uncontrolled
remainders of $O(v^6)$ in the purely PN portion of the timeslice (i.e., not too close to the
holes). This extension
also includes various other readily available higher-order terms. The addition of these
terms decreases the constraint violations in certain regions, even though it
does not increase the data's formal accuracy. The resulting data are smooth, since we
join all the metrics together by smoothly interpolating between them. We perform this
interpolation using transition functions constructed to avoid introducing
excessive additional constraint violations. Due to their inclusion of tidal
deformations and outgoing radiation, these data should substantially
reduce both the high- and low-frequency components of the initial spurious (``junk'')
radiation observed in current simulations that use conformally flat initial
data. Such reductions in the nonphysical components of the initial
data will be necessary for simulations to achieve the accuracy required
to supply Advanced LIGO and LISA with the templates necessary
for parameter estimation.

\end{abstract}


\maketitle

\section{Introduction}

\subsection{Motivation and overview of results}

At present, several years after the initial breakthroughs in the evolution of
binary black hole spacetimes~\cite{Pretorius2005, Campanellietal,
BakeretalPRL}, numerical relativity has matured to the point where successful
binary black hole simulations are now commonplace (see, e.g.,~\cite{Pretoriusrev} for
a review). Recent progress in the nonspinning case includes
simulations of systems with mass ratios of up to
$10:1$~\cite{Bakeretal, GSB} and longer, more accurate simulations of
equal-mass systems~\cite{Scheeletal, Husaetal_6o}.

It is now time to consider what
improvements need to be made to these simulations so that they are
accurate
enough to provide gravitational wave detectors such as LIGO~\cite{LIGO} with
the model waveforms they need to detect and study binary black holes. The
accuracy required of such waveforms has been studied by Lindblom, Owen, and
Brown (LOB)~\cite{LOB}. Their results imply that current simulations are
sufficiently accurate to supply the waveforms necessary for detection with
either LIGO or LISA~\cite{LISA}.
This conclusion is supported by the Samurai~\cite{Samurai} and
NINJA~\cite{NINJA} projects, which indicate that currently used data
analysis pipelines (including some not based on matched filtering) can
easily detect a wide variety of numerical relativity waveforms at
essentially the same level in stationary gaussian noise.
The Samurai project also performs a more detailed comparison of a subset of
waveforms for parameter estimation, and finds that they are all
indistinguishable if used for estimation of intrinsic parameters (i.e., not
sky position or arrival time) at a signal-to-noise ratio (SNR) of less than
$14$ ($25$ if one eliminates the code that disagrees the most with the
others).

However, Advanced LIGO may detect binary black hole signals with SNRs of order
$100$~\cite{LOB}.
According to LOB~\cite{LOB}, if a waveform's phase error (suitably averaged
in the frequency domain) is less than $0.007$ radians, then it is suitable
for use in parameter estimation with Advanced LIGO with such an SNR.
However, even the Caltech/Cornell group's simulations (which are arguably
the most accurate yet) have maximum phase errors (in the time domain) of
order $0.01$ radians or more (see~\cite{Scheeletal, Boyleetal} for some
discussion of their error budget).
Converting such error measures to the criteria of LOB is more subtle than
it looks and we do not attempt it here. (See~\cite{Lindblom} for a discussion of
some possible pitfalls, along with suggestions for successful applications of the
standards from LOB.)
However it seems likely that even the Caltech/Cornell group's simulations
may not satisfy the LOB conditions for parameter estimation, at least for
binaries whose masses place the worst phase error in the detector's most
sensitive frequency band.

All current binary black hole simulations incur some error from the initial 
data used. These data sets' lack of astrophysical realism is clearly announced by the burst of
spurious (or ``junk'') radiation present at the beginning of these simulations. This junk
radiation is also responsible for various deleterious effects on simulations. First,
one wastes the computer time required for the spurious radiation to propagate off the
computational grid.
Second, the system that remains after that time has been perturbed by the spurious radiation.
This radiation's most prominent effect is to increase the binary's
eccentricity,\footnote{In
practice, the eccentricity can be reduced by various means~\cite{Pfeifferetal_ecc, Husaetal_ecc, WBM}.} 
though it also slightly increases the masses of the holes---see~\cite{BSHH}.
Additionally, in the unequal-mass case, the initial junk radiation is emitted
anisotropically, giving the system a small
``kick'' transverse to the holes' initial orbital motion~\cite{Gonzalezetal_kick}.
The high-frequency
component of the spurious radiation is a particular problem for spectral
codes. For instance, the Caltech/Cornell group finds
that the high-frequency component of the initial pulse of spurious radiation
generates secondary spurious waves that propagate throughout the
computational domain for two light-crossing times after the initial junk
radiation has exited~\cite{Scheeletal, Boyleetal}.

Currently employed initial data's omission of significant features of the
spacetime can also be seen analytically. Except for occasional tests of initial data
(discussed in Sec.~\ref{ccdata}), all
current simulations use initial data sets that assume \emph{conformal flatness} (i.e., that
the spatial
metric is a scalar multiple of the flat $3$-metric). The
assumption of conformal flatness is convenient, since it allows one to get simple, mostly
analytic expressions for initial data that exactly solve the constraint equations and
include orbiting black holes (see, e.g.,~\cite{CookLRR, Gourgoulhon} for
reviews).

In general,
these sets are geared either towards the puncture or excision approaches. The
majority of the community uses punctures, with initial data stemming from~\cite{Puncture}.
These data are very flexible, as they contain parameters with which
one can directly set the momentum and spin of each hole. For instance,
for evolutions with spinning holes, one can simply set these parameters
using post-Newtonian (PN) results as
in~\cite{Marronetti:2007ya,Marronetti:2007wz,Tichy:2007hk,Tichy:2008du}.
For nonspinning configurations in a circular orbit, parameter choices
based on the assumption of a helical Killing 
vector~\cite{Tichy:2003zg,Tichy:2003qi} are possible as well.
The Caltech/Cornell and Princeton groups use 
excision~\cite{Scheeletal_ex,Shoemakeretal}, with initial
data constructed using the conformal thin-sandwich
method (see, e.g.,~\cite{CP}). These data are slightly harder
to construct than puncture data are, since one has to solve a larger number of
elliptic equations. However, excision data
have the advantage of a direct connection to the isolated
horizon formalism~\cite{Ashtekar:2004cn}, which allows one
to construct holes with well-defined masses and spins. Additionally, the excision
approach is applicable to a wider array of initial data construction methods: It is
used in all of the extant evolutions of superposed black hole data sets except for
one specifically tailored to the puncture approach. (See Sec.~\ref{ccdata} for a
discussion of these evolutions.) The
data we present here also require evolutions using excision or the
turducken approach~\cite{Brownetal}.

Conformally flat initial data cannot accurately represent some
features of a binary black hole spacetime, since the PN metric for a binary
system stops being spatially conformally flat at $O(v^4)$, where $v$ is the
binary's orbital velocity in units of $c$, the speed of light (see Sec.~\ref{PNCF} for
discussion). This
is the same order at which gravitational radiation enters the PN metric (see 
Sec.~\ref{PNrad} for discussion). 
The order at which this fundamental disagreement with PN
predictions first occurs gives a
rough indication of the error committed in using conformally flat initial data.
At present, the simulation for which this error is the smallest is the longest
of the Caltech/Cornell runs (in~\cite{Scheeletal}), for which
$v_\mathrm{initial} \simeq 0.24$, where $v_\mathrm{initial}$ 
is the binary's initial orbital velocity. We expect the initial data's
conformal flatness to only affect the waveform at $O(v_\mathrm{initial}^{4})$,
which for this run
is comparable to the phase error allowed for a waveform to be used for
parameter
estimation with Advanced LIGO. It is thus possible (though perhaps not likely)
that conformally flat initial data would be suitable for use in the
simulations that will generate such waveforms.

It is unlikely that conformally flat initial data can be used to
generate the waveforms required for parameter estimation with LISA. Here the
required (appropriately averaged frequency domain) phase accuracy is
$2 \times 10^{-4}$ radians~\cite{LOB}, $20$ times smaller than
$v_\mathrm{initial}^4$ for the longest of Caltech/Cornell's simulations. One
can reduce the error in the initial data by
starting the simulation with a larger separation. However,
$v_\mathrm{initial}^4 \simeq 2
\times 10^{-4}$ implies an initial (PN coordinate) separation of $\sim 71$
times the binary's mass, and thus a merger time that is over $400$ times as long as the
Caltech/Cornell group's longest simulations to date, which start from a (PN coordinate)
separation of $\sim 15.3$ times the binary's mass. It is thus necessary to improve the
accuracy of the initial data. Evolutions of more accurate initial data will also give a
direct measure of the errors introduced in using current, conformally flat
initial data.

This paper provides initial data that include more of the
physics present in the binary's spacetime than any previous constructions.
In particular, our data's accurate description of certain properties of the
spacetime should substantially
reduce \emph{both} components of the aforementioned spurious radiation.
These two components are thought to come from different physical effects. The
long-wavelength component is thought to correspond to the initial data's lack
of outgoing gravitational radiation, whose
wavelength would be somewhat longer than the orbital separation. Of course,
one expects the junk to be generated predominantly in the strong-field region
near the holes, where the binary's gravitational radiation cannot be
disentangled from the rest of its gravitational field. However, one also
expects the
pieces that one wants in the strong-field region to appear at $O(v^4)$, just as the true
gravitational waves do in the radiation (or far) zone (defined in
Sec.~\ref{MatchingIntro}). Our data include all the $O(v^4)$ terms in
the strong and weak field regions. The short-wavelength component is thought to come from
the holes' quasinormal modes
ringing down, emitting gravitational radiation with wavelengths on the order of
their masses, as they relax from their initial, close to spherical state to their desired
tidally deformed state (see, e.g.,~\cite{Boyleetal}). Our data include the
Newtonian
quadrupole and octupole tidal deformations each hole induces on the other, as
well as the $1$PN
corrections to the quadrupole deformations.

The tidal deformations are contained in perturbed Schwarzschild metrics, given (in the
horizon-penetrating coordinate system we use) in Eqs.~\eqref{hCS} and~\eqref{hCS_pieces}.
The tidal fields are fixed by asymptotically matching these Schwarzschild metrics to an
$O(v^4)$ PN metric, given in Eqs.~\eqref{gPN}; expressions for the tidal fields around
hole~$1$ are given in Eqs.~\eqref{TFs}. One also needs to introduce a coordinate
transformation in order to put the black hole metrics in the same coordinate system as
the PN metric. This transformation is also determined (perturbatively) by the matching;
instructions for putting it together around hole~$1$ are given in Sec.~\ref{CT}.
Instructions for converting all of these results to the region around hole~$2$ are given
at the beginning of Sec.~\ref{Matching}.

The PN metric mentioned above treats
retardation perturbatively and thus becomes inaccurate far from the binary. In that
region, we thus use a version of the PN metric that includes retardation explicitly (but
also uses a multipolar decomposition, so it does not provide the desired accuracy closer
to the holes). This metric is given in Eqs.~\eqref{FZ_metric}. Due to
retardation, one needs to know the binary's past history accurately in order
to obtain the far zone metric accurately far from the binary. This past history is
computed in Sec.~\ref{phasing} to the highest PN order possible with current results.
The contributions of these terms are of equal or higher order than some of
our uncontrolled remainders,
but their inclusion is necessary if one wishes to obtain, e.g., the correct phasing for the
outgoing radiation.

We have also added other formally higher-order terms to the
metrics, including all the readily available PN results, along with a
resummation of the
black hole backgrounds in the PN metric. We found that certain of these terms improved the
constraint violations in various regions; other of these additional terms are
expected to improve evolutions of the data. The resummation is given in
Sec.~\ref{BR}; all the remaining higher-order terms are discussed in
Sec.~\ref{InclRad}. The transition functions that we use to
stitch the metrics together smoothly are given in Sec.~\ref{Transitions}. We
have constructed these transition functions so that they satisfy the so-called
Frankenstein theorems~\cite{Frankenstein}. The resulting merged metric is thus
guaranteed to have constraint violations whose formal order is no larger than
the constraint violations of the individual metrics.
See
Appendix~\ref{metric_comp} for the technical details of how we compute the metrics. We have
generated {\sc{Maple}} scripts and C code that produce initial data for a
nonspinning binary of any mass ratio and initial separation in a quasicircular
orbit. These are available at~\cite{Wolfgangs_website}.

\begin{figure}[htb]
\epsfig{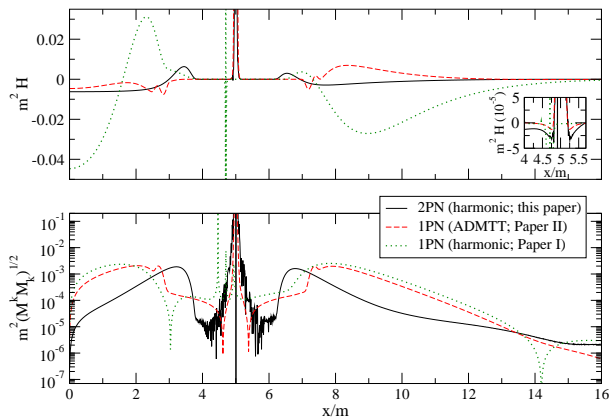} 
\caption{\label{old_data_comp} The Hamiltonian constraint and norm of the momentum
constraint along the $x$-axis around hole $1$ for this paper's data along with the data from
Papers~I and~II. (In the norm of the momentum constraint, the index is lowered using the
metric in question.) All of these were computed for an equal-mass binary with a coordinate
separation of $10$ times its mass. In the inset, we zoom in to show how the Hamiltonian
constraint violations behave close to the horizon.
Note that the data from Paper~II are in a different coordinate
system throughout, and that the black hole background in Paper~I's data is in a
different (and not horizon penetrating) coordinate system.}
\end{figure}

\begin{table}
\begin{tabular}{ccccccc}
\hline\hline
 & $m^{3/2}\|H\|_2$ & $m^2\|H\|_\infty$ &
$m^{3/2}\|\vec{M}\|_2$ & $m^2\|\vec{M}\|_\infty$\\
 & $(10^{-2})$ & $(10^{-2})$ & $(10^{-3})$ & $(10^{-3})$\\
\hline
This paper & 1.565 & 0.631 & 2.973 & 1.879\\
Paper II   & 1.566 & 0.758 & 4.627 & 2.065\\
Paper I    & 9.149 & 4.466 & 5.837 & 2.531\\
\hline\hline
\end{tabular}
\caption{\label{constraint_norm_old_data} The $L^2$ and sup norms (denoted by $\|\cdot\|_2$ and
$\|\cdot\|_\infty$, respectively) of the Hamiltonian and momentum constraints ($H$ and
$\vec{M}$) for an equal-mass binary with a coordinate separation of $10$ times its total
mass, $m$. (The norms of the momentum constraint include the $L^2$ $3$-vector norm
$\sqrt{M^kM_k}$, where the index is lowered using the metric under consideration.)
These are computed along the
$x$-axis outside the unperturbed horizons of the holes, but inside the interval
$[-16.4m,16.4m]$.}
\end{table}

The constraint violations give a measure of how much more accurate this
paper's data are
compared with those constructed in previous papers (viz.,~\cite{YTOB}
and~\cite{YT}, which we refer to as Papers~I and~II). We plot the
constraint violations of the three sets of data in Fig.~\ref{old_data_comp}.
[See, e.g., Eqs.~(14)--(15) in~\cite{CookLRR} for expressions for the constraint equations.]
N.B.: While the Hamiltonian constraint violations of this paper's data are
larger close to the hole than those of either of the previous papers' data, this is due to
our inclusion of the full time dependence of the tidal fields, as discussed in
Appendix~\ref{IZ_comp}; the motivation for doing this, even at the cost of larger constraint
violations, is given in Sec.~\ref{InclRad}. If one does not include these higher-order
terms, then the new data's constraint violations are smaller close to the hole than those
of either of the previous papers' data. See Fig.~\ref{IZ-comparison} for a comparison of the
constraint violations of the inner zone metric with and without full time dependence in the tidal fields. For a simple, quantitative comparison
of the overall constraint violations, we can consider their $L^2$ and sup
norms. These are presented for all three sets of data in
Table~\ref{constraint_norm_old_data}
and show, as expected, that the new data have smaller overall constraint violations than
either of the previous papers' data. Of course, the $L^2$ norm of the new data's Hamiltonian
constraint violations is only very slightly less than that of the data from Paper~II, but
this is probably to be expected: See Sec.~\ref{CV} for some discussion of this, as well as
further details about the comparison plot and table.

\subsection{Conformally curved initial data}
\label{ccdata}

The problems with conformally flat data have inspired a variety of
constructions of 
conformally curved binary black hole data, for which there are two general
approaches.
One approach---the one we have taken---is primarily motivated by a desire for astrophysical
realism, seeing the spurious radiation as an indicator of the failure of conformally flat
initial data to model the desired spacetime accurately enough. These constructions
so far have restricted
their attention to nonspinning binaries and used the PN approximation to include
the binary's physics. (Of course, true astrophysical binaries are expected to have
significant spins, so the consideration of nonspinning binaries is merely a technical
convenience appropriate for initial attempts at constructing astrophysically realistic
data.) The other approach is primarily concerned with reducing the junk
radiation in practice with relatively simple choices for initial data (viz., a superposition
of boosted black holes). This approach is often geared primarily towards spinning black
holes, since the amount of
spurious radiation increases with the spin of the holes. (This is due to the
nonexistence of a conformally
flat slicing even of an isolated Kerr black hole with nonzero
spin---see~\cite{PhysRevD.61.124011, Valiente_Kroon}.)

Besides our present work, other constructions in the first category are
this work's antecedents (\cite{Alvi, YTOB, YT}, discussed in the next
subsection), along with the approaches
of Nissanke~\cite{Nissanke:2005kp} and Kelly \emph{et al.}~\cite{Kellyetal}. (The
latter two only use the PN approximation in their construction, so the resulting data
cannot accurately describe the spacetime near the holes.)
Nissanke,
building on the work of Blanchet~\cite{Blanchet_ID} (who constructed initial data for a
head-on collision of initially stationary holes), obtains explicit analytic
expressions for the $3$-metric and extrinsic curvature from the $2$PN metric. (N.B.:
These constructions only use the version of the PN metric that treats retardation
perturbatively, so the resulting data rapidly lose accuracy away from the binary.)
Kelly \emph{et al.}~\cite{Kellyetal} extend the work of Tichy \emph{et al.}~\cite{Tichyetal} to give
initial data from the $2.5$PN ADMTT metric that are valid through $O(v^4)$ and thus
contain the binary's outgoing radiation in the far zone. (See Sec.~\ref{PNrad} for a
discussion of why gravitational radiation is present at that order.) They do not obtain
data that are valid through $O(v^5)$, even though the $2.5$PN metric contains the
$O(v^5)$ terms in all its components, since one would need the $O(v^6)$ terms in the
spatiotemporal components to obtain initial data valid through $O(v^5)$: See
Sec.~\ref{Comparisons} for further details. In order to obtain these data, they have to
evaluate an integral numerically, so their data are not completely analytic. The original
data in Tichy \emph{et al.}~\cite{Tichyetal} are completely analytic, though they use the
version of the ADMTT metric available in the literature, which gives expressions for the
$O(v^4)$ and $O(v^5)$ pieces of the transverse-traceless (TT) part of the metric that are
only valid in the near zone. Tichy \emph{et al.}'s data were thus only valid
through $O(v^3)$ far from the holes. Kelly \emph{et al.}\ calculated the additional terms
necessary for the TT pieces to be valid through $O(v^4)$ in the far zone.

The most recent progress in the second category for nonspinning black holes
is Lovelace's~\cite{Lovelace} construction and evolution of superposed
Schwarzschild data. (There is a companion construction and evolution of
superposed spinning black hole data in Lovelace \emph{et al.}~\cite{Lovelaceetal}.)
Lovelace uses the superposed black hole data as free data for a constraint
solver and finds that the resulting data produce less spurious radiation than
conformally flat data. There are also older
constructions in a similar vein~\cite{MHS, MM, Marronettietal}, though their 
data have only been evolved in head-on collisions without first solving the
constraints~\cite{Sperhake}. In this case, it was found that there was \emph{more}
spurious radiation with the superposed data than with Brill-Lindquist
data. The Kerr puncture data presented by Hannam \emph{et al.}~\cite{Hannametal} (who use Dain's work in~\cite{DainPRL, Dain} to construct
superposed Kerr free data in puncture form for a constraint
solver) have been shown to reduce the
junk radiation for head-on collisions of spinning black holes. However, the
given construction is only applicable to
the case of a head-on collision of initially stationary holes: It has not been extended to
give the holes linear momenta.
The freedom inherent in these superposed black hole initial data constructions has been
studied in~\cite{PCT}, and a helical Killing vector version of superposed data is
constructed in~\cite{YCSB}. But neither of those papers' data sets
have been evolved, to our knowledge.

While the use of a conformally curved Kerr metric near each of the holes is
likely responsible for most of the reduction of spurious radiation seen in the case of a
spinning binary, it is not
clear, physically, what feature of Lovelace's data is reducing the spurious radiation in the
Schwarzschild case.
Furthermore, it is known~\cite{Nissanke:2005kp} that
the superposed nonspinning Kerr-Schild metric differs from the PN metric at $2$PN, due to
the lack of interaction terms (e.g., terms that involve products of the distances from the
field point to each of the holes). Such a disagreement with the PN metric is probably
characteristic of all superposed initial data constructions, thus limiting their ability
to substantially reduce the junk radiation.

Evolutions of conformally curved binary initial data are uncommon, and most
have restricted their
attention to the computationally cheap case of head-on collisions, as seen above. Only
Lovelace~\cite{Lovelace} and Lovelace \emph{et al.}~\cite{Lovelaceetal} have evolved
superposed black hole data for an orbiting
binary. For the PN data, such head-on tests are mostly inapplicable: Only Nissanke's
data set is not already specialized to a (quasi)circular orbit. But even though her data
(which reduce to Blanchet's in the limit of a head-on collision of initially stationary
black holes) could be tested using a head-on collision, they have not been evolved, to our
knowledge.

In fact, the only evolution of PN initial data of which we are aware is that of Kelly
\emph{et al.}~\cite{Kellyetalev}.
They evolved the data they obtained in~\cite{Kellyetal} (as well as the
original version without waves from~\cite{Tichyetal}), necessarily doing so for
an orbiting binary, since their data were derived under the assumption of a
quasicircular orbit.
They found what na\"{\i}vely appears to be a slightly
better reduction of the initial spurious radiation than that seen with
Lovelace's superposed
black hole initial data, even without first solving the constraint equations. (Their data
satisfy the constraints approximately, but not exactly.)
However, the results are not directly comparable:
Most importantly, Lovelace begins his evolution with a separation of about twice that
with which Kelly \emph{et al.} begin their evolution. Thus, while Kelly \emph{et al.}\
see a slightly larger
reduction in the maximum amplitude of their junk radiation, even their smallest
amplitude is larger than that from the conformally flat data with which
Lovelace is comparing his data's performance. In addition, Kelly \emph{et al.} 
are comparing their data's performance with puncture data, while Lovelace is using
conformal thin-sandwich excision data (constructed in the manner of Cook and
Pfeiffer~\cite{CP}) as a benchmark. (There are also
other important differences, such as the extraction radius used and the mode of the
junk radiation that is reduced most substantially.)

It is also important to realize that Kelly \emph{et al.}'s data only reduce the
low-frequency component of the junk
radiation: The high-frequency component visible when using conformally flat puncture data
is still present in the evolution of Kelly \emph{et al.}'s data. This is not
surprising, since their initial data make no attempt to include
accurate tidal deformations on the holes. (And, indeed, their data's constraint
violations are largest near the holes.) In addition, they obtain nearly identical
junk radiation when evolving
the data with and without waves: The only difference is that the spurious radiation is
superposed over the outgoing wave train when they evolve the data
with waves; the spurious radiation itself appears to be unchanged. This, again, is what
should be anticipated, since one expects the junk to be generated primarily in the strong
field region near the holes, and Kelly \emph{et al.}'s data (with waves in the far
zone) have the same accuracy in the strong field region as Tichy \emph{et al.}'s original
data (without waves in the far zone).

As mentioned in the previous subsection, the data set we construct here should help reduce
\emph{both} components of the junk radiation, since it includes an accurate description of
the spacetime near the holes (including the quadrupolar and octupolar tidal deformations),
matched to the PN metric through $O(v^4)$.
We also offer an extension to these data that are accurate through $O(v^5)$ in the PN
portions of the timeslice (i.e., not too close to the holes). This extension
includes certain other
higher-order terms, as well, including higher-order corrections to the
trajectories. While these ``extra'' higher-order
terms do not increase the data's formal accuracy, even in just the PN portions,
such terms will likely improve the accuracy of the data in
practice. This extension will also allow for a more direct comparison with
Kelly \emph{et al.}'s results, since they included the $O(v^5)$ terms in the
spatial metric in the near zone (though not the matching $O(v^6)$ terms in the
extrinsic curvature, which we have). Such a comparison would
reveal how much of the spurious radiation is due to the initial data's failure
to include the correct tidal deformations.

\subsection{Specifics of our approach and its relation to other work}

With currently available technology, if one wishes to generate initial data
that include the holes' tidal deformations or the binary's outgoing radiation,
it is necessary to allow the initial data (as first constructed) to be merely
an approximate solution of the constraint equations. (One can always use these
approximate data as free data for a constraint solver and thus obtain an exact
solution of the constraints, to numerical precision.) Since the
post-Newtonian approximation has been developed to a very high order, it is
an obvious choice for the description of the binary's spacetime. Indeed, an
explicit expression for the metric to $2.5$PN order (with a perturbative
treatment of retardation) is given by Blanchet, Faye, and Ponsot
(BFP)~\cite{BFP}.

However, one cannot obtain accurate initial data throughout a timeslice of the
binary's spacetime using just the PN metric, since the PN approximation breaks down near
the holes: The PN approximation is a weak-field approximation (due to the
post-Minkowskian iteration in powers of $G$, Newton's gravitational constant), in
addition to being a slow-motion approximation (the post-Newtonian expansion proper, which
formally proceeds as an expansion in $1/c$, where $c$ is the speed of light).
Moreover, the standard PN approximation (as presented, e.g., in BFP) treats retardation perturbatively. It thus becomes inaccurate quite rapidly as
one enters the radiation zone (i.e., when one is further than about a reduced gravitational
wavelength away from the binary's center of mass).

The resolution of both of these problems is to realize that there is an appropriate
approximate description of the spacetime in each of the regions where the standard PN
metric is no longer a good approximation: Near each of the holes, in the regions known as
the \emph{inner zones}, spacetime is well described by a perturbed black hole metric. (These zones, as well as the others we mention here, are defined more
precisely in Sec.~\ref{MatchingIntro}.)
In the radiation zone (or \emph{far zone}) there is another version of the PN metric that
incorporates retardation nonperturbatively. These are all readily available in
the literature to the order we need them. For the reasons described below, we
choose to use Detweiler's perturbed black hole metric~\cite{Det05} in addition
to BFP's PN metric. We also choose to put
together the far zone metric following the recipe and ingredients supplied by
Pati and Will~\cite{PW1, PW2}. (In addition, we use Blanchet's results~\cite{BlanchetLRR}
for the evolution of the binary's phase with radiation reaction: These
secular effects are important in obtaining the far zone metric accurately, due
to its dependence upon retarded time---see Sec.~\ref{phasing}.)

One then has to stitch all these spacetimes---the far zone, inner zones, and the
\emph{near zone} where the standard PN metric is valid---together into one
global approximate metric. (Of course, this metric will be global in space, but not in
time---i.e., it will only be accurate in a temporal
neighborhood of a timeslice of the binary's spacetime.)
This stitching-together proceeds in two steps: First, one uses the technique of
matched asymptotic expansions
to match the metrics at a formal level. This puts all the metrics
in the same coordinate system (up to uncontrolled remainders) and fixes any previously undetermined parameters (e.g., the holes' tidal perturbations) so that the metrics are asymptotic to each other in their
regions of mutual validity (the \emph{buffer zones}). The final, numerical merging of the
metrics is effected by
\emph{transition functions} that smoothly interpolate between the metrics in their
mutual buffer zones. The resulting merged metric is guaranteed to satisfy the Einstein equations to the same order as its
constituent metrics because the transition functions are constructed following the so-called Frankenstein theorems~\cite{Frankenstein}. (We have checked
this scaling explicitly in Sec.~\ref{CV}.)

Once one has obtained such approximate initial data, it is, of course, possible
to use them as the input to a numerical constraint solver, and thus obtain an exact
solution, to numerical precision. In fact, it is probably desirable to do so. The
idea when doing this is that if the input to the constraint solver satisfies the
constraints to some reasonably good tolerance, and describes the desired physics,
the ``exact'' solution one obtains after solving the constraints will not differ
too much from the input in its physical content. This is probably true regardless
of how one chooses to produce the ``exact'' solution to the constraints, as long
as the procedure modifies the initial guess in a reasonable way. (For instance,
the York-Lichnerowicz decomposition~\cite{CookLRR} multiplies the initial data by
an appropriate conformal factor and modifies the extrinsic curvature by the gradient of
a vector field so that the data satisfy the constraint
equations---both Pfeiffer \emph{et al}.~\cite{Pfeifferetal_CS}
and Tichy \emph{et al}.~\cite{Tichyetal}
have implemented this numerically without the assumption of
conformal flatness.)
However, it may be preferable to project the approximate initial data onto
the closest ``point'' on the constraint hypersurface (as measured by some
appropriate norm), possibly using the results of~\cite{Holstetal}.

If one chooses to evolve without solving the constraints (as did Kelly \emph{et al.}), then
one will not have a true vacuum evolution: The constraint violations will act as
matter (which may not satisfy any of the standard energy conditions). For instance, Bode
\emph{et al.}~\cite{Bodeetal} investigated the evolution of initial data that only
approximately satisfied the Hamiltonian constraint (though the momentum constraint was
satisfied exactly). They found that the holes accreted the negative Hamiltonian
constraint violations that surrounded them, decreasing their
masses. Additionally, the initial apparent horizon masses were larger than the irreducible
masses for the constraint-violating data, but equaled them for the constraint-satisfying
data.

\subsubsection{Comparisons with similar constructions}
\label{Comparisons}

Alvi~\cite{Alvi} was the first to attempt to construct binary black hole initial data by matching perturbed black hole metrics onto a PN metric, considering,
for simplicity, two nonspinning black holes in a circular orbit. (All subsequent attempts using this procedure, including the present one, have also restricted their attention
to this simplest case.) In performing this calculation,
he fixed the perturbation (encoded in multipolar tidal fields) on the holes \emph{a priori}, using its expected Newtonian quadrupole value, instead of reading it off from
the matching as we do here. This gives the same result for the lowest-order
pieces of the tidal fields as our method, though it does not offer
the opportunity of reading off the higher-order corrections, as we can.

Numerical experiments with Alvi's initial data~\cite{JB} demonstrated that they
were not
suitable for use in evolutions. This inspired Yunes \emph{et al.}~\cite{YTOB} (Paper~I) to revisit the problem and correct
various deficiencies in Alvi's method, such as an inconsistent order counting
(due to not including the terms needed to compute the extrinsic curvature to
the appropriate order; this is discussed below) and a lack of smoothness at the
joins of
the global metric (due to not actually performing asymptotic matching). Yunes
and Tichy~\cite{YT} (Paper~II) then improved this initial data construction (in the sense of
getting better numerical agreement between the metrics) by putting the near zone metric in a form that is very
close to the inner zone metric near the holes. This was done by applying some
resummation and using
ADMTT coordinates instead of harmonic coordinates for the PN metric.
They also constructed horizon-penetrating coordinates to give the
first usable initial data generated with this method. However, in both cases the initial
data were only valid through $O(v^2)$ and the tidal fields were still just the
lowest-order ones Alvi had obtained.

The present calculation builds on all these previous attempts, computing fully matched
initial data through $O(v^4)$,
so as to include the pieces of the PN metric that break conformal flatness (and contain
gravitational radiation), and reading off the tidal fields (including the
$1$PN corrections to the quadrupole fields) from the matching. (We also
demonstrate that the mass parameters of the PN metric are equal to those of the
perturbed Schwarzschild metrics to the order considered: This is established
here more firmly than in Papers~I and~II in addition to being extended to
higher order.) In
addition, we have included the radiation zone portion of the metric to accommodate the
larger grid sizes common in current
simulations. This was done explicitly by Alvi and implicitly in Paper~II, due
to its use of the ADMTT PN metric, though neither of them included the effects
of radiation reaction on the binary's past evolution in their far zone metric.
The radiation zone was neglected completely in Paper~I, which used a
harmonic metric, as we do here. Additionally, all the previous
versions used a corotating coordinate system, while the current calculation
stays in inertial coordinates.

While we only obtain fully matched initial data through $O(v^4)$, we actually have to
carry out
the matching of the $4$-metric through $O(v^5)$ in order to do so: We need to
match the $O(v^5)$ pieces of the spatiotemporal components of the $4$-metrics
in order to obtain the extrinsic curvature consistently [see the discussion
after Eq.~(2) in Paper~I], and one needs to carry out the matching of all the
components in order to obtain the $O(v^5)$ piece of the coordinate
transformation.

Additionally, while our goal was simply to keep terms of quadrupolar order overall in the multipole
expansion, as was done previously, we found that it was necessary to match the lowest-order octupolar pieces in order to match the $1$PN corrections to the quadrupole
pieces consistently. This is discussed in Sec.~\ref{Matching1}. In fact, we have carried
out the matching of quadrupole pieces to the highest possible order to which it can be done
consistently without the inclusion of the hexadecapole tidal fields. (These
hexadecapole pieces can only be included with input from nonlinear black hole
perturbation theory, as the ``quadrupole squared'' pieces are of hexadecapole order.)
We also obtained the $1$PN correction to the electric octupole and the associated
piece of the coordinate transformation as a further application of our matching
procedure. However, we cannot obtain the other
$O(v^4)$ octupole pieces in the initial data (since they include the hexadecapole
tidal fields), so our knowledge of these
corrections does not allow us to increase the formal order to which our data
are valid.

Building on the work done in Paper~II, we have used horizon-penetrating coordinates for
the black holes from the outset. This requirement of horizon penetration is necessary for
numerical purposes. The coordinates need to be regular and the lapse positive in a
neighborhood of the horizon: Even though the spacetime near the singularity will be
excised or filled with matter, one needs to be able to evolve at least a small portion of
the spacetime inside the horizon.

At the same time, we want the coordinates for the black hole and PN metric to agree as
closely as possible before the matching has been performed: Close agreement
makes for simple matching algebraically and improves the numerical agreement of
the resulting matched metrics. Ideally, the coordinates
would agree exactly for an unperturbed black hole, though this is not compatible with the
requirement of horizon penetration, as standard PN coordinates (harmonic or ADMTT) are
not horizon penetrating. We thus attempted instead to obtain agreement between
the two coordinate systems to as high a PN order as possible.

These desiderata are satisfied if we use the fully harmonic version of Cook-Scheel coordinates~\cite{CS} for the black hole and standard (PN) harmonic
coordinates for the PN metric:\footnote{We use the term \emph{harmonic coordinates} to refer to \emph{any} coordinates $x^\alpha$ that satisfy
$\nabla_\alpha\nabla^\alpha x^\beta = 0$, not just PN harmonic coordinates. (Here $\nabla_\alpha$ is the covariant derivative associated with the metric under consideration
and indices are raised using that metric.) For an unperturbed Schwarzschild black hole of mass $M$, PN harmonic coordinates
are obtained by transforming the Schwarzschild radial coordinate $\sP$ to $R_\mathrm{PN} = \sP - M$ and thus retain the coordinate singularity at the
horizon present in Schwarzschild coordinates.} Cook-Scheel coordinates are horizon-penetrating, and in their fully harmonic version only differ from PN
harmonic coordinates for an unperturbed black hole at $O(v^4)$. See Appendix~\ref{Coord_comp} for an explicit comparison. This agreement was the best of any
of the horizon-penetrating coordinate systems present in the literature we consulted
(even if we also consider ADMTT coordinates for the PN metric). Of course, we
then adjust this coordinate system perturbatively so that it agrees with the
near zone
coordinate system to the order we have matched. However, we have checked that this
adjustment does not affect the coordinates' horizon penetration. The agreement
between the coordinate systems used in Paper~II was exact in the unperturbed case, while
it is not here. However, we decided against converting the PN metric to
horizon-penetrating coordinates for this version of the data---see Sec.~\ref{BR}
for further details.

The other choices for our ingredients were made for computational ease. We selected Detweiler's perturbed black hole metric~\cite{Det05} instead of Poisson's~\cite{PoissonPRL}
because Detweiler expresses the tidal fields in the Thorne-Hartle-Zhang (THZ) harmonic specialization of locally inertial coordinates~\cite{TH, Zhang}. This
gauge choice agrees better with the PN metric in harmonic coordinates than
does Poisson's
light-cone gauge. For the far zone, the results from the direct integration of the relaxed
Einstein equations (DIRE) approach were the obvious choice: Pati and Will~\cite{PW1, PW2} give an explicit recipe for computing the far zone metric to the order we need it,
along with all the necessary ingredients (except for a few that can be obtained from Will
and Wiseman~\cite{WW}). Even more conveniently, their expression is in the same (harmonic) coordinate
system as BFP's PN metric, so we do not have to determine a coordinate
transformation for the matching between the near and far zones. [We have
checked explicitly that the near and far zone metrics match through $O(v^5)$ in
all components.] One also needs to know the binary's past history in order to obtain the
far zone metric accurately, due to retardation: We calculate this in the PN approximation
using Blanchet's results~\cite{BlanchetLRR}.

We could have used Alvi's result for the far zone metric~\cite{Alvi} if we had
only been interested in obtaining initial data valid through $O(v^4)$: Alvi's
metric, computed following Will and Wiseman, is in the same coordinate system
as Pati and Will's
(and thus BFP's near zone metric), and is computed through $O(v^5)$ in all its
components. However, we needed higher-order results than he obtained, or
Will and Wiseman's results could give, to construct our extended data. This
is also why BFP's metric was preferable to the ADMTT
metric given by Jaranowski and Sch\"{a}fer~\cite{JS}: One needs to calculate
several integrals (one of which Kelly \emph{et al.}~\cite{Kellyetal}
had to resort to numerical techniques to evaluate) to obtain even the
$O(v^4)$ pieces of the far zone metric in this approach, while it is possible
to obtain the $O(v^5)$ pieces of the far zone metric merely by taking
derivatives using Pati and Will's results. [One is also able to obtain initial
data through $O(v^5)$ in the near zone using BFP's results, since they give the
$O(v^6)$ pieces of
the spatiotemporal components of the metric. This is not possible with Jaranowski and
Sch\"{a}fer's results.]

\subsubsection{Comparison with Taylor and Poisson's determination of the tidal fields}

Our method for determining the tidal fields can be compared and contrasted with that employed
in the recent calculation by Taylor and Poisson~\cite{TP}. Most importantly, Taylor and
Poisson's aims are slightly different and more general: They have carried out the matching of
a single black hole to an arbitrary $1$PN metric (expressed in terms of
potentials), and used the standard PN prescription (for obtaining equations of
motion) for which orders to keep in the various components, as opposed to our
``initial data prescription.'' They thus keep terms through $O(v^4)$ in the purely temporal
component of the metric, $O(v^3)$ in the spatiotemporal components, and $O(v^2)$ in the
purely spatial components, while we keep terms through $O(v^5)$ in all
components. [While
the $O(v^5)$ terms in the purely temporal and purely spatial components are needed to
obtain the $O(v^5)$ piece of the coordinate transformation, which itself is necessary
for obtaining the initial data to a formal accuracy of $O(v^4)$, those components
themselves do not increase the formal accuracy of the resulting initial data.] We also
expand the near zone metric in multipoles before matching, while Taylor and
Poisson do not.

But the
general setups have some superficial similarities: Both approaches use harmonic coordinates for the PN metric and THZ
coordinates~\cite{TH, Zhang} for the black hole perturbation. However, Taylor and Poisson
transform Poisson's perturbed black hole metric~\cite{PoissonPRL} to the THZ gauge instead of
using Detweiler's result~\cite{Det05} directly.
Additionally, they use PN harmonic coordinates for the black hole background, since they have
no need for their coordinates to be horizon penetrating.

The actual determination of the tidal fields is carried out in a very different manner in the
two calculations: Taylor and Poisson first introduce the most general coordinate
transformation that preserves the post-Newtonian form of the metric. They then specialize it
so that it transforms harmonic coordinates to harmonic
coordinates, and apply it to the PN potentials. After this, they decompose the transformed
potentials into irreducible pieces, and can then finally
use the matching conditions to determine expressions for the tidal fields in terms of the potentials.
They also obtain the black hole's
equations of motion and the previously undetermined pieces of the coordinate
transformation from the matching conditions.

We, however, have adopted a more brute force approach that assumes nothing about the
coordinate transformation \emph{a priori}, except its zeroth-order value. (In fact, even the
zeroth-order value can be shown to be constrained by the matching.) However, it
\emph{does} assume that the PN point particle trajectories are valid for black holes, as
is required by the strong equivalence principle: Taylor and
Poisson demonstrate that this is indeed the case (to the order they have matched). Our method also requires
no decomposition, though it makes heavy use of the computer algebra system {\sc{Maple}}
and the associated tensor manipulation package {\sc{GRTensorII}}~\cite{GRTensor}. We use
the gauge invariance of the linearized Riemann tensor and linear independence to
separate out the portions of the equations that determine the tidal fields from those
that determine the coordinate transformation. See Sec.~\ref{Matching1} for a detailed
presentation of our algorithm.

With this method, we obtained expressions for the tidal fields that can be compared with
those of Taylor and Poisson: The pieces that we both computed agree exactly. See
Appendix~\ref{TFTP} for a comparison, including explicit expressions for the tidal fields we
obtained. While our expressions do not have the full generality of Taylor and Poisson's,
they \emph{do} include the $1$PN corrections to the magnetic quadrupole and electric octupole
fields (for a circular orbit), neither of which Taylor and Poisson computed.

\subsection{Structure of the paper}

We begin by giving an overview of our matching procedure in Sec.~\ref{MatchingIntro} and then present expressions for the inner and near zone metrics in
Secs.~\ref{IZ} and~\ref{NZ}. In Sec.~\ref{NZ}, we also consider two relevant aspects of
the PN metric, viz., conformal flatness breaking and gravitational radiation effects.
Next we discuss the specifics of our matching procedure and read off the matching
parameters and coordinate transformation order-by-order in Sec.~\ref{Matching}.
We compute the far zone metric in Sec.~\ref{FZ}, where we also discuss the
PN results we use to obtain the effects of radiation reaction on the binary's
evolution. In Sec.~\ref{InclRad} we give an overview of the
construction of an extension of this data set that is valid through $O(v^5)$
in the near and far zones, in addition to including various other higher-order terms.
Then we stitch the metrics together
numerically in Sec.~\ref{NMatching}, first resumming the near zone metric to improve its
strong field behavior (and thus the matching), then constructing transition functions to
stitch all the metrics together smoothly, and finally considering the constraint
violations of the resulting merged metric. 
Lastly, we conclude and summarize in Sec.~\ref{Conclusions}.

We present various ancillary results and technical details in the
appendices: Appendix~\ref{Coord_comp} compares Cook-Scheel and PN harmonic
coordinates. We provide explicit expressions for the tidal fields and some related
discussion in Appendix~\ref{TidalFields}, along with the calculation of the fourth order
pieces of the octupole tidal fields and the polynomial part of the associated
coordinate transformation. In Appendix~\ref{ExtraTerms}, we give the
details of our calculation of the higher-order extension to the data, and in
Appendix~\ref{metric_comp} the precise details of how the metrics are
implemented numerically.

\subsection{Notation and Conventions}
\label{NC}
\emph{Units}: We use geometrized units with $G = c = 1$ throughout. ($G$ is
Newton's constant and $c$ is the speed of light.)

\emph{Binary parameters}: The binary's orbital velocity is $v$, its orbital angular
velocity is $\omega$, and its coordinate separation is
$b$. The masses of the holes are $m_1$ and $m_2$; their total mass is
$m := m_1 + m_2$.

\emph{Orbit terminology}: When we are only considering terms through $O(v^4)$ and the PN equations of motion
are thus conservative, we shall refer to the binary's orbit as circular. When we are
considering nonconservative terms [at $O(v^5)$ and higher] but find that their effects can
be ignored in the current portion of our calculation, we shall refer to the binary's orbit
as (quasi)circular. When we construct the extension to our data in Sec.~\ref{InclRad} and
Appendix~\ref{ExtraTerms} or are considering the binary's past history as
encoded in the far zone metric in Sec.~\ref{phasing} and can no longer ignore radiation reaction, we
refer to quasicircular orbits.

\emph{Indices}: We use the standard convention that Greek letters denote spacetime indices, while lowercase Roman letters
(here $k$, $l$, $p$, $s$, $u$, and $v$) denote spatial indices. In Sec.~\ref{IZ}, uppercase Roman letters denote indices on the $2$-sphere, but in Secs.~\ref{MatchingIntro}, \ref{NZ}, and~\ref{Transitions} they label the holes, as well as, by extension, the zones into which we divide the
data's timeslice. [In the latter context,
the notation ``$+\, (1 \leftrightarrow 2)$'' denotes that the preceding expression
is to be added to itself with the labels $1$ and $2$ switched.] In
Sec.~\ref{FZ} and Appendix~\ref{ExtraTerms}, $Q$ denotes a multi-index. Except where
otherwise noted (e.g., in Secs.~\ref{PNCF}, \ref{FZ}, and~\ref{NMatching}), spacetime
indices are raised and lowered using the Minkowski metric
$\eta_{\alpha\beta}$, so spatial indices are
raised and lowered using the Kronecker delta $\delta_{kl} := \mathrm{diag}(1,1,1)$ (the symbol ``$:=$'' indicates a definition). We may even freely raise and lower spatial
indices within expressions for notational convenience, particularly in Sec.~\ref{NZ}. The summation convention is always in force, and may even be applied to spatial indices
that are at the same level, particularly in Sec.~\ref{FZ}. Parentheses, square brackets, and angle brackets on
indices denote symmetrization, antisymmetrization, and the symmetric trace-free projection, respectively. For instance, $A_{(\alpha\beta)} :=
(A_{\alpha\beta} + A_{\beta\alpha})/2$, $A_{[\alpha\beta]} := (A_{\alpha\beta} - A_{\beta\alpha})/2$, and $A_{<kl>} :=
(A_{kl} + A_{kl})/2 - A^p{}_p\delta_{kl}/3$. We use vertical bars to exclude indices from these operations---e.g.,
$A_{(\alpha|\beta|\gamma)} := (A_{\alpha\beta\gamma} + A_{\gamma\beta\alpha})/2$.

\emph{Arrays}: In addition to the ordinary $3$-dimensional Kronecker delta defined above, we also define a ``lowered
4-dimensional Kronecker delta,'' $\Delta_{\alpha\beta} := \mathrm{diag}(1,1,1,1)$. Our
conventions for the three- and four-dimensional Levi-Civita symbols
$\epsilon_{klp}$ and $\epsilon_{\alpha\beta\gamma\delta}$ are that $\epsilon_{123} =
\epsilon_{0123} = 1$.

\emph{Metrics}: As is usual, $\eta_{\alpha\beta}$ denotes the Minkowski metric; our signature is $(-,+,+,+)$. In Secs.~\ref{NZ} and \ref{Matching} (and Appendix~\ref{TidalFields})
$g_{\alpha\beta}$ denotes the near zone (PN) metric and $h_{\alpha\beta}$ denotes the inner zone (perturbed black hole)
metric. In Sec.~\ref{BR} and Appendix~\ref{Coord_comp}, $g_{\alpha\beta}$ (sometimes
with decorations) is also used for the unperturbed Schwarzschild metric. In
Sec.~\ref{FZ} (and the associated Appendix~\ref{ExtraTerms}), $g_{\alpha\beta}$ denotes the far zone metric, and $h_{\alpha\beta}$ its associated metric
perturbation.

\emph{Coordinates}: Inner zone coordinates are $\sX^\alpha = (\sT, \sX^k)$, with $\sP := \sqrt{\sX_k\sX^k}$
(Schwarzschild); $X^\alpha = (T, X^k) = (T, X, Y, Z)$, with $R := \sqrt{X_kX^k}$ (Cook-Scheel). PN harmonic
coordinates (used in the near and far zones) are $x^\alpha = (t, x^k) = (t, x, y, z)$, with $r := \sqrt{x_kx^k}$; $r^\alpha$ denotes
just the spatial coordinates [i.e., $r^\alpha = (0, x^k)$]. Unit vectors are denoted by ``hats.''
For instance, $\hat{t}^\alpha, \hat{x}^\alpha, \hat{y}^\alpha$, and $\hat{z}^\alpha$ are the Cartesian PN coordinate
basis vectors corresponding to indices 0, 1, 2, and 3, respectively.
Because we are interested in expanding in the distance from hole~1, we
also define ``tilded'' coordinates with their origin at hole~1's position at $t = 0$, viz., $\tx^\alpha := x^\alpha - (m_2/m)b\hat{x}^\alpha$, $\tx
:= x - (m_2/m)b$, $\tr := \sqrt{\tx_k\tx^k}$, and $\tr^\alpha := r^\alpha - (m_2/m)b\hat{x}^\alpha$. Spatial vectors (or the
spatial parts of spacetime vectors) will be denoted either with an arrow or a spatial index.

\emph{Norms}: The Euclidean norm for spatial vectors is denoted by $\|\cdot\|$ (so, e.g., we could write the definition of $R$ above as $R := \|\vec{X}\|$).

\emph{Derivatives}: All partial derivatives are taken with respect to harmonic PN coordinates, so $\partial_\alpha := \partial/\partial x^\alpha$. Overdots denote differentiation
with respect to $t$ (i.e., PN harmonic time). While in Sec.~\ref{IZ}, overdots properly denote differentiation with respect to $T$
(i.e., the Cook-Scheel time coordinate), as we shall see, this is equivalent to
differentiation with respect to $t$ to the order we are considering. We use the shorthand $\partial_{\alpha\beta} := \partial_\alpha\partial_\beta$.

\emph{Order counting}: We have to deal with two expansions here, since we are
performing asymptotic matching. For our purposes, the PN
expansion can be treated as an expansion in $\sqrt{m_2/b}$. We use
$m_2$ (instead of $m_1$ or $m$) for convenience, as we concentrate on matching
around hole $1$. (We can do this without loss of generality, by the
symmetry of exchanging labels 1 and 2.) The
black hole
perturbation expansion is in multipoles, so around hole $1$ it can be treated as an expansion in $\tr/b$. Here we treat
$t/b$ as the same order as $\tr/b$, as required by the slow time variation assumption made in the derivation of the perturbed
black hole metric. This means that we only match in a neighborhood of $t = 0$.

\emph{Order notation}: It will be important for us to have a compact
notation for, e.g., the $j$th term of the expansion of some
arbitrary quantity. For this, we use the ``order projection'' notation, which
has three different flavors: Since the slow-motion PN expansion can be considered our
primary expansion (and is the expansion to which we refer when we simply call
something $j$th order), the most commonly used flavor will be
$\od{\cdot}{j}$. This denotes the coefficient of $(m_2/b)^{j/2}$
in the (asymptotic) power series expansion of its argument.
For the reasons discussed in Sec.~\ref{Matching1}, this includes
multipoles through octupolar order for $j \in \{2,3\}$ but only through quadrupole order
for $j \in \{4,5\}$. (In principle, we include all the multipoles for
$j \in \{0,1\}$, since there are only monopole contributions.) For the times
when we need to single out a particular multipole for consideration, we shall use $\od{\cdot}{j,n}$, which denotes the
$2^n$th multipole-order piece of $\od{\cdot}{j}$. Finally, when we want to include all of the multipolar orders through $2^n$
we shall use $\od{\cdot}{j,\le n}$.

\emph{Polynomial and nonpolynomial parts}: All the expressions we consider when
performing the matching of the inner and near zones can be written as the sum of a
polynomial in $\tx^\alpha$ and nonpolynomial terms of a particular form,
viz., polynomials in $\tx^\alpha$ multiplied by $\tr^n$, $n \in \mathbb{Z}
\setminus \{0, 2, 4, \ldots\}$. It will often be convenient to
divide expressions up into their polynomial and nonpolynomial parts, since these can be treated separately, by linear independence. We shall thus denote the polynomial and
nonpolynomial parts of an expression by the superscripts P and NP, respectively.

\emph{References}: \cite{YTOB} and~\cite{YT} will be referred to as Papers~I and~II,
respectively.

\section{Asymptotic Matching}\label{MatchingIntro}

The technique of asymptotic matching is a standard one in the analysis of
multiscale and singular perturbation problems, allowing one to relate and combine approximate solutions
that are valid on different scales~\cite{BO}. It has been used in general
relativity to
obtain, e.g., PN equations of motion---\cite{TP} contains the most recent of
these calculations---and the radiation zone metric of a binary
system~\cite{PB}. (See Paper~I and~\cite{TP} for further
discussion and references.) Here we specialize our discussion to the case of a binary
black hole.

\begin{figure}[htb]
\epsfig{file=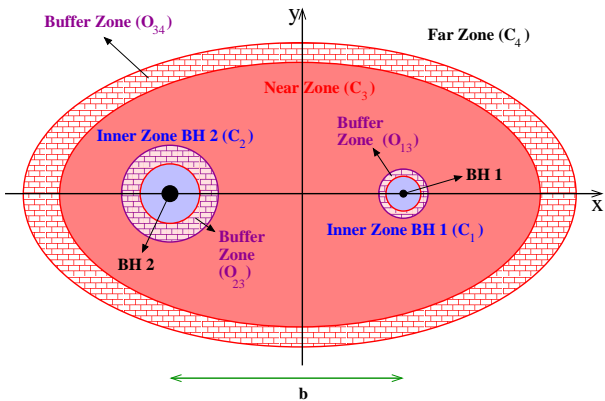,width=8cm,clip=true} 
\caption{\label{ZD} A diagram of the zones into which we divide the binary black hole timeslice. The two black
holes (BH $1$ and BH $2$, denoted by filled-in circles) lie on the $x$-axis, surrounded by their respective inner zones ($C_1$ and $C_2$) and inner-to-near buffer zones ($O_{13}$ and $O_{23}$). (In
actuality, the black holes should be tidally distorted, along with their associated inner and buffer zones. We neglect this distortion in the above diagram for
simplicity.) The near zone, $C_3$, covers the orbit of the binary and is surrounded by the far zone, $C_4$, and the near-to-far buffer zone, $O_{34}$.}
\end{figure}

A timeslice of a binary black hole spacetime
divides naturally into four primary zones and three secondary
\emph{buffer zones}: See Fig.~\ref{ZD} for an illustration. In practical work, the
boundaries of all of these zones are necessarily only given approximately,
since we do not currently possess sharp estimates for the error bars of the
approximations used
to describe this spacetime. There are two \emph{inner zones} around the black holes, given
by $r_A \ll b$, where $r_A$ is the distance from (the point particle associated with)
hole $A$: Here the spacetime is
well described by a perturbed black hole. Surrounding (and partially overlapping) these
is
the \emph{near zone}, where the standard (harmonic coordinate) PN metric is
valid---i.e., not too close to the holes, yet not so far
away that retardation cannot be treated perturbatively. This is given by
$r_A \gg m_A$ and
$r \lesssim \lambdabar$, where $m_A$ is the mass of hole $A$, $r$ is the distance
from the system's center-of-mass, and $\lambdabar = b/2v$ is the reduced
characteristic wavelength of the binary's gravitational radiation. Finally, the
remainder of the timeslice (along with a little bit of the outer portion of
the near zone) comprise the \emph{far zone}, given by $r \gtrsim \lambdabar$,
where retardation can no longer be treated perturbatively, and spacetime is
described by a separate PN metric that accounts for this. (N.B.:
Different relations were given for the outer edge
of the near zone and inner edge of the far zone in Papers~I and~II, but the ones used
here are more accurate.) See
Table~\ref{Zone_boundaries} (in Sec.~\ref{Transitions}) for numerical values for the boundaries
of the zones for a particular equal-mass binary.

The three buffer zones are the portions of the timeslice where the preceding four regions
overlap. (Due to the ``fuzziness'' inherent in $\lesssim$ and $\gtrsim$, the near and
far zone can have a substantial overlap, despite formal appearances.) We restrict our
attention to cases in which the specified zones overlap in
the manner we shall describe, but in no other fashion (e.g., we do not want the two inner zones to overlap). There are two buffer zones where the two inner zones overlap the
near zone (thus these are given by $m_A \ll r_A \ll b$), and another buffer
zone where the near zone overlaps the far zone, which is very roughly an
annulus whose radius and thickness are both of order $\lambdabar$. (The order of the
thickness of the annulus is our choice---see Sec.~\ref{Transitions}: All that is required
formally is that it increase as $v$ decreases, corresponding to a larger realm of validity
of the near zone metric's pertubative treatment of retardation. Additionally, as is
indicated in the figure, this buffer zone would not be spherical in a more
nuanced description.)
These buffer zones are where we perform the formal matching that determines the coordinate
transformation and relations between parameters as well as where we stitch the metrics together
numerically.

This matching and stitching relies on the observation that if the metrics that are valid in
the various zones are all to be different approximations to the same (unknown)
global exact metric, then, considered as abstract tensors, they should agree with each other in their realms of mutual validity. More specifically, assume that there exists a
buffer zone in which two
approximate metrics $\mathbf{g}^{(1)}$ and $\mathbf{g}^{(2)}$ are both valid.
(We write these metrics without indices to emphasize that they are currently
being considered as abstract tensors.) Take their associated small parameters to be $\epsilon_1$ and $\epsilon_2$, respectively. (The buffer zone is then
defined to be the region in which these parameters are indeed small.)
Then make a bivariate expansion of the metrics in both small parameters. That is, take $\mathbf{g}^{(1)}$, which is already an expansion in $\epsilon_1$, and expand it in
$\epsilon_2$ as well; similarly expand $\mathbf{g}^{(2)}$ in $\epsilon_1$. The coefficients of both bivariate expansions, considered as abstract tensors, should then be
equal if the metrics describe the same spacetime. The resulting equations relate the parameters of the two metrics.

While the statement of this result in terms of abstract metrics is simple, in practice one
works with the metrics' coordinate components. In general, the coordinate
systems in which the metrics' components are known will not agree, so one first
chooses the coordinate system in which one of the metrics is expressed to be the
primary coordinate system. (For us, this will be the near zone's PN harmonic
coordinate system.) One then
applies an arbitrary coordinate transformation to the other metric in order to put
it in the same coordinate system as the first (to the order one matches). Equating
coefficients of the bivariate expansions of the components of the metrics
(including the coordinate transformation) is then equivalent to equating the
coefficients of the expansions of the abstract tensors. In this case, the
resulting equations will determine not only the relations between the metrics'
parameters, but also the arbitrary coordinate transformation. In our
calculation, each order's contribution to the coordinate transformation
contains some arbitrary pieces---i.e., pieces that can be set to any value
without affecting that order's matching. We have found that most of these are
determined by the matching at two orders higher.

\section{Inner Zone Metric}\label{IZ}

Detweiler's perturbed black hole metric~\cite{Det05} is derived under the slow-motion and weak-curvature assumptions of
Thorne and Hartle~\cite{TH}. Here one characterizes the strength of the perturbation by the length scale $\sR$, defined to
be the smallest of the radius of curvature, inhomogeneity scale, and scale of time variation of the external universe. For
convenience, these are generally all taken to be formally of the same order. (In the nomenclature of Thorne and Hartle, the
``external universe'' refers to the perturbing spacetime in which the black hole is placed, here consisting of the tidal
fields generated by the hole's binary companion.) For a post-Newtonian binary we have $\sR \sim \sqrt{b^3/m_2}$ [see, e.g., the discussion following
Eq.~(1.4b) in~\cite{TH}]. Here $b$ is the coordinate separation of the holes and $m_2$ is
the mass of the hole's companion (the one responsible for the
perturbing tidal fields). (We are thus specializing to the inner zone
surrounding hole~$1$ in giving this scaling, as well as the scalings of the
tidal fields below. One can obtain the scalings for the inner zone surrounding
hole~$2$ using the substitution $1 \leftrightarrow 2$.)

With these assumptions, the time derivative of a
quadrupole field is of octupolar order. Since one is only able to include quadrupole and octupole perturbations in
linear black hole perturbation theory, the time dependence of the tidal fields
in the metric is restricted to the quadrupole fields, and even there it is
only linear. Of course, it is possible to obtain the complete time dependence
of the tidal fields (for times much less than the radiation reaction
timescale), as Taylor and Poisson~\cite{TP} did, since we can compute the tidal fields
at any point in the orbit. (See Appendix~\ref{TidalFields} for further discussion.)
However, knowing this dependence does not let us compute the metric to a
higher formal accuracy.

In Detweiler's metric, the tidal perturbations are encoded by symmetric trace-free electric and magnetic tidal fields $\sE_{kl}$ (electric quadrupole), $\sE_{klp}$ (electric
octupole), $\sB_{kl}$ (magnetic quadrupole), and $\sB_{klp}$ (magnetic octupole). These come from the Thorne-Hartle-Zhang (THZ) harmonic specialization of
locally inertial coordinates (from~\cite{TH,Zhang}), which
Detweiler uses to express the perturbation of Minkowski space that his metric
approaches in the buffer zone.

As their names suggest, the quadrupole and octupole tidal fields formally scale as $\sR^{-2}$ and $\sR^{-3}$, respectively. However, this is only a formal scaling, since
it requires that one treat the radius of curvature,
inhomogeneity scale, and scale of time variation of the external universe as being of the same order [see, e.g., Eq.~(1.4b) in~\cite{TH}]. For a post-Newtonian
binary, these are not of the same order: The magnetic moments are a factor of $v$ smaller than the corresponding electric moments; the octupole moments are a factor of $v$
\emph{larger} than the formal scaling would suggest---see Eqs.~(2.1) and~(2.2) in \cite{TP} and the surrounding discussion. Specifically, the magnetic quadrupole tidal field on hole~1
is actually $\sim (m_2/b)^{1/2}(m_2/b^3)$ [cf.\ the expression given in Eqs.~(5.45b) and~(5.56b) of~\cite{BHTMPV}], and the
electric octupole tidal field on hole~1 is actually $\sim m_2/b^4$ [cf.\ Eqs.~(5.50) and~(5.56a) in~\cite{BHTMPV}]. In order to
simplify later expressions and make their expansion in $\tr/b$ explicit ($\tr$ denotes the distance from hole $1$), we define ``barred''
versions of the tidal fields around hole~1 with the scaling taken out, viz.,
\begin{align}
\bE_{kl} &:= (b^3/m_2)\sE_{kl},& \bB_{kl} &:= (b/m_2)^{1/2}(b^3/m_2)\sB_{kl},\nonumber\\
\label{TF_scaling}
\bE_{klp} &:= (b^4/m_2)\sE_{klp},& \bB_{klp} &:= (b/m_2)^{1/2}(b^4/m_2)\sB_{klp}.
\end{align}

Detweiler~\cite{Det05} presents the metric perturbation in Schwarzschild coordinates,
giving the portion involving the quadrupole fields, including their time derivatives
(which are actually of octupolar order), in his Eqs.~(G.6)--(G.11), and the
portion with the
octupole fields in his Eq.~(58).
Putting this all together, along with the background unperturbed Schwarzschild metric, we obtain a line element of
\begin{widetext}
\<\label{hSchw}
\begin{split}
h^\Schw_{\alpha\beta}d\sX^\alpha d\sX^\beta& =
-H^\Schw_{\sT^2} d\sT^2 + H^\Schw_k d\sT d\sX^k
- \frac{2}{3}\left[2\sP + M - \frac{4M^2}{\sP - 2M} + \frac{6M^3}{\sP(\sP - 2M)}\right]\sE'_{kl}\sX^k\sX^ld\sT d\sP\\
&\quad + \frac{\sP^2}{3(\sP - 2M)}\sC'_{klp}\sX^l\sX^pd\sX^kd\sP + H^\Schw_{\sP^2} d\sP^2 + H^\Schw_\sigma\sP^2\sigma_{AB}d\sX^Ad\sX^B + O(\sP^4/\sR^4),
\end{split}
\?
where the ``$\Schw$'' acts as a reminder that this is in Schwarzschild coordinates $\sX^\alpha = (\sT, \sX^k)$, with a radial coordinate of $\sP := \sqrt{\sX_k\sX^k}$.
Additionally, $M$ is the mass of the hole, $\sigma_{AB}$ denotes the metric on the
$2$-sphere, and we have defined
\<
\begin{split}
H^\Schw_{\sT^2} &:= 1 - 2\frac{M}{\sP} + \left[1 - 2\frac{M}{\sP}\right]^2\left[\sE_{kl} - 2M\log\left(\frac{\sP}{2M}\right)\sE'_{kl}\right]\sX^k\sX^l
- \left[\sP - 4M + 12\frac{M^3}{\sP^2} - \frac{16}{3}\frac{M^4}{\sP^3} - \frac{8}{3}\frac{M^5}{\sP^4}\right]\sE'_{kl}\sX^k\sX^l\\
&\quad\; + \frac{1}{3}\left[1 - 2\frac{M}{\sP}\right]^2\left[1 - \frac{M}{\sP}\right]\sE_{klp}\sX^k\sX^l\sX^p,\\
H^\Schw_k &:= \frac{4}{3}\left[1 - 2\frac{M}{\sP}\right]\left[\sC_{klp} - 2M\log\left(\frac{\sP}{2M}\right)\sC'_{klp}\right]\sX^l\sX^p -
\frac{4}{3}\left[\sP - 2M - 4\frac{M^2}{\sP} + 4\frac{M^3}{\sP^2} + \frac{8}{3}\frac{M^4}{\sP^3} + \frac{8}{3}\frac{M^5}{\sP^4}\right]\sC'_{klp}\sX^l\sX^p\\
&\quad\; + \frac{2}{3}\left[1 - 2\frac{M}{\sP}\right]\left[1 - \frac{4}{3}\frac{M}{\sP}\right]\sC_{klps}\sX^l\sX^p\sX^s,\\
H^\Schw_{\sP^2} &:= \left[1 - 2\frac{M}{\sP}\right]^{-1} - \left[\sE_{kl} - 2M\log\left(\frac{\sP}{2M}\right)\sE'_{kl}\right]\sX^k\sX^l
- \frac{1}{3}\left[1 - \frac{M}{\sP}\right]\sE_{klp}\sX^k\sX^l\sX^p\\
&\quad\; + \left[\sP - \frac{4M^2}{\sP - 2M} + \frac{4M^3}{(\sP - 2M)^2} - \frac{16}{3}\frac{M^4}{\sP(\sP - 2M)^2} - \frac{8}{3}\frac{M^5}{\sP^2(\sP - 2M)^2}\right]\sE'_{kl}\sX^k\sX^l,\\
H^\Schw_\sigma &:= 1 - \left[1 - 2\frac{M^2}{\sP^2}\right]\left[\sE_{kl} - 2M\log\left(\frac{\sP}{2M}\right)\sE'_{kl}\right]\sX^k\sX^l
+ \left[\sP - 6\frac{M^2}{\sP} - 4\frac{M^3}{\sP^2} + \frac{8}{3}\frac{M^4}{\sP^3}\right]\sE'_{kl}\sX^k\sX^l\\
&\quad\; - \frac{1}{3}\left[1 - 2\frac{M}{\sP} + \frac{4}{5}\frac{M^3}{\sP^3}\right]\sE_{klp}\sX^k\sX^l\sX^p.
\end{split}
\?
\end{widetext}
We have also defined $\sC_{klp} := \epsilon_{kls}\sB^s{}_p$ and $\sC_{klps}
:= \epsilon_{klu}\sB^u{}_{ps}$, for convenience, as the magnetic tidal fields only
appear in the perturbation in these dual forms. (The ``barred'' versions of these
quantities are defined analogously to $\bB_{kl}$ and $\bB_{klp}$.) Here $\epsilon_{klp}$
is the 3-dimensional Levi-Civita symbol, with $\epsilon_{123} = 1$. While we have not
indicated this explicitly, $\sE_{kl}$ and $\sC_{klp}$ both depend (linearly) on the null
ingoing Eddington-Finkelstein coordinate $V := \sT + \sP + 2M\log(\sP/2M - 1)$. (The
octupole fields are treated as constants, since their time derivatives are of
hexadecapole order.) Thus, primes denote derivatives with respect to $V$. We include the
$\sB'_{kl}$ contributions
here, even though they do not increase the formal accuracy of our initial data: We
fix the lowest-order piece of $\sB'_{kl}$ when we read off the $1$PN correction
to the electric octupole (in Appendix~\ref{TFHO}) as an application of our
matching procedure. We also are able to obtain the full time dependence of the
tidal fields (up to radiation reaction effects) \emph{a posteriori}---see the
discussion in Appendix~\ref{TFHO}. (Including such
formally higher-order pieces can improve the accuracy of the data in practice.)

There is one gauge subtlety here that Detweiler~\cite{Det05} does not address in his paper: If
one compares the expression for the THZ tidal perturbation in his Eq.~(53) to
that of the metric perturbation in his Eqs.~(56)--(57), it appears that the portions
of the metric perturbation involving time derivatives of the tidal fields do not
approach the analogous portions of the THZ tidal perturbation. The resolution of
this apparent discrepancy is that Detweiler made a tacit gauge transformation
(away
from the pure THZ gauge, though only affecting the time derivatives of the tidal
fields, to the order considered) to obtain the given compact expressions for the
metric perturbation~\cite{DetPC}.

We now transform this metric to the quasi-Cartesian form of Cook-Scheel harmonic
coordinates (from~\cite{CS}), which we denote by
$X^\alpha = (T, X^k)$, with $R := \sqrt{X_kX^k}$. The transformation is given by
\<\label{CS_trans}
\begin{aligned}
\sT   &= T - 2M\log\left\lvert\frac{R - M}{R + M}\right\rvert,\\
\sX^k &= \left[1 + \frac{M}{R}\right]X^k,
\end{aligned}
\?
so $\sP = R + M$. This comes from Cook and Scheel's Eqs.~(20), (41), and (43) upon noting that Boyer-Lindquist coordinates reduce to
Schwarzschild coordinates for a Schwarzschild hole.
In order to simplify our transformation, we use the relation $\sP^2\sigma_{AB}d\sX^Ad\sX^B = (1 + M/R)^2(dX_pdX^p - dR^2)$, obtained from
$d\sX_kd\sX^k = d\sP^2 + \sP^2\sigma_{AB}d\sX^Ad\sX^B$, to eliminate $\sigma_{AB}$, thus obtaining a full transformed line element of
\begin{widetext}
\<
\begin{split}\label{hCS}
h_{\alpha\beta}dX^\alpha dX^\beta &= -H_{T^2}dT^2 + H_{RT}dRdT +
\frac{16}{3}\frac{M^2}{R}\left[1 + \frac{M}{R} - \frac{2}{3}\frac{M^2}{R^2} -
\frac{2}{3}\frac{M^3}{R^2(R + M)}\right]\dot{\sC}_{klp}X^lX^pdX^kdT\\
&\quad + H^{[1]}_kdX^k\left[\left(1 - \frac{M^2}{R^2}\right)dT - 4\frac{M^2}{R^2}dR\right]
+ H^{[2]}_kdX^kdR + H_{R^2}dR^2 + H_\mathrm{trc}dX_sdX^s + O(R^4/\sR^4),
\end{split}
\?
where
\<\label{hCS_pieces}
\begin{split}
H_{T^2} &:= \frac{R - M}{R + M} + \left[1 - \frac{M}{R}\right]^2\left[(\sE_{kl} + T\dot{\sE}_{kl})X^kX^l +
\frac{1}{3}\sE_{klp}X^kX^lX^p\right] + \frac{4M^2}{(R + M)^2}\biggl[R - \frac{5}{3}\frac{M^2}{R}\biggr]\dot{\sE}_{kl}X^kX^l,\\
H_{RT} &:= \frac{8M^2}{(R + M)^2} + 8\frac{M^2}{R^2}\frac{R - M}{R + M}\left[(\sE_{kl} + T\dot{\sE}_{kl})X^kX^l  + \frac{1}{3}\sE_{klp}X^kX^lX^p\right]\\
&\quad\; - \biggl[\frac{4}{3}R + \frac{14}{3}M + \frac{8}{3}\frac{M^2}{R} - 2\frac{M^3}{R^2}
- \frac{104}{3}\frac{M^4}{R^2(R + M)} + \frac{80}{3}\frac{M^5}{R^2(R + M)^2} + \frac{32}{3}\frac{M^6}{R^2(R + M)^3}\biggr]\dot{\sE}_{kl}X^kX^l,\\
H^{[1]}_k &:= \frac{2}{3}\left[1 + \frac{M}{R}\right]\left[2(\sC_{klp} + T\dot{\sC}_{klp})X^lX^p +
\left(1 - \frac{1}{3}\frac{M}{R}\right)\sC_{klps}X^lX^pX^s\right],\\
H^{[2]}_k &:= \left[\frac{R}{3} + 2M + \frac{16}{3}\frac{M^2}{R} + \frac{26}{3}\frac{M^3}{R^2} - 11\frac{M^4}{R^3} - \frac{32}{3}\frac{M^5}{R^3(R + M)} -
\frac{64}{9}\frac{M^6}{R^3(R + M)^2}\right]\dot{\sC}_{klp}X^lX^p,\\
H_{R^2} &:= \sum_{n=1}^{3}\left(\frac{2M}{R + M}\right)^n - \frac{2M}{R} - \frac{M^2}{R^2} + \biggl[2\frac{M}{R} + 3\frac{M^2}{R^2} - \frac{M^4}{R^4}
- \frac{16M^4}{R^2(R + M)^2}\biggr](\sE_{kl} + T\dot{\sE}_{kl})X^kX^l\\
&\quad\; + \biggl[\frac{1}{3}\frac{M}{R} + \frac{1}{3}\frac{M^2}{R^2} - \frac{2}{5}\frac{M^3}{R^3} -
\frac{7}{15}\frac{M^4}{R^4} - \frac{1}{15}\frac{M^5}{R^5} - \frac{16}{3}\frac{M^4}{R^2(R + M)^2}\biggr]\sE_{klp}X^kX^lX^p + \biggl[\frac{16}{3}\frac{M^2}{R} + \frac{80}{3}\frac{M^3}{R^2}\\
&\quad\; + 28\frac{M^4}{R^3} + \frac{40}{3}\frac{M^5}{R^4} - \frac{176}{3}\frac{M^6}{R^4(R + M)} + \frac{72M^7}{R^4(R + M)^2} - \frac{32}{3}\frac{M^8}{R^4(R + M)^3} - \frac{32}{3}\frac{M^9}{R^4(R + M)^4}\biggr]\dot{\sE}_{kl}X^kX^l,\\
H_\mathrm{trc} &:= \left[1 + \frac{M}{R}\right]^2\biggl[1 - \left(1 + 2\frac{M}{R} - \frac{M^2}{R^2}\right)(\sE_{kl} + T\dot{\sE}_{kl})X^kX^l
- \frac{1}{3}\left(1 + \frac{M}{R} - \frac{M^2}{R^2} - \frac{1}{5}\frac{M^3}{R^3}\right)\sE_{klp}X^kX^lX^p\\
&\quad\; - 4\frac{M^2}{R^2}\left(R + 2M - \frac{2}{3}\frac{M^2}{R + M}\right)\dot{\sE}_{kl}X^kX^l\biggr].
\end{split}
\?
We have used the fact that the time dependence of $\sE_{kl}$ and $\sC_{klp}$ is only
linear (to the multipolar order we are considering) to write $\sE'_{kl} =
\dot{\sE}_{kl}$ and $\sE_{kl}(V) = \sE_{kl} +
[T + R + M + 2M\log(\sP/2M)]\dot{\sE}_{kl}$, with analogous expressions involving
$\sC_{klp}$. [This expansion removes the logarithms present in the expression for the
metric in Eq.~\eqref{hSchw}.] Here an overdot denotes a derivative with respect to $T$,
and the tidal fields on the right are constant, so $\sE_{kl}$ and $\dot{\sE}_{kl}$ are
treated formally as independent tidal fields, despite the notation. That is,
$\sE_{kl}$ just denotes the constant part of $\sE_{kl}(T) = \sE_{kl} + T\dot{\sE}_{kl}$.
Finally, despite appearances, this expression for the metric is in fact in a
quasi-Cartesian form: $dR$ should just be considered a shorthand for $X_kdX^k/R$.

\section{Near Zone Metric}\label{NZ}

We take the harmonic coordinate metric to the order needed from Eqs.~(7.2) in Blanchet, Faye, and Ponsot~\cite{BFP} and specialize it to a circular orbit, obtaining
\begin{subequations}
\label{gPN}
\<
\label{g00PN}
g_{00} + 1
= \frac{2m_1}{r_1} + \frac{m_1}{r_1}[4v_1^2 -  (\hat{n}_1\cdot\vec{v}_1)^2] - 2\frac{m_1^2}{r_1^2}
- m_1m_2\biggl[\frac{2}{r_1r_2} + \frac{r_1}{2b^3} - \frac{r_1^2}{2r_2b^3} + \frac{5}{2r_2b}\biggr]
+ (1 \leftrightarrow 2) + O(v^6),
\?
\<
\begin{split}\label{g0kPN}
g_{0k}& = -\frac{4m_1}{r_1}v_1^k - \biggl\{\frac{m_1^2}{r_1^2}(\hat{n}_1\cdot\vec{v}_1) + \frac{m_1m_2}{S^2}[
16(\hat{n}_2\cdot\vec{v}_1) - 12(\hat{n}_2\cdot\vec{v}_2)]\biggr\}n_1^k
- m_1m_2\biggl\{
4\frac{\hat{n}_1\cdot\vec{v}_1}{b^2} - 12\frac{\hat{n}_1\cdot\vec{v}_1}{S^2}
\biggr\}n_{12}^k\\
&\quad + \biggl\{\frac{m_1}{r_1}[2(\hat{n}_1\cdot\vec{v}_1)^2 - 4v_1^2] + \frac{m_1^2}{r_1^2} + \frac{m_1m_2}{b^3}[3r_1 - 2r_2] -
m_1m_2\biggl[\frac{r_2^2}{r_1b^3} + \frac{3}{r_1b} - \frac{8}{r_2b} + \frac{4}{bS}\biggr]\biggr\}v_1^k\\
&\quad + (1 \leftrightarrow 2) + O(v^6),
\end{split}
\?
\<
\begin{split}\label{gklPN}
g_{kl} - \delta_{kl}& = \biggl\{2\frac{m_1}{r_1} - \frac{m_1}{r_1}(\hat{n}_1\cdot\vec{v}_1)^2 + \frac{m_1^2}{r_1^2} + m_1m_2\biggl[\frac{2}{r_1r_2} -
\frac{r_1}{2b^3} + \frac{r_1^2}{2r_2b^3} - \frac{5}{2r_1b} + \frac{4}{bS}\biggr]
\biggr\}\delta_{kl}\\
&\quad + 4\frac{m_1}{r_1}v_1^kv_1^l + \frac{m_1^2}{r_1^2}n_1^kn_1^l - m_1m_2\biggl[\frac{4}{S^2} + \frac{4}{bS}
\biggr]n_{12}^kn_{12}^l + 4\frac{m_1m_2}{S^2}\left[n_1^{(k}n_2^{l)} + 2n_1^{(k}n_{12}^{l)}\right]
+ 8\frac{m_1m_2}{b^2}n_{12}^{(k}v_{12}^{l)}\\
&\quad + (1 \leftrightarrow 2) + O(v^6).
\end{split}
\?
\end{subequations}
\end{widetext}
Here $\vec{x}_A(t)$, $A \in \{1,2\}$ denotes the position of (the point particle
associated with) hole $A$.\footnote{We shall henceforth refrain from belaboring
the distinction between the PN point particles and true black holes, except where
we feel that it is important to emphasize this point.} Thus
$\vec{r}_A := \vec{x} - \vec{x}_A(t)$ gives the displacement from hole $A$, with 
$r_A := \|\vec{r}_A\|$ giving the distance from hole $A$ and $\hat{n}_A := \vec{r}_A/r_A$
the associated unit vector. Similarly,
$\vec{v}_A := \dot{\vec{x}}_A$ denotes the velocity of hole $A$. The displacement
vector from hole $B$ to hole $A$ is given by $\vec{r}_{AB} := \vec{r}_A -
\vec{r}_B$, with an associated unit vector of $\hat{n}_{AB} :=
\vec{r}_{AB}/\|\vec{r}_{AB}\|$. Thus, specializing to a (quasi)circular orbit,
$\|\vec{r}_{12}\| = b$, where $b$ is the separation of the holes and is therefore
constant up to orbital shrinkage (which we can neglect when performing the
matching, given the order to which we are calculating, though we include it
when implementing the metrics numerically---see Sec.~\ref{phasing}).
Additionally, we shall
usually use the shorthand $\vec{b} := \vec{r}_{12}$. (We do not use it in the above
expression for the metric since $\vec{r}_{12}$ changes sign under $1 \leftrightarrow 2$,
while $\vec{b}$ does not.) Similarly,
$\vec{v}_{AB} := \vec{v}_A - \vec{v}_B$; we also have $S := r_1 + r_2 + b$. Finally,
the notation ``$+\, (1 \leftrightarrow 2)$'' denotes that the preceding expression is to
be added to itself with the labels $1$ and $2$ switched.

For a circular orbit, we have $\Ddot{\vec{b}} = -\omega^2\vec{b}$, by
definition, where
\<\label{omega}
\omega = \sqrt{\frac{m}{b^3}}\left[1 + \frac{m}{2b}(\eta - 3) + O\left(\frac{m^2}{b^2}\right)\right]
\?
is the (harmonic coordinate) angular velocity [obtained from Eq.~(8.6) in \cite{BFP}], and
$\eta := m_1m_2/m^2$ is the symmetric mass ratio. Assuming a (quasi)circular orbit [so that the separation vector of the point particles is orthogonal to their
velocities up to $O(v^5)$ radiation reaction effects] also allows us to make
the simplifications $\hat{n}_{12} \cdot \hat{v}_1 =
\hat{n}_{12} \cdot \hat{v}_2 = \hat{n}_{12} \cdot \hat{v}_{12} = O(v^5)$ in obtaining the expression for the metric given in Eqs.~\eqref{gPN}.
We take the explicit expression for the orbit to be
$\vec{b} = b(\hat{x}\cos\omega t + \hat{y}\sin\omega t)$. From the definition of the center-of-mass coordinates used in the PN metric, the positions of (the
point particles associated with) the holes are given, to the order we need them, by
\begin{align}\label{rel-to-COM}
\vec{x}_1& = \frac{m_2}{m}\vec{b},& \vec{x}_2 = -\frac{m_1}{m}\vec{b}.
\end{align}
[For a circular orbit, the first PN corrections to these are $O(v^4)$ and are thus not
needed here.] The expressions for the holes' velocities are
obtained by taking a time derivative.

The explicit expansions of the near zone metric at various orders given in the next section were obtained using the expansions
\<
\begin{split}
&\frac{1}{r_2^n} = \frac{1}{b^n}\biggl\{1 - n\frac{\vec{r}_1\cdot\hat{b}}{b} + \frac{1}{2b^2}[n(n + 2)(\vec{r}_1\cdot\hat{b})^2 - nr_1^2]\\
&\; + \frac{1}{6b^3}[3n(n + 2)r_1^2(\vec{r}_1\cdot\hat{b}) - n(n + 2)(n + 4)(\vec{r}_1\cdot\hat{b})^3]\\
&\; + O\left(\left[\frac{r_1}{b}\right]^4\right)\biggr\} 
\end{split}
\?
and
\<\label{S^n}
\begin{split}
\frac{1}{S^n}& = \biggl[2b + r_1 + \vec{r}_1\cdot\hat{b} + \frac{r_1^2 - (\vec{r}_1\cdot\hat{b})^2}{2b} +
O\left(\left[\frac{r_1}{b}\right]^3\right)\biggr]^{-n}\\
& = \frac{1}{(2b)^n}\biggl\{1 - \frac{n}{2}\biggl[\frac{r_1}{b} + \frac{\vec{r}_1\cdot\hat{b}}{b}\biggr] + \frac{n(n - 1)}{8}\frac{r_1^2}{b^2}\\
&\quad + \frac{n(n + 1)}{4}\frac{r_1(\vec{r}_1\cdot\hat{b})}{b^2} + \frac{n(n + 3)}{8}\frac{(\vec{r}_1\cdot\hat{b})^2}{b^2}\\
&\quad + O\left(\left[\frac{r_1}{b}\right]^3\right)\biggr\}.
\end{split}
\?
Here we include the octupole pieces in the expression for $1/r_2^n$ because they become of quadrupole order when, e.g.,
multiplied by $m_1/r_1$ in the $-2m_1m_2/r_1r_2$ piece of $g_{00}$. We do not need to expand $1/S^n$ to octupolar order as
it is not multiplied by $m_1/r_1$ at the PN order we are working.

\subsection{The PN metric and conformal flatness}
\label{PNCF}

At $O(v^{4})$, the spatial PN metric is no longer conformally
flat: That is, there does not exist a coordinate system in which the $3$-metric can
be written as $g_{kl} = \Psi\delta_{kl}$. This can be guessed from a cursory
inspection of the PN metric, as nondiagonal spatial components first appear at $O(v^4)$,
but was established more firmly in~\cite{Rieth}, and then further studied
in~\cite{Damour:2000we}. In both cases, this result was an offshoot of a comparison of
the predictions of a post-Newtonian analysis of the Isenberg-Mathews-Wilson (IMW)
approximation~\cite{Isenberg, WMM} with those of the full PN approximation: They were
found to differ starting at $O(v^4)$. Such a comparison gives the desired result
because the IMW approximation assumes spatial conformal flatness (along with
maximal slicing---i.e., a vanishing trace of the extrinsic curvature) in an
attempt to remove the dynamical degrees of freedom from the gravitational field.

However, the comparison with the IMW
approximation is not exact. The post-Newtonian analyses use the ADMTT
gauge, for which we have $\delta_{kl}\pi^{kl} = 0$, where $\pi^{kl}$ is the
canonical momentum conjugate to the $3$-metric, $g_{kl}$, while maximal
slicing sets $g^{lp}K_{lp} = 0$, where $K_{lp}$ is the extrinsic curvature of
the slice. Of course, these two slicings agree as long as $g_{kl}$ is a
multiple of the flat space
metric, since then $\delta_{kl}\pi^{kl} = 0$ $\Leftrightarrow$
$g_{kl}\pi^{kl} = 0$ $\Leftrightarrow$ $g^{lp}K_{lp} = 0$.
[See, e.g., Eq.~(E.2.31) in Wald~\cite{Wald} for the relation between
$\pi^{lp}$ and $K^{lp}$.]
But in general, the two slicings will differ at the higher, non-conformally flat orders
we are interested in.

It is also possible to demonstrate this lack of spatial conformal flatness directly,
and we shall do so here for the harmonic slicing we are using.
In four or more dimensions, the Weyl tensor settles questions of conformal flatness: It vanishes if and only if a
space is conformally flat~\cite{Stephani:2003tm}. However, in three dimensions
the Weyl tensor vanishes identically, and its
analogue for settling questions of conformal flatness is the Bach or Cotton-York tensor,
$C_{kl}$. This is defined (with indices raised by the $3$-metric) by
\begin{equation}
C_{kl} := 2 \epsilon_{k}{}^{p s} \nabla_sR_{l p} -
\frac{1}{2} \epsilon_{kl}{}^{p} \nabla_p R.
\end{equation}
Here $\nabla_k$, $R_{kl}$, and $R$ are, respectively, the (three-dimensional) covariant
derivative, Ricci tensor, and Ricci scalar associated with $g_{kl}$. In fact, the
nonvanishing of the Cotton-York tensor is a
necessary and sufficient condition to render its associated $3$-metric non-conformally flat~\cite{Stephani:2003tm,PhysRevD.61.124011} (for a proof, see Chap.~VI, \S5 in~\cite{Schouten}).

\begin{figure}[htb]
\epsfig{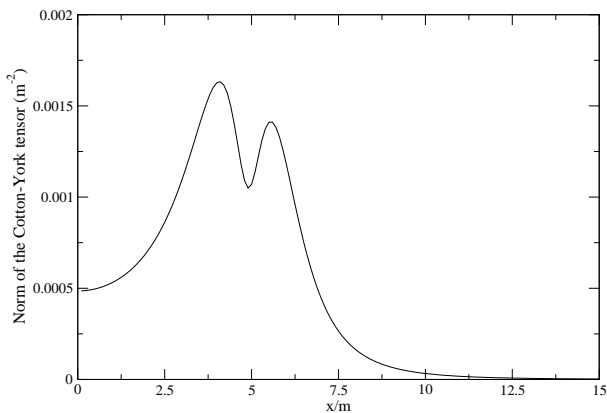} 
\caption{\label{C-Y} The norm of the Cotton-York tensor for the test system
$m_1 = m_2 = m/2$, $b = 10m$ along the $x$-axis (i.e., the axis passing through
both holes). (We only show
it around hole $1$ because it is symmetric about $x = 0$.)}
\end{figure}

We have computed the formal $O(c^{-4})$ portions of this
tensor [i.e., those that correspond to the $O(c^{-4})$ $\Leftrightarrow$ $O(v^4)$ pieces of
the spatial metric] symbolically using {\sc{Maple}} and {\sc{GRTensorII}} and verified that certain
components are nonvanishing at various points of the timeslice. As an illustration, we have
plotted the lowest-order piece of the norm of the Cotton-York tensor,
$\sqrt{\delta^{kp}\delta^{ls}C_{kl}C_{ps}}$ along the axis passing through
the holes in Fig.~\ref{C-Y} for the standard
equal-mass test system of $m_1 = m_2 = m/2$, $b = 10m$. As expected, the values are largest
in the region around the holes, showing that this is the region in which the largest
perturbation would be required to make the $2$PN metric conformally flat. 

There is, however, a more intuitive way of understanding the breaking of spatial conformal
flatness of the PN metric (here returning to the ADMTT slicing). This approach relates explicitly to the failure of the PN metric
to be manifestly conformally flat at $O(v^4)$ and comes from the work
of Nissanke~\cite{Nissanke:2005kp}.\footnote{This argument was suggested to one of us by Luc
Blanchet~\cite{BlanchetPC}. Additional intuition can be obtained from Valiente Kroon's
result~\cite{Valiente_Kroon} that stationary spacetimes only admit conformally flat slices
if certain ``obstructions,'' constructed out of Geroch-Hansen quadrupoles, vanish.}
From
the study of PN theory in the canonical ADM formalism~\cite{1985AnPhy.161...81S}, one knows
that the PN metric can be rewritten as
\begin{equation}
g_{kl} = \Psi \delta_{kl} + h_{kl},
\end{equation}
where $\Psi$ is some conformal factor, while $h_{kl}$ is not proportional to the flat
metric. In fact, $h_{kl}$ must be symmetric and trace-free and
contains a piece that is proportional to
\begin{equation}
\label{TT-piece}
\frac{\partial^2}{\partial{x_{1}^{<k}}\partial{x_{2}^{l>}}} \log S,
\end{equation}
where $S$ was defined in the previous subsection and the angle brackets stand for the symmetric trace-free
projection.
[See, e.g., Eq.~(3.1) in~\cite{Nissanke:2005kp}; we have left off pieces of the form
$v_1^{<k}v_1^{l>}$.] This is clearly not a manifestly conformally flat contribution to
the metric, so we can use it to connect with the heuristic that ``the 2PN metric is not
spatially conformally flat because it contains various pieces that are not manifestly
conformally flat.''

To do so, we shall demonstrate how the
presence in $h_{kl}$ of the term given in Eq.~\eqref{TT-piece} prevents the obvious sort of coordinate transformation from rendering $g_{kl}$ conformally flat. For the
purposes of illustration, we take our coordinate transformation to be of the form $x^{k} \to x'^{k} = x^{k} + \xi^{k}$, where $\xi^k = O(v^4)$, and ask that it remove the
$h_{kl}$ piece from the $3$-metric. (Of course, in general the coordinate transformation would just need to turn $h_{kl}$ into a scalar multiple of $\delta_{kl}$, and
would not have to be of the form above, but our assumptions suffice for a heuristic argument.) Thus, $\xi^k$ should satisfy
\begin{equation}\label{diffeo}
\partial_{(k}\xi_{l)} = - \frac{1}{2} h_{kl},
\end{equation}
but this equation has no solution when the right-hand side is given by
Eq.~\eqref{TT-piece}. To show that this is the case, it is sufficient
[by the gauge invariance of the linearized Riemann tensor---see the discussion
surrounding Eq.~\eqref{IC}] to show that the
(three-dimensional) flat space linearized Riemann tensor associated with $h_{kl}$ does
not vanish. We have computed this tensor [with $h_{kl}$ given by Eq.~\eqref{TT-piece}]
and verified that several components are indeed generically nonzero.

\subsection{Gravitational radiation in the PN metric}
\label{PNrad}

The order at which the effects of gravitational radiation appear in the PN metric
is different for different effects, which can easily lead to confusion. We thus
offer a brief discussion of these orders here. To avoid even further confusion,
since we shall discuss some pieces that are not dimensionless,
we shall describe all orders in terms of the formal slow motion expansion in
$1/c$, as do Blanchet, Faye, and Ponsot (BFP)~\cite{BFP}. [Since the metric is
dimensionless, an $O(c^{-n})$ contribution to it can be
unambiguously identified as being $O(v^n)$.] It is well known that the
effects of gravitational radiation reaction first enter the equations of motion
at $O(c^{-5})$ for a circular orbit. [See, e.g., Eq.~(189) in
Blanchet~\cite{BlanchetLRR}.] This is also the leading order of the binary's
gravitational wave luminosity. [See, e.g., Eq.~(171) in Blanchet.] However,
the lowest-order ``quadrupole formula'' piece of the gravitational waveform
appears in the PN far zone metric at one order lower, viz., $O(c^{-4})$. This
can be seen in Blanchet's Eq.~(238); his $x$ variable is $O(c^{-2})$.
These terms can also be seen in Eqs.~(6.10) and~(6.11a) of~\cite{WW}. Here they are
presented in a form that allows for a more direct comparison with our expression for the
far zone metric, though this expression does not show the factors of $c^{-1}$ explicitly.
The terms in question appear as the final two terms in the curly brackets in our
expression for the far zone spatial metric in
Eq.~\eqref{FZ_gkl}. Explicitly, they are $4(m_1/r)(v_1^{kl} - \omega^2x_1^{kl}) +
(1 \leftrightarrow 2)$.

Our initial data should thus contain the binary's outgoing gravitational radiation
(whether one uses the extension that adds on various higher-order terms or not),
since we have included all the $O(v^4)$ pieces in the spatial metric, along with the
matching $O(v^5)$ pieces in the extrinsic curvature. This explains why Kelly
\emph{et al.}\ obtain the outgoing wave train when they
evolve~\cite{Kellyetalev}, since their data set~\cite{Kellyetal} contains only terms in the
far zone through $O(v^4)$ [in the spatial metric, with the matching $O(v^5)$
pieces in the extrinsic curvature], even though they do not have the $O(v^5)$ terms in
the spatial metric.

The factor of $1/c$ one obtains in going from the terms in the near zone that give the
lowest-order piece of the waveform when expanded in the far zone (i.e., in $b/r$),
which are $O(c^{-4})$, to the lowest-order $O(c^{-5})$ radiation reaction contribution
to the equations of motion is explained by Blanchet in~\cite{Blanchet93}: Radiation
reaction effects arise from antisymmetric waves---i.e., those that are time-odd---and
such waves will come from expressions involving at least one more time derivative (and
thus factor of $1/c$) than their symmetric (and time-even) counterparts. Explicitly,
these near zone terms that match onto the lowest-order gravitational
wave terms in the far zone are $4(m_1/r_1)v_1^kv_1^l - 4(m_1m_2/bS)n_{12}^kn_{12}^l +
(1 \leftrightarrow 2)$, since $r_A = r[1 + O(b/r)]$ and $S = 2r[1 + O(b/r)]$. The
(multipolar expansion of the) first time derivative of these terms gives the
lowest-order radiation reaction contribution to the metric given in~\cite{Blanchet93}.
These terms themselves are also some of the contributions to the near zone metric that
cause it to be conformally curved.

\section{The matching calculation}\label{Matching}

\subsection{The setup}\label{Matching1}

By symmetry, we can concentrate on performing the matching around hole~1: All our results in this case can be directly translated
to the matching around hole~2 by taking $1 \leftrightarrow 2$, $(t,x,y,z) \to (t,-x,-y,z)$, and $(T,X,Y,Z) \to (T,-X,-Y,Z)$, along with $\hat{x}_\alpha \to -\hat{x}_\alpha$
and $\hat{y}_\alpha \to -\hat{y}_\alpha$ in the tidal fields. Here $\hat{x}_\alpha$ and $\hat{y}_\alpha$ denote the unit vectors in the $x_1$- and $x_2$-directions,
respectively. (Similarly, $\hat{t}_\alpha$ and $\hat{z}_\alpha$ denote unit vectors in the $x_0$- and $x_3$-directions---we shall use these later.) In other words, we switch the masses to turn
$m_1$ into $m_2$, and then rotate by $\pi$ radians around the $z$-axis to
move the new $m_2$ into the same position as the old $m_2$. However, we also need to rotate the inner zone coordinates so that they have the same relation to the new near zone
coordinates that they did when we performed the original matching around hole~1. Finally, the tidal fields transform as Cartesian tensors under this rotation, and this transformation
is taken care of by the final two substitutions.

Since our matching calculation will determine the coordinate transformation, the
relations between the metrics' mass parameters, and the inner zone metric's tidal
fields, we need to posit expansions for all of these. For the coordinate transformation,
we make nearly the same ansatz as in Papers~I and~II, viz.,
\<\label{Xexpansion}
X^\alpha(x^\beta) = \sum_{j=0}^{5}\left(\frac{m_2}{b}\right)^{j/2}\od{X^\alpha}{j}(x^\beta) +
O(v^6).
\?
(This is slightly more general than the ansatz used in the previous papers because we do not fix the zeroth order coordinate
transformation from the outset.) We choose $\sqrt{m_2/b}$ as our expansion parameter
because it makes for slightly simpler notation than either $v = \sqrt{m/b}$ or
$\sqrt{m_1/b}$ when expanding in the buffer zone around hole~1, as we are doing here. We can make this choice without loss of generality: The resulting coordinate
transformation will be the same regardless of which of these three possibilities we
choose to use as our expansion parameter, though the coefficients of the expansion
parameter will differ by ratios of the masses.
As in the previous papers, we implicitly assume that $\od{X_\alpha}{j}$ is a power series in $\tr/b$, including negative powers---i.e., a Laurent series---so we can write,
e.g., $m_2/r = (m_2/b)(b/r)$. However, $\od{X_\alpha}{j}$ should not depend on
$m_2/b$, by definition.

We also need to worry about the multipole expansion of each $\od{X_\alpha}{j}$. This
would seem to be straightforward, since we only want to keep terms through quadrupole
order overall. However, the structure of the inner zone metric creates some
complications: In order to obtain data that include all the quadrupole [$O([\tr/b]^2)$]
pieces at fourth and fifth orders,\footnote{We count orders using our primary
expansion in $v$ (or, equivalently, $\sqrt{m_2/b}$).} one needs to obtain the octupole
[$O([\tr/b]^3)$] pieces of
the coordinate transformation when matching at second and third orders. This is due
to the appearance of $b/\tr$ terms in $\od{h_{\alpha\beta}}{j}$ for $j \ge 2$.
These enter the fourth and fifth order coordinate transformation equations, where they
multiply the second and third order pieces of the coordinate transformation and thus
produce quadrupolar contributions from octupolar pieces of the coordinate transformation.
The octupole fields themselves also enter, as they are multiplied by $b/\tr$ in the
fourth  and fifth order pieces of the inner zone metric. However, we shall see that the
octupole piece of the second order coordinate transformation
vanishes, and that of the third order one only appears in the time coordinate. Thus,
the only place where the octupole pieces of the coordinate transformation
appear is in the $\od{h_{(\alpha|\gamma}}{2}\od{A_{|\beta)}{}^\gamma}{3}$
piece of the fifth order coordinate transformation equations [given in full in
Eq.~\eqref{A5}].

This increase in the number of multipoles
that have to be kept as one proceeds to higher and higher orders in $v$ is a
general feature of the matching of these two metrics. It is thus a source of
significant technical
difficulty: One would need to include the hexadecapole pieces in the matching
calculation if one wanted to include all of the quadrupole pieces at sixth and higher
orders. This follows since at sixth order the hexadecapole pieces start to be
multiplied by $b^2/\tr^2$, making them of quadrupole order. (It would still be possible
to obtain all the dipole pieces at fifth and sixth orders if one had an
expression for the near zone metric through seventh order, but one would not be
able to include any new higher-order corrections to the tidal fields this way.)

We posit the same expansion in $\sqrt{m_2/b}$ for the mass parameter of the
inner zone metric, $M_1$, as we did for the coordinate transformation, so
\<
M_1 = \sum_{j=0}^{3}\left(\frac{m_2}{b}\right)^{j/2}(M_1)_j + O(v^4). 
\?
However, it will turn out that we did not need to allow this freedom, as we
shall find that the mass parameters of the two metrics agree to the highest
order to which our matching fixes them---i.e., $M_1 = m_1 + O(v^4)$.
Similarly, we asymptotically expand the tidal fields in $\sqrt{m_2/b}$, so
\<
\begin{split}\label{TFexpansion}
\sE_{kl} & = \frac{m_2}{b^3}\sum_{j=0}^{3}\left(\frac{m_2}{b}\right)^{j/2}\od{\bE_{kl}}{j} +
O(v^6),\\
\dot{\sE}_{kl} & = \frac{m_2}{b^3}\sum_{j=1}^{2}\left(\frac{m_2}{b}\right)^{j/2}\ods{\dot{\bE}_{kl}}{j} + O(v^5),\\
\sE_{klp} & = \frac{m_2}{b^4}\sum_{j=0}^{2}\left(\frac{m_2}{b}\right)^{j/2}\od{\bE_{klp}}{j} + O(v^5),\\
\sB_{kl} & = \left(\frac{m_2}{b}\right)^{3/2}\frac{1}{b^2}\sum_{j=0}^{2}\left(\frac{m_2}{b}\right)^{j/2}\od{\bB_{kl}}{j} + O(v^6),\\
\dot{\sB}_{kl} & = \left(\frac{m_2}{b}\right)^2\frac{1}{b^2}\ods{\dot{\bB}_{kl}}{1} + O(v^5),\\
\sB_{klp} & = \left(\frac{m_2}{b}\right)^{3/2}\frac{1}{b^3}\sum_{j=0}^{1}\left(\frac{m_2}{b}\right)^{j/2}\od{\bB_{klp}}{j} + O(v^5).
\end{split}
\?
[The expansions of the duals of the magnetic fields---e.g., $\sC_{klp}$---are defined
analogously. Also recall that the ``barred'' tidal fields are defined in
Eq.~\eqref{TF_scaling}. Additionally, the tidal fields and their time
derivatives---e.g., $\sE_{kl}$ and $\dot{\sE}_{kl}$---are treated as formally
independent.] We do not include the $O(v^5)$ pieces of the octupole tidal
fields here because we did not fix them in the matching: We only had to match
the octupole fields through $O(v^3)$ to obtain initial data with formal
uncontrolled remainders of $O(v^5)$ and $O([r/b]^3)$ (i.e., octupolar order).
We also chose to read off the $O(v^4)$ pieces of the octupole fields
separately (in Appendix~\ref{TFHO}), but did not do so for the $O(v^5)$ parts.

In order to read off the matching parameters (and any undetermined pieces of
the lower-order coordinate transformations) as efficiently as possible, we
note that all of our equations for the coordinate transformation at orders
beyond the zeroth will be of the form
\<\label{SPDE}
\partial_{(\alpha}X_{\beta)} = S_{\alpha\beta} \equiv S_{(\alpha\beta)}.
\?
Here $\partial_\alpha := \partial/\partial x^\alpha$ (i.e., all partial
derivatives are taken with respect to PN harmonic coordinates), $X^\alpha$ is
a function of $x^\alpha$, and
$S_{\alpha\beta}$, the equation's source, is some
(symmetric) matrix function of $x^\alpha$ (either explicitly, or implicitly
though $X^\alpha$) which is $C^2$ in the buffer zone. The integrability
condition for this equation is that the flat-space linearized Riemann tensor
associated with $S_{\alpha\beta}$ vanish, i.e., that we have
\<\label{IC}
\sI_{\alpha\beta\gamma\delta} := \partial_{\alpha\beta}S_{\gamma\delta} +
\partial_{\gamma\delta}S_{\alpha\beta} - 
\partial_{\alpha\delta}S_{\gamma\beta} -
\partial_{\gamma\beta}S_{\alpha\delta} = 0.
\?
[N.B.: For convenience, we have defined $\sI_{\alpha\beta\gamma\delta}$ with
a slightly different index ordering than the linearized Riemann tensor---given
in, e.g., Eq.~(5.44) of~\cite{BHTMPV}---and without the factor of $1/2$.]
This follows from the gauge invariance of the linearized Riemann tensor. (See,
e.g., \S4.1 in Straumann~\cite{Straumann} for a proof and discussion of that
result. In~\cite{BD}, Blanchet and Damour use this gauge invariance for the
same purpose we do.)

For future use, we note that the homogeneous equation,
$\partial_{(\alpha}X_{\beta)} = 0$, is the
flat space Killing equation, and its most general solution is given by
\<\label{hom_sol}
X_\alpha = F_{\beta\alpha}x^\beta + C_\alpha,
\?
where $F_{\alpha\beta} \equiv F_{[\alpha\beta]}$ is some constant
(antisymmetric) $4 \times 4$ matrix and $C_\alpha$ is some constant
$4 \times 1$ matrix. See, e.g., \S13.1 in Weinberg~\cite{Weinberg} for a
proof. (Our $X_\alpha$, $F_{\alpha\beta}$, and $C_\alpha$ correspond to
Weinberg's $\xi_\alpha$, $b_{\beta\alpha} = -b_{\alpha\beta}$, and
$a_\alpha$, respectively.) We shall primarily employ this result tacitly at
each order beyond the zeroth to ensure that we have the most general
expression for that order's contribution to the coordinate transformation.

Our general approach to the nontrivial matching that occurs at second order
and beyond will be as follows: We first use the above integrability condition
to read off the matching parameters, exploiting the linear
independence of various terms to simplify the process and justify our claims
of uniqueness. We start with the nonpolynomial terms, which determine many of
the previously undetermined parts of the coordinate transformation from two
orders lower, as well as the inner zone mass parameter; the polynomial part
then determines the tidal fields.\footnote{It is intuitively reasonable that
the polynomial and nonpolynomial parts should determine the parameters that
they do: If one neglects all gauge subtleties and the like, the nonpolynomial
terms can be thought of as being those associated with hole~1, and the
polynomial terms with (the tidal fields of) hole~2. We discuss this more fully
in Sec.~\ref{CT}.} The nonpolynomial part consists of all the terms that are
not polynomials in $\tx^\alpha := x^\alpha - (m_2/m)b\hat{x}^\alpha$. For this
calculation, these are all of the form of a polynomial in $\tx^\alpha$
multiplied by $\tr^n$, where $n \in \mathbb{Z} \setminus
\{0, 2, 4, 6, \cdots\}$. After we have fixed all the parameters that can
be fixed at a given order, we can then solve for that order's contribution to
the coordinate transformation (and the polynomial part can be solved for
separately from the nonpolynomial part that
first appears at fourth order). In all of this, {\sc{Maple}} and {\sc{GRTensorII}} proved very helpful: They work extremely well for all aspects of the polynomial part, while
requiring more care when applied to the nonpolynomial part.

\subsection{Zeroth Order [$O([m_2/b]^0)$]}

To lowest (zeroth) order in $(m_2/b)^{1/2}$, we have
\<
\od{g_{\alpha\beta}}{0} = \od{A_\alpha{}^\gamma}{0}\od{h_{\gamma\delta}}{0}\od{A_\beta{}^\delta}{0},
\?
where
we have defined $A_\alpha{}^\beta := \partial_\alpha X^\beta$. Since $\od{g_{\alpha\beta}}{0} =
\od{h_{\alpha\beta}}{0} = \eta_{\alpha\beta}$,
matching just tells us that $\od{A_\alpha{}^\beta}{0}$ must be a (general) Lorentz transformation---i.e., not necessarily
one continuously connected to the identity---which
we call $L_\alpha{}^\beta$. We also pick up a $4 \times 1$ matrix ``constant of integration'' $\od{C^\alpha}{0}$ in
obtaining $\od{X^\alpha}{0}$, so we are finally left with
\<
\od{X^\alpha}{0} = L_\beta{}^\alpha x^\beta + \od{C^\alpha}{0}.
\?
For simplicity, we shall take this lowest-order piece to be the
expected translation due to the position of $m_2$ at $t = 0$, viz.,
\<
L_\alpha{}^\beta = \delta_\alpha{}^\beta,\qquad \od{C_\alpha}{0} = -\frac{m_2}{m}b\hat{x}_\alpha.
\?

In fact, one can show that matching through third order requires that the
spatial part of $\od{C_\alpha}{0}$ be as given above (though the temporal part
can still be freely specified). Similarly, that matching requires
$L_\alpha{}^\beta$ to differ from the identity only by a possible rotation
about the $y$-axis, along with possible spatial and temporal reflections.
These are combined with a rotation that takes $+y$ to $-y$ if we have an odd
number of reflections. We thus conjecture that matching at higher orders will
further constrain this lowest-order coordinate transformation to be as given
above. At the very least, the matching at fourth and fifth orders is
independent of the remaining freedom.

\subsection{First Order [$O(\sqrt{m_2/b})$]}

We have $\od{g_{\alpha\beta}}{1} = \od{h_{\alpha\beta}}{1} = 0$.
Thus, the first order matching gives us
\<\label{A1}
\od{A_{(\alpha\beta)}}{1} = 0,
\?
using the fact that $\od{h_{\alpha\beta}}{0} = \eta_{\alpha\beta}$.
As was given in Eq.~\eqref{hom_sol}, the most general solution of this
equation is
\<\label{X1}
\od{X_\alpha}{1} = \od{F_{\beta\alpha}}{1}\tx^\beta + \od{C_\alpha}{1},
\?
where we have written this in terms of $\tx^\alpha := x^\alpha - (m_2/m)b\hat{x}^\alpha$
because $\tr/b := \sqrt{\tx_k\tx^k}/b$ is one of our small parameters. We can do this
without loss of generality, as it simply entails a different value for $C_\alpha$.

This result differs from that given in Eq.~(21) of Paper~I; the latter suffers from some
sign errors introduced during transcription. However, this does not affect that paper's
final coordinate transformation, as the relevant constants were all taken to be zero. This
is appropriate for Papers~I and~II, since they were using a corotating coordinate system:
The boost encoded in our $\od{F_{\alpha\beta}}{1}$ [seen in Eq.~\eqref{F1}] would thus
not be expected to appear in the coordinate transformation.

\subsection{Second Order [$O(m_2/b)$]}

Proceeding to the next order, we have, recalling that $\ods{\dot{\bE}_{kl}}{0} = 0$,
\<
\begin{split}
\od{h_{\alpha\beta}}{2}& = \biggl[\frac{2\od{M_1}{0}}{m_2}\frac{b}{\od{R}{0}} -
\frac{\od{\bE_{kl}}{0}}{b^2}\od{X^k}{0}\od{X^l}{0}\\
&\quad - \frac{\od{\bE_{klp}}{0}}{3b^3}\od{X^k}{0}\od{X^l}{0}\od{X^p}{0}\biggr]\Delta_{\alpha\beta}.
\end{split}
\?
Similarly, noting that $\ods{\dot{b}_k}{0} = 0$,
\<
\begin{split}
&\od{g_{\alpha\beta}}{2} = \Biggl[\frac{2m_1}{m_2}\frac{b}{\od{r_1}{0}} + 2 -
\frac{2}{b}\{\od{\vec{r}_1}{0} \cdot {(\hat{b})}_0\}\\
&\quad + \frac{1}{b^2}\{3[\od{\vec{r}_1}{0} \cdot {(\hat{b})}_0]^2 - [\od{r_1}{0}]^2\}\\
&\quad + \frac{1}{b^3}\{3[\od{r_1}{0}]^2[\od{\vec{r}_1}{0} \cdot {(\hat{b})}_0] - 5[\od{\vec{r}_1}{0} \cdot {(\hat{b})}_0]^3\}\Biggr]\Delta_{\alpha\beta}.
\end{split}
\?
Here $\Delta_{\alpha\beta} := \mathrm{diag}(1,1,1,1)$ is the ``lowered
$4$-dimensional Kronecker delta.'' Also, we have $\ods{\hat{b}_k}{0} = \hat{x}_k$
and, with our choice for the zeroth order coordinate transformation,
$\od{r_1^k}{0} = \tr^k$.
The equations for the coordinate transformation at this order are thus
\<
\begin{split}
\od{g_{\alpha\beta}}{2}& =
\od{A_\alpha{}^\gamma}{0}\od{h_{\gamma\delta}}{2}\od{A_\beta{}^\delta}{0} +
\od{A_\alpha{}^\gamma}{1}\od{h_{\gamma\delta}}{0}\od{A_\beta{}^\delta}{1}\\
&\quad + 2\od{A_{(\alpha|}{}^\gamma}{0}\od{h_{\gamma\delta}}{0}\od{A_{|\beta)}{}^\delta}{2},
\end{split}
\?
since $\od{h_{ab}}{1} = 0$, or, using our previous results,
\<
\od{g_{\alpha\beta}}{2} = \od{h_{\alpha\beta}}{2} +
\od{F_\alpha{}^\gamma}{1}\od{F_{\beta\gamma}}{1} + 2\od{A_{(\alpha\beta)}}{2}.
\?
Thus, at this order, the source of the differential equation [cf.\ Eq.~\eqref{SPDE}] is given by
\<
2\od{S_{\alpha\beta}}{2} = 2\od{A_{(\alpha\beta)}}{2} = \od{g_{\alpha\beta}}{2} - \od{h_{\alpha\beta}}{2} - \od{F_\alpha{}^\gamma}{1}\od{F_{\beta\gamma}}{1}.
\?

We now apply the integrability condition from Eq.~\eqref{IC} and focus on the
nonpolynomial piece of
$\od{S_{\alpha\beta}}{2}$; here this is the one that diverges [as $\od{R}{0} = \od{r_1}{0} = \tr \to 0$]. It must satisfy the integrability condition
independently of the other pieces, by linear independence, and [considering,
e.g., $\od{\sI_{kl00}}{2}$] gives $\od{M_1}{0} = m_1$, as expected.
The polynomial piece of the integrability conditions tells us that
\begin{align}
\od{\bE_{kl}}{0}& = \delta_{kl} - 3\hat{x}_k\hat{x}_l,& \od{\bE_{klp}}{0} = 15\hat{x}_k\hat{x}_l\hat{x}_p - 9\delta_{(kl}\hat{x}_{p)},
\end{align}
using linear independence to read off the quadrupole and octupole tidal fields
separately. Solving for the coordinate transformation, we obtain
\begin{widetext}
\<
\label{X2}
\od{X_\alpha}{2} = \left[1 - \frac{\tx}{b}\right]\Delta_{\alpha\beta}\tx^\beta +
\frac{\Delta_{\beta\gamma}\tx^\beta\tx^\gamma}{2b}\hat{x}_\alpha
- \frac{1}{2}\od{F_\alpha{}^\gamma}{1}\od{F_{\beta\gamma}}{1}\tx^\beta + \od{F_{\beta\alpha}}{2}\tx^\beta + \od{C_\alpha}{2}.
\?

\subsection{Third Order [$O([m_2/b]^{3/2})$]}

At this order, the inner zone metric is
\begin{subequations}
\<
\begin{split}
\od{h_{00}}{3}& = \frac{2\od{M_1}{0}}{m_2}b\od{\frac{1}{R}}{1} + \frac{2\od{M_1}{1}}{m_2}\frac{b}{\od{R}{0}}
- 2\frac{\od{\bE_{kl}}{0}}{b^2}\od{X^k}{0}\od{X^l}{1} - \frac{\od{\bE_{kl}}{1}}{b^2}\od{X^k}{0}\od{X^l}{0}\\
&\quad - \frac{\ods{\dot{\bE}_{kl}}{1}}{b^2}\od{T}{0}\od{X^k}{0}\od{X^l}{0} - \frac{\od{\bE_{klp}}{0}}{b^3}\od{X^k}{0}\od{X^l}{0}\od{X^p}{1} -
\frac{\od{\bE_{klp}}{1}}{3b^3}\od{X^k}{0}\od{X^l}{0}\od{X^p}{0},
\end{split}
\?
\<
\od{h_{0k}}{3} = \frac{2}{3}\frac{\od{\bC_{klp}}{0}}{b^2}\od{X^l}{0}\od{X^p}{0} +
\frac{1}{3}\frac{\od{\bC_{klps}}{0}}{b^3}\od{X^l}{0}\od{X^p}{0}\od{X^s}{0} -
\frac{2}{3}\frac{\ods{\dot{\bE}_{lp}}{1}}{b^2}\od{X^l}{0}\od{X^p}{0}\od{X_k}{0},
\?
\<
\od{h_{kl}}{3} = \od{h_{00}}{3}\delta_{kl},
\?
\end{subequations}
and the near zone metric is
\begin{subequations}
\<
\od{g_{00}}{3} = \frac{2m_1}{m_2}b\od{\frac{1}{r_1}}{1} - \frac{2}{b}[\od{\vec{r}_1}{0} \cdot {(\hat{b})}_1]
+ \frac{2}{b^2}\{3[\od{\vec{r}_1}{0} \cdot {(\hat{b})}_1][\od{\vec{r}_1}{0} \cdot {(\hat{b})}_0] - \od{\vec{r}_1}{1} \cdot \od{\vec{r}_1}{0}\}
+ \frac{6}{b^3}[\od{\vec{r}_1}{0} \cdot \od{\vec{r}_1}{1}][\od{\vec{r}_1}{0} \cdot {(\hat{b})}_0],
\?
\<
\begin{split}
\od{g_{0k}}{3}& = 4\frac{m_1}{m}\biggl\{1 - \frac{b}{\od{r_1}{0}} - \frac{1}{b}[\od{\vec{r}_1}{0} \cdot {(\hat{b})}_0] +
\frac{1}{2b^2}[3\{\od{\vec{r}_1}{0} \cdot {(\hat{b})}_0\}^2 - \{\od{r_1}{0}\}^2]\\
&\quad + \frac{1}{2b^3}[3\{\od{r_1}{0}\}^2\{\od{\vec{r}_1}{0} \cdot {(\hat{b})}_0\} - 5\{\od{\vec{r}_1}{0} \cdot {(\hat{b})}_0\}^3]\biggr\}\ods{\dot{b}_k}{1},
\end{split}
\?
\<
\od{g_{kl}}{3} = \od{g_{00}}{3}\delta_{kl}.
\?
\end{subequations}
\end{widetext}
Here we have ${(\vec{b})}_1 = \sqrt{m/m_2}t\hat{y}$, so $\od{\vec{r}_1}{1} = -(m_2/m){(\vec{b})}_1 = -\sqrt{m_2/m}t\hat{y}$, and thus
$\od{\vec{r}_1}{1} \cdot {(\hat{b})}_0 = 0$.
Additionally, ${(\hat{b})}_1 = \sqrt{m/m_2}(t/b)\hat{y}$ and $\ods{\dot{b}_k}{1} = \sqrt{m/m_2}\hat{y}_k$. We also have
\<
\begin{split}\label{1/R1}
\od{\frac{1}{R}}{1} & = -\frac{1}{\od{R}{0}}\frac{\od{X_k}{0}\od{X^k}{1}}{[\od{R}{0}]^2}\\
& = -\frac{\tr^k[\od{F_{\alpha k}}{1}\tx^\alpha + \od{C_k}{1}]}{\tr^3},
\end{split}
\?
and
\<\label{1/r1}
\od{\frac{1}{r_1}}{1} = -\frac{1}{\od{r_1}{0}}\frac{\od{\vec{r}_1}{0} \cdot \od{\vec{r}_1}{1}}{[\od{r_1}{0}]^2} =
\sqrt{\frac{m_2}{m}}\frac{yt}{\tr^3},
\?
where $\tr^\alpha := r^\alpha - (m_2/m)b\hat{x}^\alpha$.

At this order, the equations for the coordinate transformation are
\begin{multline}
2\od{S_{\alpha\beta}}{3} = 2\od{A_{(\alpha\beta)}}{3} = \od{g_{\alpha\beta}}{3} - \od{h_{\alpha\beta}}{3}\\
- 2\od{h_{(\alpha|\gamma}}{2}\od{F_{|\beta)}{}^\gamma}{1} - 2\od{F_{(\alpha|}{}^\gamma}{1}\od{A_{|\beta)\gamma}}{2},
\end{multline}
recalling that $\od{h_{\alpha\beta}}{1} = 0$ and utilizing our lower-order
results. To obtain $\od{F_{\alpha\beta}}{1}$, we look at the $\tr^{-7}$ piece
of $\od{\sI_{kl00}}{3}$. Such a piece can only come from
two spatial derivatives both acting on $\tr^{-3}$ in the $\tr^{-3}$ pieces of $\od{S_{00}}{3}$; those pieces, in turn, only come
from $\od{1/R}{1}$ and $\od{1/r_1}{1}$. Therefore, using Eqs.~\eqref{1/R1} and
\eqref{1/r1}, the integrability conditions require that we have
\<
\tr^k[\od{F_{\alpha k}}{1}\tx^\alpha + \od{C_k}{1}] = -\sqrt{\frac{m_2}{m}}yt,
\?
from which we immediately see that
\begin{align}\label{F1}
\od{C_k}{1}& = 0,& \od{F_{\alpha\beta}}{1} = 2\sqrt{\frac{m_2}{m}}\hat{t}_{[\alpha}\hat{y}_{\beta]},
\end{align}
using the antisymmetry of $F_{\alpha\beta}$. (Recall that $\hat{t}_0 = -1$.) By similar
logic, the $\tr^{-5}$ piece of $\od{\sI_{kl00}}{3}$ only comes from the
$\od{M_1}{1}$ piece of $\od{h_{00}}{3}$ and gives us $\od{M_1}{1} = 0$. The remaining nonpolynomial pieces cancel, so we have extracted all the information we can from the nonpolynomial part of the integrability condition.

From the polynomial part of the integrability conditions, we first read off the octupole parts of the tidal fields, which are
\begin{gather}
\ods{\dot{\bE}_{kl}}{1} = -\frac{6}{b}\sqrt{\frac{m}{m_2}}\hat{x}_{(k}\hat{y}_{l)},\qquad \od{\bE_{klp}}{1} = 0, \nonumber\\
\od{\bB_{klp}}{0} = \frac{9}{2}\sqrt{\frac{m}{m_2}}\left[5\hat{x}_{(k}\hat{x}_l\hat{z}_{p)} - \delta_{(kl}\hat{z}_{p)}\right],
\end{gather}
and then the quadrupole parts, which are
\begin{align}
\od{\bE_{kl}}{1}& = 0,& \od{\bB_{kl}}{0} = -6\sqrt{\frac{m}{m_2}}\hat{x}_{(k}\hat{z}_{l)}.
\end{align}
The third order coordinate transformation is thus
\begin{widetext}
\<
\begin{split}\label{X3}
\od{X_\alpha}{3}& = \sqrt{\frac{m}{m_2}}\biggl\{-\frac{yt}{b^2}\Delta_{\alpha\beta}\tx^\beta +
\left[\frac{\tx_\mu\tx^\mu - 4\tx^2}{2b^2} + \left(2 - \frac{m_2}{m}\right)\frac{\tx}{b} + \left(2 +
\frac{1}{2}\frac{m_2}{m}\right)\frac{m_2}{m}\right]y\hat{t}_\alpha
+ 2\left[1 - \frac{m_2}{m}\right]\frac{yt}{b}\hat{x}_\alpha\\
&\quad+ \left[\frac{3\tr^2 + t^2}{6b^2} + \left(\frac{m_2}{m} - 2\right)\frac{\tx}{b} +
\frac{1}{2}\left(\frac{m_2}{m}\right)^2 + 4\right]t\hat{y}_\alpha\biggr\}
+ \left\{\sqrt{\frac{m_2}{m}}\left[\hat{y}_{(\alpha}\od{F_{\beta)0}}{2} - \hat{t}_{(\alpha}\od{F_{\beta)2}}{2}\right] +
\od{F_{\beta\alpha}}{3}\right\}\tx^\beta\\ 
&\quad + \od{C_\alpha}{3} + \frac{1}{2b^3}\sqrt{\frac{m}{m_2}}\tx y(4\tx^2 - y^2 - z^2)\hat{t}_\alpha,
\end{split}
\?
\end{widetext}
where the octupole part is the final term.

\subsection{Fourth and Fifth Orders [$O([m_2/b]^2)$ and $O([m_2/b]^{5/2})$]}
\label{4O5O}

The matching at fourth and fifth orders proceeds in the same way as it did at lower orders,
though the algebraic complexity increases substantially. We shall thus give far fewer
details of the calculations than we did before, and mostly concern ourselves with pointing
out the new features of the calculation that arise at these orders. The most prominent new
feature, and the one responsible for much of the algebraic complexity, is the presence of a
nonpolynomial part in the coordinate transformation. We know to expect this at fourth order
because the transformation between Cook-Scheel and PN harmonic coordinates is nonpolynomial,
and its lowest-order piece is $O(v^4)$---see Sec.~\ref{CT}. However, there are
various other nonpolynomial pieces present in the coordinate transformation at
fourth and fifth orders. We have to solve for these nonpolynomial parts of the
coordinate transformation by inspection (though we can still use {\sc{Maple}}
to obtain the polynomial part). It is reasonably easy to do so if one first
breaks the source into pieces by multipolar order.

The other subtleties involve the multipole expansion and are best illustrated
by giving two examples: First looking at the fourth order piece of the
inner zone metric component, we have
\begin{widetext}
\<
\begin{split}
\od{h_{00}}{4}& = 2\frac{\od{M_1}{2}}{m_2}\frac{b}{\od{R}{0}} + 2\frac{\od{M_1}{0}}{m_2}b\od{\frac{1}{R}}{2} -
2\frac{\left[\od{M_1}{0}\right]^2}{m_2{}^2}\frac{b^2}{\left[\od{R}{0}\right]^2} 
- \frac{\od{\bE_{kl}}{2}}{b^2}\od{X^k}{0}\od{X^l}{0} - \frac{\od{\bE_{kl}}{0}}{b^2}\od{X^k}{1}\od{X^l}{1}\\
&\quad - 2\frac{\od{\bE_{kl}}{0}}{b^2}\od{X^k}{0}\od{X^l}{2,0} +
2\frac{\od{M_1}{0}}{m_2}\frac{\od{\bE_{kl}}{0}}{b\od{R}{0}}\od{X^k}{0}\od{X^l}{0}
+ \frac{2}{3}\frac{\od{M_1}{0}}{m_2}\frac{\od{\bE_{klp}}{0}}{b^2\od{R}{0}}\od{X^k}{0}\od{X^l}{0}\ods{X^p}{0}.
\end{split}
\?
\end{widetext}
The terms to notice are the ones involving $\od{X^k}{2,0}$ and $\sE_{klp}/R$. The first of
these reflects the necessity of avoiding ``hidden octupole'' pieces (i.e.,
pieces of octupolar order that arise when multiplying together pieces of
lower multipolar order) when looking at corrections to lower-order terms:
We only want to include $\od{X^k}{2,0}$, the monopole
(zeroth-order-in-$\tx/b$) piece of $\od{X^k}{2}$, in
$\od{\bE_{kl}}{0}\od{X^k}{0}\od{X^l}{2,0}/b^2$; the dipole contribution to
$\od{X^k}{2}$ would give an octupole contribution to $h_{00}$. Contrariwise,
we do not want to leave out any terms of quadrupolar order, even if they arise
from, e.g., octupolar tidal fields. The $\sE_{klp}/R$ contribution is
such a term.

These subtleties arise in a slightly different form in the equation for the fifth order
piece of the coordinate transformation:
\begin{widetext}
\<\label{A5}
\begin{split}
&2\od{S_{\alpha\beta}}{5} = 2\od{A_{(\alpha\beta)}}{5} = \od{g_{\alpha\beta}}{5} - \od{h_{\alpha\beta}}{5} - 2\od{h_{(\alpha|\gamma}}{4}\od{F_{|\beta)}{}^\gamma}{1} -
\od{F_\alpha{}^\gamma}{1}\od{h_{\gamma\delta}}{3}\od{F_\beta{}^\delta}{1} - 2\ods{h^\rP_{(\alpha|\gamma}}{3}\od{A_{|\beta)}{}^\gamma}{2,0}\\
&\quad - 2\ods{h^\rNP_{(\alpha|\gamma}}{3}\od{A_{|\beta)}{}^\gamma}{2} - 2\od{F_{(\alpha|}{}^\gamma}{1}\od{h^\rP_{\gamma\delta}}{2}\od{A_{|\beta)}{}^\delta}{2,0}
- 2\od{F_{(\alpha|}{}^\gamma}{1}\od{h^\rNP_{\gamma\delta}}{2}\od{A_{|\beta)}{}^\delta}{2} - 2\ods{h^\rP_{(\alpha|\gamma}}{2}\od{A_{|\beta)}{}^\gamma}{3,0}\\
&\quad - 2\ods{h^\rNP_{(\alpha|\gamma}}{2}\od{A_{|\beta)}{}^\gamma}{3,\le3} - 2\od{A_{(\alpha|\gamma}}{2}\od{A_{|\beta)}{}^\gamma}{3,\le1} -
2\od{A_{(\alpha|\gamma}}{2,0}\od{A_{|\beta)}{}^\gamma}{3,2} - 2\od{F_{(\alpha|\gamma}}{1}\od{A_{|\beta)}{}^\gamma}{4}.
\end{split}
\?
\end{widetext}
Here we have to avoid ``hidden octupole'' terms in many of the contractions: For instance,
we only want the quadrupole-and-lower pieces of
$\od{A_{(\alpha|\gamma}}{2}\od{A_{|\beta)}{}^\gamma}{3}$, which are given by
$\od{A_{(\alpha|\gamma}}{2}\od{A_{|\beta)}{}^\gamma}{3,\le1} +
\od{A_{(\alpha|\gamma}}{2,0}\od{A_{|\beta)}{}^\gamma}{3,2}$. However, it is also
necessary to split up the inner zone metric's contributions into polynomial and
nonpolynomial parts to keep from excluding quadrupole pieces as well [since the
nonpolynomial parts of the second and third order pieces of the inner zone metric are all
$O(b/R)$]. In fact, this behavior means that we need to include the octupole
part of the third order piece of coordinate transformation [in
$\ods{h^\rNP_{(\alpha|\gamma}}{2}\od{A_{|\beta)}{}^\gamma}{3,\le3}$]. (One
would also need to include the octupole part of the second order piece of the
coordinate transformation, but it vanishes.)

The rest of the calculation proceeds as before, with the same general results: We obtain the next two orders' contributions to the matching parameters, with
the nonpolynomial pieces giving $\od{F_{\alpha\beta}}{2} = 0$ and
\<
\label{F3}
\od{F_{\alpha\beta}}{3} = \left[\left(\frac{m_2}{m}\right)^{3/2} + 3\sqrt{\frac{m_2}{m}} + 5\sqrt{\frac{m}{m_2}}\right]\hat{t}_{[\alpha}\hat{y}_{\beta]},
\?
along with $\od{C_k}{j} = 0$ and $\od{M_1}{j} = 0$ for $j \in \{2, 3\}$.
Similarly, the polynomial pieces give
\<
\begin{split}
\od{\bE_{kl}}{2} &= \frac{1}{2}\left[3\hat{x}_k\hat{x}_l - \delta_{kl} + \frac{m}{m_2}(4\hat{x}_k\hat{x}_l - 5\hat{y}_k\hat{y}_l + \hat{z}_k\hat{z}_l)\right],\\
\od{\bB_{kl}}{2} &= \left[5\left(\frac{m}{m_2}\right)^{3/2} + 7\sqrt{\frac{m}{m_2}} - 3\sqrt{\frac{m_2}{m}}\right]\hat{x}_{(k}\hat{z}_{l)},
\end{split}
\?
along with $\od{\bB_{kl}}{1} = \od{\bE_{kl}}{3} = 0$.
The fourth order piece of the coordinate transformation is
\begin{widetext}
\<
\begin{split}\label{X4}
\od{X_\alpha}{4}& =
-\od{\sA_t}{4}\hat{t}_\alpha +
\frac{1}{2}\frac{m_1}{m_2}\frac{\tx}{b}\left[5\frac{\tx^2}{b\tr} -
6\frac{\tx}{\tr} - \frac{\tr}{b}\right]\tr_\alpha
+ \od{\sA_x}{4}\hat{x}_\alpha + \od{\sA_y}{4}\hat{y}_\alpha + \od{\sA_z}{4}\hat{z}_\alpha\\
&\quad +
\left\{\sqrt{\frac{m_2}{m}}\left[\hat{y}_{(\alpha}\od{F_{\beta)0}}{3} -
\hat{t}_{(\alpha}\od{F_{\beta)2}}{3}\right] + \od{F_{\beta\alpha}}{4}\right\}\tx^\beta +
\od{C_\alpha}{4},
\end{split}
\?
where
\<\label{sA4}
\begin{split}
\od{\sA_t}{4} &:= 4\frac{m_1^2}{m_2^2}\frac{b^2}{\tr} + t\biggl[\frac{t^2}{6b^2} +
\frac{5(y^2 - \tx^2) + z^2}{2b^2} + 2\frac{m}{m_2}\frac{\tx^2 - y^2}{b^2} -
\left(\frac{1}{2}\frac{m_2}{m} + \frac{m}{m_2} - 2\right)\frac{\tx}{b}\\
&\quad\; + 1 + \frac{m}{2m_2}
+ \frac{3}{2}\frac{m_2}{m} + \frac{5}{8}\left(\frac{m_2}{m}\right)^2\biggr],\\
\od{\sA_x}{4} &:= \frac{1}{2}\frac{m_1}{m_2}\left[\frac{\tr^2}{b^2} - 5\frac{\tx^2}{b^2} + 6\frac{\tx}{b} + 4\right]\tr + \biggl[\left(\frac{t^2 - z^2}{b^2} + \frac{m}{m_2}\frac{z^2 -
y^2}{b^2} + 4 - \frac{7}{2}\frac{m}{m_2}\right)\tx + \left(\frac{m}{2m_2} - 1\right)\frac{\tx^2}{b}\\
&\quad\; + \left(\frac{m_2}{m} - \frac{m}{2m_2}\right)\frac{y^2}{b} - \left(1 + \frac{3}{2}\frac{m}{m_2} -
\frac{m_2}{m}\right)\frac{t^2}{b} + \left(2 - \frac{3}{2}\frac{m}{m_2}\right)\frac{z^2}{b}\biggr],\\
\od{\sA_y}{4} &:= y\biggl[\frac{m_1}{m_2}\frac{3\tx^2 - y^2}{b^2} - \left(\frac{1}{2}\frac{m_2}{m} + 2\frac{m}{m_2} - 2\right)\frac{\tx}{b} + \frac{5}{2}\frac{m}{m_2} - \frac{9}{2} - \frac{m_2}{2m} -
\frac{5}{8}\left(\frac{m_2}{m}\right)^2\biggr],\\
\od{\sA_z}{4} &:= z\biggl[\frac{t^2 + z^2 + 3y^2 - 5\tx^2}{2b^2} +
\frac{m}{m_2}\frac{\tx^2 - y^2}{b^2} + \frac{m}{2m_2} - \frac{1}{2}\biggr].
\end{split}
\?
The fifth order piece is
\<
\begin{split}\label{X5}
\od{X_\alpha}{5}& = \od{\sA_t}{5}\hat{t}_\alpha -
\left\{\sM_0\left(6,\frac{1}{2}\right)\frac{\tx yt}{b^2\tr} + \bM\left[3 -
\frac{5}{2}\frac{\tx}{b}\right]\frac{\tx^2yt}{b\tr^3}\right\}\tr_\alpha
+ \frac{5}{2}\bM\frac{\tx^2t}{b^2\tr}\epsilon_{0\alpha k3}\tr^k +
\od{\sA_x}{5}\hat{x}_\alpha + \od{\sA_y}{5}\hat{y}_\alpha\\
&\quad + \frac{yzt}{2b^2}\sM(3,3,-5)\hat{z}_\alpha +
\left\{\sqrt{\frac{m_2}{m}}\left[\hat{y}_{(\alpha}\od{F_{\beta)0}}{4} -
\hat{t}_{(\alpha}\od{F_{\beta)2}}{4}\right] +
\od{F_{\beta\alpha}}{5}\right\}\tx^\beta + \od{C_\alpha}{5}.
\end{split}
\?
Here $\epsilon_{\alpha\beta\gamma\delta}$ is the $4$-dimensional Levi-Civita
symbol, with $\epsilon_{0123} = 1$. Additionally, for notational convenience,
we have defined
\<
\begin{split}
\od{\sA_t}{5} &:= \sM_0(1,0)\left[1 + \frac{14}{3}\frac{\tx}{b}\right]\frac{y}{b}\tr
- 4\sM_0(1,1)\frac{ytb^2}{\tr^3} -
\frac{y}{4}\biggl[\sM(1,11,-10)\frac{t^2}{b^2} +
\sM\left(\frac{35}{3},-\frac{35}{3},10\right)\frac{\tx^2}{b^2}\\
&\quad\; - \sM\left(3,-\frac{17}{3},6\right)\frac{y^2}{b^2}
- \sM\left(\frac{5}{3},\frac{7}{3},2\right)\frac{z^2}{b^2}
- \sM(4,-6,6)\frac{\tx}{b} - \sM\left(4,\frac{17}{3},-\frac{17}{3},10\right)\frac{m_2}{m}\biggr],\\
\od{\sA_x}{5} &:= \sM_0\left(3,\frac{3}{2}\right)\frac{yt}{b^2}\tr -
\bM\left[2 + 3\frac{\tx}{b}\right]\frac{yt}{\tr} -
\frac{yt}{b}\left[\frac{1}{2}\sM(-11,9,1)\frac{\tx}{b} + \sM(8,-11,2,1)\right],\\
\od{\sA_y}{5} &:= \left[\sM_0(2,0) +
\sM_0\left(3,-\frac{1}{2}\right)\frac{\tx}{b}\right]\frac{t}{b}\tr +
3\bM\frac{\tx^2t}{b\tr} -
\frac{t}{4}\biggl[\sM\left(3,\frac{7}{3},-\frac{4}{3}\right)\frac{t^2}{b^2} -
\sM_0(7,12)\frac{\tx^2}{b^2} +  \sM_0(7,8)\frac{y^2}{b^2}\\
&\quad\; + \sM(9,-5,0)\frac{z^2}{b^2} + \sM_0(16,6)\frac{\tx}{b} -
\sM\left(0,\frac{47}{3},-\frac{17}{3},10\right)\frac{m_2}{m}\biggr] -
\sM_0(8,0)\tx\biggr\},
\end{split}
\?
and
\<
\begin{split}\label{M}
\sM(A,B,C,D) &:= A\left(\frac{m}{m_2}\right)^{3/2} + B\sqrt{\frac{m}{m_2}} + C\sqrt{\frac{m_2}{m}} + D\left(\frac{m_2}{m}\right)^{3/2},\\
\sM(A,B,C) &:= A\left(\frac{m}{m_2}\right)^{3/2} + B\sqrt{\frac{m}{m_2}} + C\sqrt{\frac{m_2}{m}},\\
\sM_0(A,B) &:= \sM(A,-A-B,B) = A\left(\frac{m}{m_2}\right)^{3/2} - (A + B)\sqrt{\frac{m}{m_2}} + B\sqrt{\frac{m_2}{m}},\\
\bM &:= \sM_0(0,-1) = \sqrt{\frac{m}{m_2}} - \sqrt{\frac{m_2}{m}},
\end{split}
\?
where $\sM_0(A,B)$ and $\bM$ vanish in the limit $m_1 \to 0$ ($\Rightarrow m \to m_2$).
\end{widetext}

\subsection{Summary of matching results}
\label{CT}

The final output of the matching is a set of expressions for the tidal fields, which
are given explicitly in Eqs.~\eqref{TFs}; a relation between the mass parameters $M$
and $m_1$, which we found to be equal [up to uncontrolled remainders of $O(v^4)$]; and
the coordinate transformation necessary to place the inner zone metric in the same
coordinate system as the near zone metric to the order we matched.
To obtain the full coordinate transformation, we start from Eq.~\eqref{Xexpansion} and
insert the various pieces we have read off or chosen. We shall take anything we were
unable to fix by matching to be zero---here this will be $C_0$ (to all orders), along
with $\od{F_{\alpha\beta}}{j}$ and $\od{C_k}{j}$ for $j \in \{4,5\}.$
With the results of our matching, this means that we have $C_\alpha = 0$ (to
all orders). We took the zeroth-order piece of the coordinate transformation
to simply be the expected translation of the origin from the binary's
center-of-mass to hole~1, so 
\<
\od{X_\alpha}{0} = \tx_\alpha := x_\alpha - (m_2/m)b\hat{x}_\alpha.
\?
With $\od{C_\alpha}{1} = 0$, we also have
\<
\od{X_\alpha}{1} = \od{F_{\beta\alpha}}{1}\tx^\beta,
\?
where
\<
\od{F_{\alpha\beta}}{1} = 2\sqrt{\frac{m_2}{m}}\hat{t}_{[\alpha}\hat{y}_{\beta]}.
\?
Continuing onward, $\od{X_\alpha}{2}$ is given by Eq.~\eqref{X2}, where $\od{F_{\alpha\beta}}{2} = 0$ and $\od{C_\alpha}{2} = 0$; $\od{X_\alpha}{3}$ comes from Eq.~\eqref{X3}, where
\<
\od{F_{\alpha\beta}}{3} = \sM(0,5,3,1)\hat{t}_{[\alpha}\hat{y}_{\beta]},
\?
and $\od{C_\alpha}{3} = 0$. [$\sM(A,B,C,D)$ is defined in Eq.~\eqref{M}.] Similarly, $\od{X_\alpha}{4}$ can be obtained from Eq.~\eqref{X4}, and we take $\od{F_{\alpha\beta}}{4} = 0$ and $\od{C_\alpha}{4} = 0$. Finally, $\od{X_\alpha}{5}$ is given in Eq.~\eqref{X5}; again, we set $\od{F_{\alpha\beta}}{5} = 0$ and $\od{C_\alpha}{5} = 0$.

Looking back over this coordinate transformation, it is possible to gain some
physical intuition about what it is accomplishing: The expected Lorentz boost
due to the holes' orbital motion is present (through third order, which is the
highest order at which we have fixed all the coordinate transformation, up to
a possible temporal shift). We also have the lowest-order piece of the
transformation between Cook-Scheel and harmonic PN coordinates for an
unperturbed Schwarzschild black hole. [This is given by the first term in
$\od{\sA_t}{4}$ in Eq.~\eqref{sA4}.] The remainder
of the coordinate transformation is probably mostly concerned with effecting the
transformation from locally inertial coordinates centered on the black hole to PN
barycentric coordinates. In addition, as we noted previously, the polynomial and
nonpolynomial parts of the full coordinate transformation are related to the
individual holes in the expected manner: The nonpolynomial parts are 
associated with hole~1 and vanish in the limit $m_1 \to 0$. The polynomial
pieces are associated with hole~2---indeed, everything except
for the piece of the background Cook-Scheel-to-PN-harmonic transformation
vanishes in the limit $m_2 \to 0$.

\section{Far zone metric}\label{FZ}

The direct integration of the relaxed Einstein equations (DIRE) approach
\cite{PW1,PW2} can be used to compute the full $4$-metric $g_{\alpha\beta}$
(in harmonic coordinates) in both the near and far zones. The resulting far
zone metric is expressed in
terms of derivatives of multipole moments obtained by integrating the
``effective'' stress-energy pseudotensor over the near zone. One also obtains
nonlinear contributions from integrating over the far zone (known as the
\emph{outer integral} in the DIRE approach), though only two of the resulting
terms appear in the metric perturbation $h^{\alpha\beta}$ [defined in Eq.~(2.2)
of~\cite{PW1}] to the order we are considering.

In this formalism, the metric perturbation in the far zone can be expressed in terms of
the source multipoles $\sI^Q$ and $\sJ^Q$ via [Eqs.~(5.12)
in~\cite{PW1}]\footnote{We have corrected a sign
error in Pati and Will's expression for $h^{kl}$: They give coefficients of
$+2/3$ and $+8/3$ for the second and third terms. The correct signs can be
obtained from Pati and Will's Eqs.~(2.13) and~(4.7b) in~\cite{PW1}.}
\begin{subequations}
\begin{align}
\label{metric-perturbation}
\begin{split}
h^{00} &= 4 \frac{{\cal{I}}}{r} + 2 \partial_{kl} \left[ \frac{{\cal{I}}^{kl}(u)}{r} \right] - \frac{2}{3} \partial_{klm} \left[ \frac{{\cal{I}}^{klm}(u)}{r} \right]\\
&\quad + 7  \frac{{\cal{I}}^{2}}{r^{2}} + O(v^{6}),
\end{split}\\
\begin{split}
h^{0k} &= - 2 \partial_{l} \left[ \frac{\dot{\cal{I}}^{kl}(u)}{r} \right] + 2 \epsilon^{lkp} \frac{n^{l} {\cal{J}}^{p}}{r^{2}} + \frac{2}{3} \partial_{lp} \left[ \frac{\dot{\cal{I}}^{klp}(u)}{r} \right]\\
&\quad + \frac{4}{3} \epsilon^{lkp} \partial_{ls} \left[ \frac{{\cal{J}}^{ps}(u)}{r} \right] + O(v^{6}),
\end{split}\\
\begin{split}
h^{kl} &= 2  \frac{\ddot{\cal{I}}^{kl}(u)}{r} - \frac{2}{3} \partial_{p} \left[ \frac{\ddot{\cal{I}}^{klp}(u)}{r} \right]\\
&\quad - \frac{8}{3} \epsilon^{ps(k|} \partial_{s} \left[ \frac{\dot{\cal{J}}^{p|l)}(u)}{r} \right] + \frac{{\cal{I}}^{2}}{r^{2}} \hat{n}^k\hat{n}^l + O(v^{6}).
\end{split}
\end{align}
\end{subequations}
Here
$r := \|\vec{x}\|$ is the distance from the binary's center-of-mass to the
field point and $\hat{n}^{k} := x^{k}/r$ is its associated unit vector. The
$(\sI/r)^2$ terms are the two
contributions from the outer integrals mentioned previously. We have included
all the terms Pati and Will give, even though some of the ones in the purely temporal
and spatial components only contribute terms that are of a higher order than
we need here. We do this for
completeness and also because we shall need these higher-order terms when we
construct the extension to this data set (in Appendix~\ref{ExtraTerms}).
  
The source multipoles $\sI^Q$ and $\sJ^Q$ ($Q$ is a multi-index) are defined in Eqs.~(4.5)
of~\cite{PW1} With these definitions, the mass monopole $\sI$ is simply (a PN corrected
version of) the total mass of the system, the dipole moment $\sI^k$ is the center
of mass vector (so it vanishes in our coordinate system), and the current
dipole $\sJ^k$ is the total angular momentum. One can
show that these three quantities are conserved up to radiative losses.

The source multipoles can be expanded in the PN approximation to find
\begin{subequations}
\begin{align}
{\cal{I}} &=  m_{1} \left(1 + \frac{1}{2} v_{1}^{2} - \frac{m_{2}}{2b}\right) + (1 \leftrightarrow 2) + O({b} v^{4}),
\\
{\cal{J}}^{k} &= \epsilon^{klp} m_{1} x_{1}^{l} v_{1}^{p}  + (1 \leftrightarrow 2) + O({b}^{2} v^{3}),
\\
\begin{split}\label{Ikl}
{\cal{I}}^{kl} &= m_{1} x_{1}^{kl} \left(1 + \frac{1}{2} v_{1}^{2} - \frac{m_{2}}{2 b} \right) + \frac{7}{4} m_{1} m_{2}  b \delta^{kl}\\
&\quad + (1 \leftrightarrow 2) + O({b}^{3}v^{4}),
\end{split}\\
{\cal{J}}^{kl} &= \epsilon^{kps} m_{1} v_{1}^{s} x_{1}^{pl} + (1 \leftrightarrow 2) + O({b}^{3} v^{3}) ,
\\
{\cal{I}}^{klp} &= m_{1} x_{1}^{klp} + (1 \leftrightarrow 2) + O({b}^{4}v^2).
\end{align}
\end{subequations}
The first two come from Will and Wiseman's Eqs.~(4.16)~\cite{WW}, and the
remainder from Pati and Will's Eq.~(D1)~\cite{PW2}. (Even though we only need
the lowest-order piece of $\sI^{kl}$ here, we include the $1$PN corrections
that Pati and Will give since we will need them in our construction of the
higher-order extension in Appendix~\ref{ExtraTerms}.) Here the notation is
mostly the same as for the near zone metric and was defined in Sec.~\ref{NZ}.
One new definition is $x_{1}^{kl} := x_{1}^{k} x_{1}^{l}$ (with similar
definitions holding for different vectors, as well as larger collections of
indices).

Even though this structure is somewhat obscured in the expression Pati and
Will give for it, the far zone PN expansion of the metric perturbation is
clearly a bivariate expansion. There is a post-Minkowskian expansion in powers
of $G$,
given by radiative multipole moments, and a post-Newtonian expansion in $v$ (or, more
formally, $1/c$)
of these radiative multipole moments in terms of source moments computed in
the near zone (see, e.g.,~\cite{BlanchetLRR}). Although formally the near and
far zone expansions are independent, in practice there is a relation between
their expansion parameters. This is given by the definition of
the far zone, $r \gtrsim \lambdabar = b/2v$, from which one finds that
 \be
 \label{ordering}
\frac{b}{r} \lesssim 2v.
 \ee
As explained in Sec.~IV~C of~\cite{PW1}, formal consistency requires that we treat each {\emph{additional}} inverse power of $r$ as raising the effective order by {\emph{at least}} one factor of $v$. [This order counting is also equivalent to the order counting
Alvi~\cite{Alvi} uses in computing the far zone metric, though he keeps terms through
$O(v^5)$ in all components.] For example, with this far zone order counting
\begin{align}
\frac{{\cal{I}}}{r}& \propto \;\,\frac{m}{r}\;\;\, =  O(v^{2}), \nonumber\\
\sI^{kl}\partial_{kl}r^{-1} & \propto \frac{m}{r} \frac{x_{1}^{2}}{r^{2}} =  O(v^{4}),\\
\sI^{klp} \partial_{klp}r^{-1} & \propto \frac{m}{r} \frac{x_{1}^{3}}{r^{3}} = O(v^{5}),\nonumber
\end{align}
%
since $m/r$ comes with a factor of $G$ due to the post-Minkowskian expansion
while $x_{1}/r$ does not. This also means that we treat the $(\sI/r)^2 \propto
(m/r)^2$ term as $O(v^4)$, not $O(v^5)$.

With this order counting in mind, we now determine
the order of each source multipole moment. This leads to the following
orders for the components of the metric perturbation in the far zone: $h^{00} =
O(v^2)$, $h^{0k} = O(v^{4})$, and $h^{kl} = O(v^{4})$.
Note that this is not the standard order counting of the metric perturbation
in the near zone, since there we have $b/r = O(v^0)$, which thus leads to
$h^{00} = O(v^{2})$, $h^{0k} = O(v^{3})$, and $h^{kl} = O(v^{4})$.  The far
zone order counting allows us to expand the full $4$-metric in the far zone to
obtain [from Eqs.~(4.2) in~\cite{PW1}]
\begin{subequations}
\label{metric}
\begin{align}
g_{00} &= -\left[1 - \frac{1}{2} h^{00} + \frac{3}{8} \left(h^{00}\right)^{2}  \right]
+ \frac{1}{2} h^{kk},
\\
g_{0k} &= - \left[ 1 - \frac{1}{2} h^{00} \right]h^{0k},
\\
g_{kl} &= \left[1 + \frac{1}{2} h^{00} - \frac{1}{8}\left(h^{00}\right)^{2} -
\frac{1}{2} h^{pp}\right]\delta^{kl} + h^{kl},
\end{align}
\end{subequations}
with remainders of $O(v^{6})$. Here we
have kept the $-(1/2)h^{00}$ term in the expression for $g_{0k}$ for formal
consistency. It only gives terms of
$O(v^6)$ with our order counting (which we neglect here), but would give terms
of $O(v^5)$ if we had used the standard near zone order counting, where
$h^{0k} = O(v^3)$. This is the only $O(v^6)$ term in the expression for $g_{0k}$, so the
uncontrolled remainder in that expression is thus actually $O(v^7)$.

We can now easily compute the full metric by carrying out all the differentiation in the
expression for the metric perturbation, Eqs.~\eqref{metric-perturbation}. In
doing this, it is
important to realize that the multipoles depend on retarded time, which must be
carefully accounted for when differentiating. Eqs.~\eqref{metric} then give the full metric. In performing this calculation,
we assume a (quasi)circular orbit, so the acceleration and the trajectories are
parallel or antiparallel to each other and the velocity is perpendicular
to either of them to $O(v^{5})$. We thus have
$\vec{x}_{1} \cdot \vec{v}_{1} = O(v^5) = \vec{a}_{1} \cdot \vec{v}_{1}$ and
$\vec{a}_A = -\omega^2\vec{x}_A + O(v^5)$, where $a_A^k := \dot{v}_A^k$ is the acceleration
of point particle $A$. [The latter leads to expressions such as
$(\vec{v}_{1} \cdot \vec{j}_{1} ) = - \omega^{2} v_{1}^{2} + O(v^5)$, where $j_A^k :=
\dot{a}_A^k$ is the jerk of point particle $A$.]
After much algebra, we finally obtain the full metric in the far zone:
\begin{widetext}
\begin{subequations}\label{FZ_metric}
\begin{align}
\begin{split}
g_{00} + 1 &=  \frac{2 m_{1}}{r} + \frac{m_{1}}{r} \biggl\{v_{1}^{2} - \frac{m_{2}}{b} +
2 \left(\vec{v}_{1} \cdot \hat{n}\right)^{2} - \frac{2 m}{r} +
6 \frac{\left(\vec{x}_{1} \cdot \hat{n} \right)}{r} \left(\vec{v}_{1} \cdot \hat{n} \right) - \frac{x_{1}^{2}}{r^{2}} + \frac{\left(\vec{x}_{1} \cdot \hat{n} \right)^{2}}{r^{2}} \left(3 - 2 r^{2} \omega^{2} \right)\biggr\}\\
&\quad + (1 \leftrightarrow 2) + O(v^{5}),
\end{split}
\\
\begin{split}
g_{0k} &=
-\frac{m_1}{r}\biggl\{4{\left(\vec{v}_1 \cdot \hat{n} \right)} + 4\,{ \left(\vec{v}_1 \cdot \hat{n} \right)^{2}} - 6\,{\left(\vec{x}_1 \cdot \hat{n} \right)^{2}{\omega}^{2}} + 4\,{\frac {\left(\vec{x}_1 \cdot \hat{n} \right)}{r}} + 12\,{\frac {\left(\vec{x}_1 \cdot \hat{n} \right)}{r}}{\left(\vec{v}_1 \cdot \hat{n} \right)} - 2\,{\frac {x_1^{2}}{{r}^{2}}} + 6\,{\frac {{\left(\vec{x}_1 \cdot \hat{n} \right)}^{2}}{{r}^{2}}}   \biggr\} v_1^{k}\\
&\quad +  
\frac{m_1}{r}\biggl\{ 4\left(\vec{x}_{1} \cdot \hat{n} \right) + 8\,{\left(\vec{x}_1 \cdot \hat{n} \right)}\,{\left(\vec{v}_1 \cdot \hat{n} \right)} - 2\,{\frac {x_1^{2}}{r}} + 6\,{\frac {{\left(\vec{x}_1 \cdot \hat{n} \right)}^{2}}{r}} 
 \biggr\} {\omega}^{2}x_1^{k}
+ (1 \leftrightarrow 2) + O(v^{6}),
\end{split}
\\
\begin{split}\label{FZ_gkl}
g_{kl} - \delta_{kl} &= \frac{2 m_{1}}{r} \delta_{kl} + \frac{m_{1}}{r} \biggl\{
\biggl[ v_{1}^{2} - \frac{m_{2}}{b} + 2 \left(\vec{v}_{1} \cdot \hat{n}\right)^{2} + \frac{m}{r}
+ 6 \frac{\left(\vec{x}_{1} \cdot \hat{n}\right)}{r} \left(\vec{v}_{1} \cdot \hat{n}\right)  
- \frac{x_{1}^{2}}{r^{2}} +
\frac{\left(\vec{x}_{1} \cdot \hat{n}\right)^{2}}{r^{2}} \left(3 - 2 \omega^{2} r^{2} \right)
  \biggr]\delta_{kl}\\
&\quad + \frac{m}{r} \hat{n}^{kl} + 4 v_{1}^{kl} - 4 \omega^{2} x_{1}^{kl}
\biggr\}
+ (1 \leftrightarrow 2) + O(v^{5}),
\end{split}
\end{align}
\end{subequations}
\end{widetext}
where everything is evaluated at the retarded time $u := t - r$. [This
expression agrees with Alvi's result, given in his Eq.~(2.17)~\cite{Alvi},
though he also includes the $O(v^5)$ terms in the purely temporal and spatial
components. We include these pieces---as well as even higher-order ones---when
we construct the extension to the data in Sec.~\ref{InclRad} and have checked
that we agree with Alvi about all the $O(v^5)$ terms.]

\subsection{Evolution of the binary's phase and separation}\label{phasing}

Even though the effects of radiation reaction on the binary's orbital
separation and phase are formally small, only beginning at $O(v^5)$,
they produce large corrections to the
far zone metric when one is far away from the binary, since the retarded time
at which one is evaluating the binary's parameters becomes large. For instance,
even as close as $r = 50m$, which is inside the outermost extraction radius
(usually well inside) for all the simulations used in the Samurai
project~\cite{Samurai}, the phase difference between a
circular orbit [using the $3$PN expression for $\omega$ given in Eq.~(190)
of~\cite{BlanchetLRR}] and $3.5$PN inspiral (the
computation of which is detailed below) is $\sim 0.015$ radians for an equal-mass
binary with a separation of $10m$. (This phase difference should be compared with
the averaged frequency domain phase accuracy required for parameter estimation with
Advanced LIGO, viz., $0.007$
radians, from~\cite{LOB}. While such a comparison is not really
warranted---see~\cite{Lindblom} for some discussion---it gives a rough idea of the
required accuracy.) See Sec.~IV~A in Kelly
\emph{et al.}~\cite{Kellyetal} for further discussion of the necessity of using
formally higher-order PN results in obtaining the far zone metric.

We thus use the most accurate ($3.5$PN) expression for the inspiral, as given
by Blanchet~\cite{BlanchetLRR}, who obtains it from an energy balance argument.
The phase itself is given by Blanchet's Eq.~(234)~\cite{BlanchetLRR}, where it is expressed in
terms of the dimensionless time variable $\Theta$, defined in his Eq.~(232)~\cite{BlanchetLRR}.
Since we
are evaluating everything at the retarded time $u$, we have $\Theta =
(\eta/5m)(t_c - u).$
Here the ``coalescence time'' $t_c$ is defined as the time at which the binary's
frequency goes to infinity (or, equivalently, its separation goes to zero). One
can calculate this in the PN approximation by using the energy balance relation
$dE/dt = -\mathcal{L}$, where $E$ and $\mathcal{L}$ are the binary's energy and
gravitational wave luminosity, respectively. These are given in terms of $\gamma
:= m/b$ though $3.5$PN in Blanchet's Eqs.~(191) and (230)~\cite{BlanchetLRR}, respectively.
We can then compute $t_c$ by integrating $dt/db = (dE/db)(dt/dE) =
-(dE/db)/\mathcal{L}$ from $b = 0$ to $b = b_0$ (where $b_0$ is the binary's separation at
$t = 0$). Here we expand the
quotient as a power series (to $3.5$PN) rather than using a Pad\'{e} approximant (or
performing any resummation of the energy or luminosity), as is
sometimes done in the literature (see, e.g.,~\cite{DIS}).

\begin{figure}[htb]
\epsfig{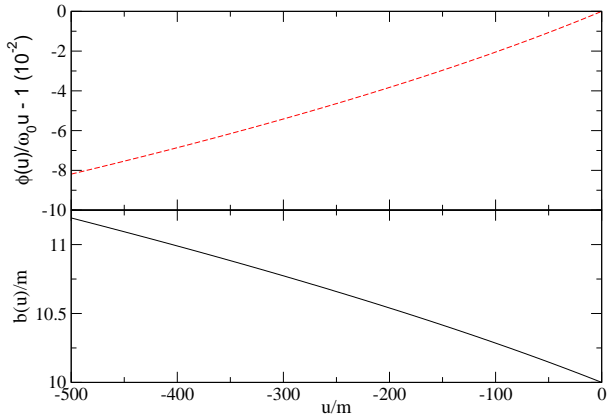} 
\caption{\label{separation_and_phase} The $3.5$PN results for the past history of an
equal-mass binary's separation, starting from $10m$ at $u = 0$, along with the fractional
deviations of its phase from $\omega_0u$.}
\end{figure}

With the $3.5$PN expression for $t_c$ in hand, we can simply substitute it
into Blanchet's Eq.~(234)~\cite{BlanchetLRR} to obtain the phase as a function of $u$, making sure to
expand to $3.5$PN order after substituting. We add a constant to the phase so that
it is zero when $t = 0$, to be consistent with our choice of initial phase in the
matching (i.e., so the holes initially lie on the $x$-axis, with our expression for
the orbit). We also take the freely specifiable gauge constant $r_0'$
and constant of integration $\Theta_0$ to be $m$ and $1$, respectively.
(The dependence of $\omega$ on $u$ is then obtained by
differentiating the phase with respect to time.) To obtain the retarded time
dependence of $b$, we use $b = m/\gamma$, along with the expressions
for $\gamma$ in terms of $x$ [Blanchet's Eq.~(193)~\cite{BlanchetLRR}] and $x$ in terms of
$\Theta$ [Blanchet's Eq.~(233)~\cite{BlanchetLRR}], expanding the quotient consistently to
$3.5$PN. (We take the appearance of $b$ in a logarithm in the expression
for $\gamma$ to consist only of $b$'s lowest-order dependence on $\Theta$,
viz., $4m\Theta^{1/4}$.) N.B.: The final expansions of the expressions for
$\phi$ and $b$ are important. If they are not performed, then one does not
recover the expected
values for $b$ and $\omega$ at $u = 0$ [viz., $b_0$ and $\omega_0$,
respectively---here $\omega_0$ is the binary's $3$PN angular velocity for
$b = b_0$ obtained from Blanchet's Eq.~(190)~\cite{BlanchetLRR}]. We do not display the
resulting expressions, as they
are quite lengthy, and best handled entirely within a computer algebra system.
(We have carried out the calculations in {\sc{Maple}} and our scripts are
available at~\cite{Wolfgangs_website}.) Our results can be seen graphically in
Fig.~\ref{separation_and_phase}, where we plot the past history of an equal-mass binary's
separation (starting from $b_0 = 10m$) along with the fractional deviations of
its phase from $\omega_0u$.

Since we are effectively using higher-order equations of motion, due to the
higher-order phasing relations, we would have liked to include higher-order terms in the
relative-to-center-of-mass (relative-to-COM) variable relation as well, for consistency.
However, the resulting expressions
for the far zone metric components are algebraically too complex for {\sc{Maple}} to
handle, so we had to forego including these terms. If one expands the contribution of
the relative-to-center-of-mass relation to the (far or near zone) metric components
consistently, then, to the order we have considered, all that one needs is the
lowest-order Newtonian version of this
relation, given in Eq.~\eqref{rel-to-COM}, and this is all we have used in the far zone
metric. Fortunately, unlike the secular radiation reaction effects in the binary's
phase and separation considered above, the neglected PN corrections to the
relative-to-COM relation are numerically small in addition to being
formally small: They first enter the metric at $O(v^8)$ (in the purely temporal
component). Moreover, maximizing over mass ratio (all the PN corrections to the
relative-to-COM relation vanish for an equal-mass binary), the magnitude of the largest
of these is $\sim 5 \times 10^{-7}$ at the inner edges of the near-to-far buffer zone
for a separation of $b = 10m$;
for comparison, the uncorrected $(m_1/r)v_1^2 + (1 \leftrightarrow 2)$ term is
$\sim 10^{-4}$ in this situation.

We \emph{were} able to include these additional terms in the relative-to-center-of-mass
relation in the near zone metric: For consistency, we include the same higher-order
terms in the phasing and separation in the near zone metric as in the far zone metric,
so it makes sense to attempt to include the higher-order relative-to-COM
relation there, even
though we were unable to do so in the far zone metric. In fact, these corrections
contribute to the near zone metric at one order lower than to the far zone metric
[viz., $O(v^7)$ in the spatiotemporal components] and are numerically quite a bit
larger: Maximizing over mass ratio, the magnitude of the largest of these corrections
is $\sim 2 \times 10^{-3}$ at the inner edge of the inner-to-near buffer
zone for a separation of $10m$; for comparison, the magnitude of the uncorrected
$-4(m_1/r_1)v_1^k$ piece is $\sim 0.2$ in this situation.

Blanchet and Iyer give these corrections through $3$PN in Eqs.~(3.11)--(3.14)
of~\cite{BI}. We have specialized their result to a circular orbit
by using the $2$PN expression for $\omega$ to express $v$ in terms of $m$,
$b$, and $\eta$. Blanchet gives this relation specialized to
a circular orbit through $2.5$PN in Eq.~(187) of~\cite{BlanchetLRR}, so we
shall just quote the $3$PN contribution to $\vec{x}_A$ here:
\<
\vec{x}_A^\mathrm{3PN} = -\eta\frac{(m_1 - m_2)m^2}{b^3}\biggl[\frac{7211}{1260} + \eta -
\frac{22}{3}\log\left(\frac{b}{r_0''}\right)\biggr]\vec{b}.
\?
In this expression, $r_0''$ is another freely specifiable gauge constant which,
though \emph{a priori} different from $r_0'$, we shall take to have the same
value, viz., $m$. [This is equivalent to taking the related gauge constants
$r_1'$ and $r_2'$ to both be $m$---see Eqs.~(3.15) and (3.19) in~\cite{BI}.]
The expression for the binary's separation vector that we
substitute into the resulting relative-to-center-of-mass relation to obtain the
trajectories of the point particles is $\vec{b} = b(\hat{x}\cos \phi +
\hat{y}\sin \phi),$ where $b$ and $\phi$ are functions of $t$ (in the near zone) or
$u$ (in the far zone).

\section{Including higher-order terms}\label{InclRad}

\begin{table*}
\begin{tabular}{ccccccc}
\hline\hline
\multirow{2}{*}{Zone} & & \multirow{2}{*}{Attribute} &\multicolumn{4}{c}{Versions}\\
 & & & \verb,O4_NoOct, & \verb,O4, & \verb,O5, & \verb,all,\\
\hline
\multirow{2}{*}{Inner} & \multirow{2}{*}{$\Big\{$} & Time dependence & Perturbative & Perturbative & Full & Full\\
& & Fourth order octupoles & No & Yes & Yes & Yes\\
Near & \multirow{2}{*}{$\Big\}$} & \multirow{2}{*}{Metric order} & \multirow{2}{*}{$O(v^4)$} & \multirow{2}{*}{$O(v^4)$} & \multirow{2}{*}{$O(v^5)$} & \multirow{2}{*}{Full extended}\\
Far & & & & & &\\
\hline\hline
\end{tabular}
\caption{\label{Versions} An overview of the contents of the various data sets we
considered.}
\end{table*}

Here we construct an extension to our data using various readily
available higher-order terms. This extension includes all the $O(v^5)$ terms in
the near and far zones, but also includes even higher-order terms that do not
improve the data's formal accuracy. [We have to change our far zone
order counting somewhat to be able to claim that we have all the $O(v^6)$
pieces of the spatiotemporal components of the far zone metric; these
components are necessary to obtain initial data valid through $O(v^5)$.] The
general
philosophy is that adding higher-order terms can often improve the quality of
the data in practice, even if it does not improve their formal accuracy.
As we have seen in the previous section, this is particularly true in the far
zone, where the binary's phase evolution depends sensitively on the inclusion
of quite high-order radiation reaction terms.

There are also more specific reasons for including certain of these terms: We would like
for a putative evolution of our data to be able to be compared directly with 
Kelly \emph{et al.}'s evolution of their data~\cite{Kellyetalev}. (Such a comparison will
give an indication of how much of the junk radiation is due to the failure of the
initial data to include the correct tidal deformations.) Kelly \emph{et al.}\ include
the $O(v^5)$ pieces of the spatial metric in
the near zone [though not the matching $O(v^6)$ pieces of the extrinsic
curvature]. The extension we have constructed includes these
terms, as well. It also includes (as noted above) the $O(v^5)$ terms in the far zone,
along with the $O(v^6)$ terms in the extrinsic curvature in the near and far
zones, so the extended data are valid through $O(v^5)$ in those zones. However,
we have coded
our data in such a way that one can easily produce a data set that
only includes the pieces that Kelly \emph{et al.}\ have, or some other subset of the
pieces in the extension.

We would have liked to have included the $O(v^5)$ terms in
the inner zone as well, for completeness, but it is not possible to obtain
initial data in that zone that is formally $O(v^5)$ while still including the
quadrupole pieces: We would need the $O(v^6)$ pieces
of (the spatiotemporal components of) the inner zone metric. The quadrupole
parts of these include hexadecapole tidal fields, and knowledge of how
those fields enter the inner zone metric would require nonlinear black hole
perturbation theory. However, we \emph{are} able to calculate the polynomial
parts of the fourth order octupole pieces---the results are presented in
Appendix~\ref{TFHO}---and include them in the extension. (These pieces
include the $1$PN correction to the electric octupole tidal field.)

Moreover, we are also able to include the full time dependence of the tidal
fields: See Appendix~\ref{TFHO}. It seems desirable to include these terms,
since some of them are necessary for obtaining the pieces of the extrinsic
curvature that describe how the tidal fields evolve: We already have the
necessary, linear-in-$t$ pieces for the lowest-order quadrupole fields, but not
for their $1$PN corrections, the explicit
appearance of their time derivatives, or any of the octupole fields. As is discussed
more fully in Appendix~\ref{metric_comp}, the additional time dependence we include
in the tidal fields is only the lowest-order dependence on $t$ alone, not the
additional space-dependence that comes from the tidal fields' dependence on the
Eddington-Finkelstein coordinate: The neglected dependence would generate
contributions to the metric that combine with the unknown contributions from
time derivatives of the tidal fields, while the $t$-dependent contributions
are not entangled in this manner. We use $t$ instead of $T$ since the full
time dependence of the tidal fields comes, in
effect, from performing the matching at different (near zone coordinate) times
$t$. We thus want these higher-order contributions to the tidal fields to depend
on that time, instead of $T$ (the inner zone time coordinate).

We have \emph{not} attempted
to obtain the full time dependence of the coordinate transformation, since this is a rather
more involved task than obtaining that of the tidal fields. Moreover, our rationale for
including the tidal fields' full time dependence was to improve the evolution of the tidal
perturbations: Including the full time dependence of the coordinate transformation
would only improve the agreement of the inner and near zone metrics in the buffer zone, while
the largest effect of the tidal perturbations (e.g., in reducing the high-frequency junk
radiation) presumably comes from closer to the horizons.

The other higher-order pieces we have added are all in the purely temporal
components of the near and far zone metrics. We have included these since
Blanchet, Faye, and Ponsot~\cite{BFP} give an explicit expression for the
purely temporal component of the near zone metric through $O(v^7)$ and it is
easy enough to calculate the matching terms in the far zone.
The specifics of where we obtained all the extra terms are given in
Appendix~\ref{ExtraTerms}, along with the accompanying caveats and new order
counting in the far zone. We refer to these ``state-of-the-art'' versions of the near
and far zone metrics as the full extended versions.

We have considered the effects of these additional
terms on how well the metrics stitch together by putting together four versions of the
data, as shown in Table~\ref{Versions}. While the versions in the table are given the
labels we use for the corresponding {\sc{Maple}} scripts and C code (available
at~\cite{Wolfgangs_website}), we shall usually refer to them by the order of their near
and far zone metrics [i.e., as $O(v^4)$, $O(v^5)$, and full extended]. This will not
cause confusion, since we will almost always consider just the right-most three. While
the \verb,O4_NoOct, version is
important, since it contains only the pieces that can be included consistently in
the matching (except for the higher-order terms necessary to obtain the phasing in
the far zone accurately and the analogous terms in the near zone), we only consider
it in Fig.~\ref{IZ-NZ-oct_constraint_comp}. Thus, when we refer to the $O(v^4)$
version without any qualifiers, we mean \verb,O4,, the version including the fourth order
octupole pieces. See Appendix~\ref{metric_comp} for the specific details of how all
these metrics are calculated. (N.B.: The near zone metrics in all of
these versions contain the background resummation detailed in Sec.~\ref{BR}.)

In the next section, we compare the extended data with their lower-order counterparts: The
volume elements of the $O(v^4)$, $O(v^5)$, and full extended sets of data are compared in the near zone in
Figs.~\ref{IZ-NZ} and~\ref{IZ-NZ-vol-el-diffs} and in the far zone in
Figs.~\ref{NZ-FZ-merged} and~\ref{FZ-oscillations}. Their constraint
violations are
compared in Figs.~\ref{IZ-NZ-version-constraint_comp}, \ref{IZ-NZ-version-constraint_comp_q3}, and~\ref{NZ-FZ-version-constraint_comp}. We compare the
constraint violations of the different versions of the inner zone metric in
Figs.~\ref{IZ-comparison} and~\ref{IZ-NZ-oct_constraint_comp}.
In general, the additional terms reduce the constraint violations and improve the
overall matching, or at least do not affect them adversely. The only exception to
this is the addition of the full time dependence in the inner zone, which increases
the constraint violations in the inner zone. However, it does not increase them by
much, and they were originally quite small: Our philosophy is thus to include these
terms, which we think will improve the data's evolution, even at the cost of slightly
higher constraint violations.

\section{Numerical considerations}\label{NMatching}

\begin{figure}[htb]
\epsfig{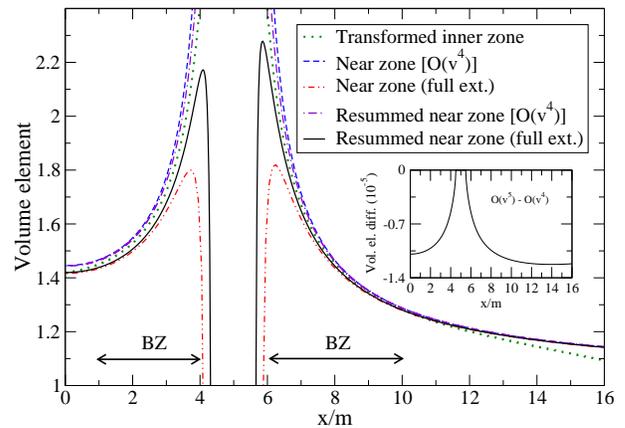} 
\caption{\label{IZ-NZ} The volume elements of the transformed inner zone
metric (dotted line) along with those of the $O(v^4)$ and extended versions of the near
zone metric. These are plotted both with [$O(v^4)$: dot-dashed line; extended: solid line] and
without [$O(v^4)$: dashed line; extended: dot-dot-dashed line] background resummation. We have also displayed the differences between the volume
elements of the $O(v^4)$ and $O(v^5)$ resummed metrics in the inset. We calculated
all of these volume elements for our equal-mass test system and have plotted
them along the $x$-axis near hole $1$. (The associated point particle is at $x/m =
5$.) We have indicated the rough locations of the intersection of the $x$-axis with hole $1$'s
inner-to-near buffer zone (BZ) using double-headed arrows.}
\end{figure}

\begin{figure}[htb]
\epsfig{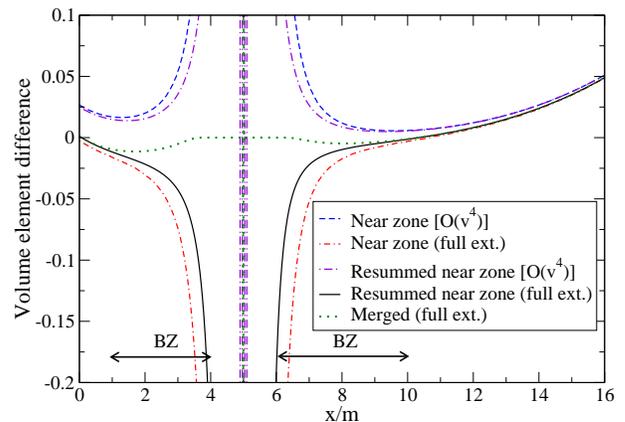} 
\caption{\label{IZ-NZ-vol-el-diffs} The differences between the volume elements of
various metrics with that of the full extended inner zone metric. The
various metrics considered are the near zone metrics plotted in Fig.~\protect\ref{IZ-NZ} 
(using the same colors and line styles as in that figure) as well as
the merged metric constructed in Sec.~\protect\ref{Transitions}. We have indicated the rough
locations of the intersection of the $x$-axis with hole $1$'s
inner-to-near buffer zone (BZ) using double-headed arrows.}
\end{figure}

\begin{figure}[htb]
\epsfig{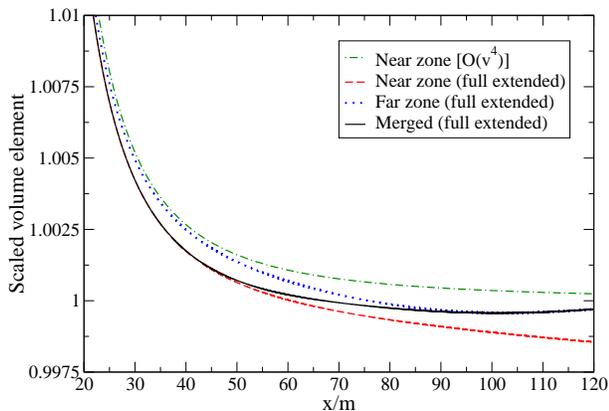} 
\caption{\label{NZ-FZ-merged} The scaled volume elements of the (resummed) near
zone [both $O(v^4)$ and extended], far zone, and merged metrics. These are calculated for our
equal-mass test system and plotted along
the $x$-axis in the portion of the near-to-far transition region that lies to the right
of hole $1$. To obtain a less crowded plot, we have scaled these by the
contribution to the volume element of the lowest-order nontrivial pieces of the far zone
metric, viz., $(1 + 2m/r)\sqrt{1 - 4m^2/r^2}$. The buffer zone is roughly the region
in which we transition, which, in turn, is roughly the region plotted.
}
\end{figure}

To get an idea of how well the matching is working numerically, we plot the
volume elements of the various 4-metrics in the vicinity of their buffer zones.
(In all these plots, we use the full extended version of the data discussed in the
previous section, unless otherwise noted.)
Of course, we first transform the inner zone metric using the coordinate transformation
obtained from the algebraic matching above. To obtain a simple, easily
interpreted plot, we choose the test system $b = 10m$, $m_{1} = m_{2}$
(the mass is in arbitrary units) and restrict our attention to the $t = 0$ timeslice
and the $x$-axis (i.e., we consider the spatial slice along the separation between the
holes), concentrating on the portion near hole~1. (We expect this slice to provide the most
stringent test of the matching, since it contains the portions of the buffer zones where
the field is strongest and changes most rapidly.) We shall use this
setup (or slight modifications thereof) for all of our later examples. The
matching in the inner-to-near transition is
shown in Figs.~\ref{IZ-NZ} and~\ref{IZ-NZ-vol-el-diffs} (along with a comparison of
different near zone volume elements). The first of these shows the
volume elements themselves [along with the differences between the volume
elements of the $O(v^4)$
and $O(v^5)$ versions of the near zone metric] and the second shows the differences between the volume
elements of the various near zone metrics (along with the merged metric) and the inner
zone metric. The volume elements in the near-to-far transition are shown in
Fig.~\ref{NZ-FZ-merged} (along with the volume element of the merged metric).

\begin{figure}[htb]
\epsfig{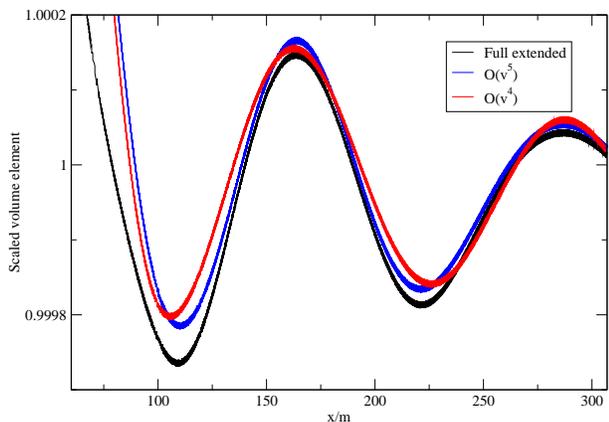} 
\caption{\label{FZ-oscillations} The scaled volume element of the far zone metric for a
binary with a separation of $10m$ and a mass ratio of $3:1$ along the $x$-axis to the
right of hole $1$ (the more massive hole). We display this for the extended data (the
darkest curve---black in color) as well as the $O(v^5)$ and $O(v^4)$ versions of the
data (the medium and lightest curves---blue and red in color, respectively) to illustrate
their differences. (We considered an unequal mass binary to make these differences more
pronounced.) As before, we have scaled these volume elements
by the contribution from the lowest-order nontrivial piece of the metric, viz.,
$(1 + 2m/r)\sqrt{1 - 4m^2/r^2}$.}
\end{figure}

The inner zone metric we display here contains the fourth-order octupole pieces discussed
in Appendix~\ref{TFHO}. To avoid clutter, we did not plot the ``plain'' version without
these additional pieces: It agrees very closely with the version we have plotted near and
inside the horizon, but bends away from the near zone metric further away from the hole.
The differences between the
versions of the inner zone with full and perturbative time dependence would not
appear in this plot, since we are looking at the $t = 0$ timeslice. We display
the differences between
the $O(v^4)$, $O(v^5)$, and full extended far zone metrics in Fig.~\ref{FZ-oscillations},
considering a binary with mass ratio $3:1$ so that these differences are more pronounced.
[In this plot, we have concentrated on the
portion of the $x$-axis to the right of the more massive hole, for clarity: The relations
between the various metrics are qualitatively the same to the left of the less massive
hole, except for differences in relative amplitude.]

In the near zone metric, the extension adds
certain terms that become large when evaluated close to the holes (e.g.,
$4m_1^2m_2/r_1^3$ in $g_{00}$, which equals $1/2$ at $r_1 = m$ for our test system) and
others that grow rapidly as one moves away from the holes (e.g.,
$-m_1^2m_2r_1^3/4b^6$, also in $g_{00}$, which is about $-0.03$ at $r_1 = 100m$ for our
test system). The terms that become large near the holes cause the $t = 0$
timeslice of the near zone metric to no longer be spacelike
in a region that extends outside the horizon.
This can be seen in the rapid decrease of the volume element around the hole
in Fig.~\ref{IZ-NZ}. (Even though we do not display this in the plot, the
volume element in fact decreases to zero.) However, for this separation, the
timeslice is still spacelike in the buffer
zone, so its bad behavior closer to the holes does not cause any problems in
the merged metric.

N.B.: The
unperturbed horizon is $m_1[1 - m_2/b + O(m_2^2/b^2)]$ away from the point
particle associated with hole $1$ in the new coordinates. For the test system,
the correction is small, so hole $1$'s unperturbed horizon intersects
the $x$-axis at $\sim 4.5m$ and $\sim 5.5m$ in the new coordinates. There are also
corrections due to the tidal distortion, but for the test system these
are even smaller than those due to the new coordinates: Taylor and
Poisson~\cite{TP} study
the effect of the tidal distortion on the horizon (in their Sec.~VIII) and find the
expected quadrupolar deformation at lowest order.

In the inner-to-near transition, the volume elements generally behave in the
expected manner: They differ when considered either too near or too far from
the hole and approach each other in the buffer zone. Things look a bit more
unusual in the near-to-far transition, since the two volume elements
agree better in the region between $20m$ to $30m$ than they do further away from the
holes. However, this should not be surprising: The reduced wavelength of the
gravitational radiation is $\sim 16m$ for the
equal-mass test system, and the difference between the near zone metric's perturbative
treatment of
retardation and the far zone metric's full treatment should become quite apparent beyond
that radial distance. In fact, the oscillations we see in the far zone metric's volume
element in Fig.~\ref{FZ-oscillations} are due to the far zone metric's
dependence on retarded time. (Even though the metric contains
gravitational radiation, this is not the only source of these oscillations.
However, it does contribute to them, as expected.) This plot also illustrates
the phase differences between the various versions of data. Additionally, we
expect the volume elements of
the near and far zone metrics to differ dramatically close to the binary.
This is indeed the case (particularly for $r < 10m$), though we did not
show this in the plot, preferring instead to concentrate on displaying the
details of the volume elements' behavior in the transition region.

\subsection{Background resummation}
\label{BR}

In the inner-to-near transition, the volume elements do not agree as closely as we might
like. In fact, the agreement is \emph{worse} near the hole with the higher-order extended
version of the near zone metric than it is with the original $O(v^4)$ version (though
the agreement further away from the hole and outside the orbit is slightly better). However,
even the
agreement with the original version is not much better than that found in Paper~II (see
its Fig.~2), even though we have matched to higher order.\footnote{The figures are not directly
comparable, since the one in Paper~II plots the $xx$ components of the metrics, not their
volume elements. However, the plot of just the $xx$ components of this paper's inner and near
zone metrics displays the same behavior as that of the volume elements, including roughly the
same numerical values for the difference between the metrics.} Part of the resolution of this
apparent problem is that we have not yet used one of the ``tricks'' from Paper~II, namely
background resummation.

The idea of background resummation is to add higher-order terms to the near
zone metric in order to 
improve its strong field behavior. If one considers the limit $m_{2} \to 0$
($\Rightarrow v_{1} \to 0$) of the 
near zone metric, then one finds that it reduces to the far field
asymptotic expansion of the 
unboosted Schwarzschild metric (with mass parameter $m_1$) in (PN)
harmonic coordinates. This expansion, however, lacks the 
causal structure of the Schwarzschild metric---it has no horizon.

One method to restore this causal structure is to resum the PN metric. This resummation
consists of adding an infinite number of higher-order terms, such that the metric reduces
identically to the {\emph{full}} (not the expanded) unboosted Schwarzschild metric in (PN) harmonic coordinates
in the limit $m_{2} \to 0$. \emph{A priori}, there is no reason to suspect that adding such higher-order terms would increase the accuracy of the PN metric.
\emph{A posteriori}, however, it is usually
the case that such resummed metrics are indeed closer to the exact solution, as was seen in Paper~II. Moreover, we should stress that these higher-order terms are not arbitrary, but guided
by the physical requirement of restoring the apparent horizons. Such physically informed
resummations have met with great success in general relativity, notably in the effective
one-body formalism (see, e.g.,~\cite{Damour2008}).

It would also be possible to resum
the far zone metric: One can calculate its Newtonian part without making a multipole
decomposition by proceeding in the same manner as in the calculation of
the Li\'{e}nard-Wiechert $4$-potential in electrodynamics (as given in,
e.g., \S14.1 in Jackson~\cite{Jackson}). However, we have
not pursued this line of investigation further. Simone \emph{et al}.~\cite{SLPW} performed
a related resummation of the Newtonian pieces of the luminosity of an extreme mass-ratio
binary, though they did this by first calculating the multipole expansion and then
resumming directly, while the resummation of the metric we are suggesting here would
involve computing the integral directly, with no multipole expansion of the Newtonian
part.

With these points in mind, let us now describe the procedure in detail, exemplifying it using the 
purely temporal component of the metric. In PN harmonic coordinates, this component of the Schwarzschild
metric takes the form $-(R - M)/(R + M)$.
Here $R$ is the harmonic radial coordinate and $M$ is the
hole's mass. (As our use of ``$R$'' indicates, this is the
same radial coordinate as in Cook-Scheel harmonic coordinates---see
Appendix~\ref{Coord_comp} for a comparison of the two coordinate systems.)
We expect to have $m_1 \to M$ and $r_1 \to R$ as $m_{2} \to 0$, which 
suggests that the purely temporal component of the PN metric should approach
$-(r_{1} - m_{1})/(r_{1} + m_{1})$ as $m_2 \to 0$.
The far field asymptotic expansion of this metric component (i.e., for $r_1 \gg m_1$) is given by
\begin{equation}
-1 + \frac{2 m_{1}}{r_{1}}  - \frac{2 m_{1}^{2}}{r_{1}^{2}} + O\biggl(\biggl[\frac{m_{1}}{r_{1}}\biggr]^3\biggr).
\end{equation}
This identically reproduces all the terms in the PN near zone metric in
the limit $m_{2} \to 0$. We then resum the PN near zone $g_{00}$ by taking
\<\label{g00resum}
\begin{split}
g_{00} - g_{00}^\mathrm{old}& = - \frac{r_{1} - m_{1}}{r_{1} + m_{1}} - \left( -1 + \frac{2 m_{1}}{r_{1}}  - \frac{2 m_{1}^{2}}{r_{1}^{2}}\right)\\
&\quad + (1 \leftrightarrow 2),     
\end{split}
\?
where $g_{00}^\mathrm{old}$ is the version of this component without resummation, given in Eq.~\eqref{g00PN}.

A similar procedure can be applied to the spatial sector of the metric. Carrying this out, we obtain
\<
\begin{split}
g_{kl} - g_{kl}^\mathrm{old} &=  \frac{r_{1} + m_{1}}{r_{1} - m_{1}} n_{kl}^{(1)} + \left(1 + \frac{m_{1}}{r_{1}}\right)^{2} \left[\delta_{kl} - n_{kl}^{(1)} \right]\\
&\quad  - 
\left[
\left(1 + \frac{2 m_{1}}{r_{1}} + \frac{m_{1}^{2}}{r_{1}^{2}} \right)\delta_{kl}  
+ \frac{m_{1}^{2}}{r_{1}^{2}} n_{kl}^{(1)}   \right]\\
&\quad + (1 \leftrightarrow 2).
\end{split}
\?
where $n^{(1)}_{kl} := x_1^{kl}/r_1^2$ and $g_{kl}^\mathrm{old}$ is given by Eq.~\eqref{gklPN}. One can check that $g_{kl}$ reduces to $g_{kl}^\mathrm{old}$
identically as $m_2 \to 0$.

We have used the Schwarzschild metric in PN harmonic coordinates to resum
the PN metric here, since it is this background that the PN metric
approaches in the $m_{2} \to 0$ limit. We thus cannot resum the
spatiotemporal components of the metric, since they already match the
background in this limit. If we had first transformed the PN metric to
Cook-Scheel coordinates, then we would have been able to resum the
background so that it exactly matched that of the inner zone metric. This
would have guaranteed a better agreement, and would have probably also
given a merged metric with smaller constraint violations, since there
would be no coordinate singularity at the horizons in the resummed near
zone metric.

However, if we had chosen this route, we
would have had to pick some region surrounding the buffer zone for each hole in which to
perform this transformation. This would have introduced further complications that we
thought it best to avoid in this implementation, even at the possible cost of somewhat
poorer matching. Moreover, the background resummation procedure that we have
implemented has indeed improved the matching, as can be seen in Fig.~\ref{IZ-NZ}: The
improvement is particularly striking for the extended version of the near zone metric, where the region in which the $t = 0$ slice is no longer spacelike
moves closer to the horizon and the graph of its volume element now crosses
that of the near zone in the buffer zone outside of the orbit. But resummation
also improves the matching of the original data: The resummed version of the
$O(v^4)$ near zone metric agrees more closely with the inner zone
metric than does the unresummed version. 

\subsection{Transition functions}
\label{Transitions}

We now turn to the process of stitching the inner and near zone metrics
together numerically. It is necessary to interpolate between the various
metrics in \emph{transition regions} (located inside the buffer zones) in
order to
stitch the metrics together with no discontinuities. A simple way to do this
is to use a weighted average of the metrics, where the precise way this average is
carried out is determined by a $C^\infty$
\emph{transition function} $F_{AB}: \R^3 \to [0,1]$. This function should have
the property that $F_{AB}(\vec{x}) = 0$ if $\vec{x} \in C_A \cap O_{AB}^\mathsf{C}$
and $F_{AB}(\vec{x}) = 1$ if $\vec{x} \in C_B \cap O_{AB}^\mathsf{C}$. Here
$O_{AB}^\mathsf{C}$ is the complement of the buffer zone between zones $C_A$
and $C_B$. (These conditions guarantee that the transition takes
place completely inside the buffer region.) If we just consider two metrics,
$g_{\alpha\beta}^{(A)}$ and $g_{\alpha\beta}^{(B)}$, for simplicity, then the resulting
merged metric is given by $[1 - F_{AB}]g_{\alpha\beta}^{(A)} +
F_{AB}g_{\alpha\beta}^{(B)}$. (We have suppressed the position dependence
of the transition functions and metrics, for notational convenience.)

In principle, these conditions are all one would impose on possible
transition functions. (One might also want to stipulate that $F_{AB}$ be ``increasing as one moves from $C_A$ to $C_B$,'' where this would have to be interpreted in some appropriate
sense.) One could then contemplate minimizing some appropriate norm of the
constraint violations of the resulting merged metric (outside, say, the
apparent horizons) over all possible transition functions satisfying these
requirements~\cite{PullinPC}. Our approach will be significantly less ambitious, leaving
a systematic study of transition functions to later work, probably that
accompanying an evolution of the data.

However, as Yunes~\cite{Frankenstein}
realized, one can exclude from consideration \emph{a priori} any transition
functions that would induce larger (formal) constraint violations in the data than those
due to the uncontrolled remainders of the individual metrics. The so-called
Frankenstein theorems enunciated in~\cite{Frankenstein} provide sufficient
conditions on the transition functions to keep this from occuring:
The first Frankenstein theorem tells us that the first and second
derivatives of $F_{AB}$ must be $O(\epsilon_A, \epsilon_B)$ and
$O(\epsilon_A^0, \epsilon_B^0)$, respectively, if the merged metric is to
satisfy the Einstein equations to the same order as the individual metrics.
(Here $\epsilon_A$ represents the small parameters associated with zone $A$.)
We have constructed our transition functions to respect the conditions of this
theorem---we check this explicitly below.

We shall take our transition region
and functions to be spherically symmetric, even though the optimal ones
would surely be distorted---neither the binary nor the holes are spherically
symmetric. Moreover, we shall
only consider one particular form for the transition functions, viz., the
same form used in Papers~I and~II:
\begin{widetext}
\<
f(r, r_0, w, q, s) :=
\begin{cases}
0,& r \le r_0,\\
\frac{1}{2}\{1 + \tanh[(s/\pi)\chi(r,r_0,w) - q^2/\chi(r,r_0,w)]\},& r_0 < r < r_0 + w,\\
1,& r \ge r_0 + w,
\end{cases}
\?
where $\chi(r, r_0, w) := \tan[\pi(r - r_0)/2w]$. The full merged metric is
thus
\<
\begin{split}
g_{\alpha\beta} &= \{1 - f_\mathrm{far}(r)\}\Bigl\{f_\mathrm{near}(x)\bigl[f_{\mathrm{inner},1}(r_1)g_{\alpha\beta}^{(3)} +
\{1 - f_{\mathrm{inner},1}(r_1)\}g_{\alpha\beta}^{(1)}\bigr]\\
&\quad + [1 - f_\mathrm{near}(x)]\bigl[f_{\mathrm{inner},2}(r_2)g_{\alpha\beta}^{(3)} +
\{1 - f_{\mathrm{inner},2}(r_2)\}g_{\alpha\beta}^{(2)}\bigr]\Bigr\}
+ f_\mathrm{far}(r)g_{\alpha\beta}^{(4)},
\end{split}
\?
\end{widetext}
where $g_{\alpha\beta}^{(A)}$ denotes the metric that lives in zone $A$ (see Fig.~\ref{ZD} for the numbering system) and
\<\label{trans-pars}
\begin{split}
f_\mathrm{far}(r) &:= f(r,\lambda/5, \lambda, 1, 2.5),\\
f_\mathrm{near}(x) &:= f(x, 2.2m_2 - m_1 b/m, b - 2.2m, 1, 2.5),\\
f_{\mathrm{inner},A}(r_A) &:= f(r_A, 0.256r_A^\mathrm{T}, 3.17(m^2b^5)^{1/7}, 0.2, b/m).
\end{split}
\?
[N.B.: We shall refer to the value of a parameter appearing in a
particular transition function by transferring that transition function's
subscript to the parameter's name, e.g., $w_{\mathrm{inner},A} := 3.17(m^2b^5)^{1/7}$.]
Here $\lambda = \pi\sqrt{b^3/m}$ is the Newtonian wavelength
of the binary's gravitational radiation. We have also used the ``transition
radius'' $r_A^\mathrm{T}$,
given by taking the leading orders of the uncontrolled remainders of the
approximations in the inner and near zones to be
comparable. This gives $(m/b)(r_A^\mathrm{T}/b)^4 = (m_A/r_A^\mathrm{T})^3$,
and thus $r_A^\mathrm{T} = (m_A^3b^5/m)^{1/7}.$
With these choices for the parameters, the transition functions satisfy the
hypotheses of the first Frankenstein theorem~\cite{Frankenstein}, given above: For instance,
the $n$th spatial derivatives of $f_\mathrm{near}$ and $f_\mathrm{far}$
scale as $w_\mathrm{near}^{-n} \propto b^{-n} \propto v^{2n}$ and
$w_\mathrm{far}^{-n} = 1/\lambda^n \propto v^n$, respectively. Matters are a
bit more complicated for $f_{\mathrm{inner},A}$: Its spatial derivatives
decrease rapidly as $v \to 0$ because they involve $\sech$ with an argument
that goes to infinity as $v \to 0$ (since $s_{\mathrm{inner},A} \propto
v^{-2}$) and $\sech$ is rapidly decreasing at infinity.

We determined the parameters we use for the transition functions by experimenting with different
choices: We found that the values given in Eq.~\eqref{trans-pars} produced the smallest overall
constraint violations along the $x$-axis for our equal-mass test system (with
$b = 10m$) of all the choices we tried. Except for the near-to-far zone
transition, which is completely new, these choices are very similar to those in
Paper~II. (In fact, despite the way it is written, our $f_\mathrm{near}$ agrees exactly
with its analogue from Paper~II, the function $G$.) The most important difference is the
scaling of the transition width,
$w_{\mathrm{inner},A}$. Both Papers~I and~II took $w_{\mathrm{inner},A}$ to
scale with $r_A^\mathrm{T}$, so it depends on the system's mass ratio; in
particular, it goes to zero as $m_A \to 0$. (Note that our
$f_{\mathrm{inner},A}$ corresponds to $F_A$ in Papers~I and~II.) We have found
that this choice for $w_{\mathrm{inner},A}$ results in large
transition-induced constraint violations near the smaller black hole for
unequal mass ratios. These
constraint violations appear to increase without bound as the mass of the
smaller hole goes to zero. This is as one would expect, since the gradient
of the transition function blows up as the transition width goes to
zero, leading to an unbounded increase in constraint
violations---see~\cite{Frankenstein} and the above discussion of the
Frankenstein theorems.

We can obtain a better-behaved transition function by freezing the dependence
of $w_{\mathrm{inner},A}$ on the mass ratio, as we have done here. (We also
have a slightly different coefficient of $r_A^\mathrm{T}$ than in Paper~II, even for an
equal-mass system.) The other
differences between our $f_{\mathrm{inner},A}$ and Paper~II's $F_A$ are a
slightly different coefficient of the transition radius in
$r_{0,\mathrm{inner},A}$, in
addition to a new transition radius that reflects the higher-order matching that
we have performed here. (The transition radius, $r_A^\mathrm{T}$, was
slightly misleadingly called the matching radius and referred to as $r_A^M$ in
Paper~II.)

The choices for the transition functions'
parameters determine effective boundaries for the various zones: These are given in
Table~\ref{Zone_boundaries} for our equal-mass test binary. This table gives both the
formal boundaries (i.e., the numerical values of the boundaries given in
Sec.~\ref{MatchingIntro}) and the effective boundaries (i.e., the boundaries determined by
our choices of parameters for the transition functions).
What we call the complete effective boundaries are determined by the entire region in which
we use a given metric, even if the coefficient of the metric (due to the transition
functions) is very small in a portion of the region. What we refer to as the practical
effective boundaries are cut off when the coefficient
of the metric becomes smaller in magnitude than $10^{-4}$ [i.e., much smaller than the
magnitude of the uncontrolled remainders, which are $\sim (b/m)^{-5/2} \simeq 3 \times 10^{-3}$].
\begin{table}
\begin{tabular}{ccccc}
\hline\hline
Zone & & Formal boundaries & \multicolumn{2}{c}{Effective boundaries}\\
 & & & Complete & Practical\\
\hline
Inner & & $r_A \ll 10m$ & $r_A \le 17.4m$ & $r_A \le 11.2m$\\
\multirow{2}{*}{Near} & \multirow{2}{*}{$\Big\{$} & $r_A \gg 0.5m,$ & $r_A \ge 0.985m,$ & $r_A \ge 1.27m,$\\
 & & $r \lesssim 15.8m$ & $r \le 119m$ & $r \le 109m$\\
Far   & & $r \gtrsim 15.8m$ & $r \ge 19.9m$ & $r \ge 30.4m$\\
\hline\hline
\end{tabular}
\caption{\label{Zone_boundaries} The zone boundaries for our equal-mass test binary
($b = 10m$).}
\end{table}

The effective inner zone boundaries given in Table~\ref{Zone_boundaries} are not quite correct:
Since even the practical effective boundaries of the inner zones are greater than half the
distance between the holes for $b = 10m$ (as they are for an equal-mass binary, with our choice
of transition functions, for $b \lesssim 165m$), one
needs to introduce a third
transition function, here called $f_\mathrm{near}$, to effect the transition between the holes.
(See Sec.~VI~B of Paper~I for further discussion; this function is referred to as $G$ in
Papers~I and~II.) With our choice of parameters, $f_\mathrm{near}$ cuts off
the complete effective inner zone
when the $x$-coordinate is closer than $1.1m$ to (the $x$-coordinate of the point particle
associated with) the other hole. The practical effective inner zone is cut
off when the $x$-coordinate is $1.93m$ away from the other hole.)

In $f_\mathrm{far}$, one might be concerned that even the practical effective
transition region extends well
outside of the standard outer boundary of the near zone, viz.,
$r \simeq \lambdabar$. Indeed,
it is quite possible that blending in the near zone metric in a
region where its perturbative treatment of retardation is not warranted will
introduce significant phase errors in the binary's outgoing wave train.
However, our choice of transition region is justified (at least for this
preliminary construction of transition functions) by the (relatively)
large constraint violations of the far zone metric at the
inner edge of our transition region, as is shown in 
Fig.~\ref{NZ-FZ-constraint_comp}.

\begin{figure}[htb]
\epsfig{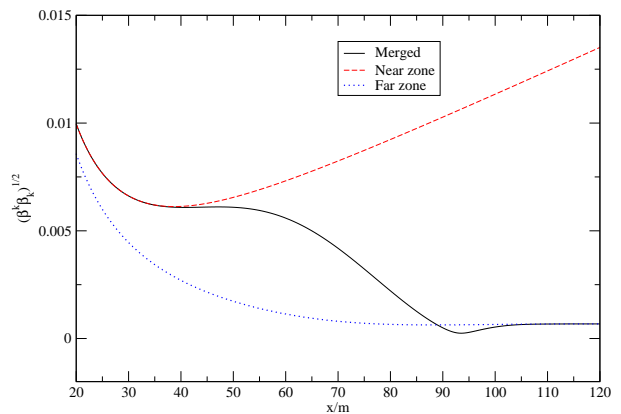}
\caption{\label{NZ-FZ_shift_matching} The norm of the shift ($\sqrt{\beta^k\beta_k}$, with
indices lowered by the metric under consideration) of the near zone, far
zone, and merged metrics to the right of hole $1$ along the axis between the holes. As usual, these are all calculated for our equal-mass test system.
}
\end{figure}
\begin{figure}[htb]
\epsfig{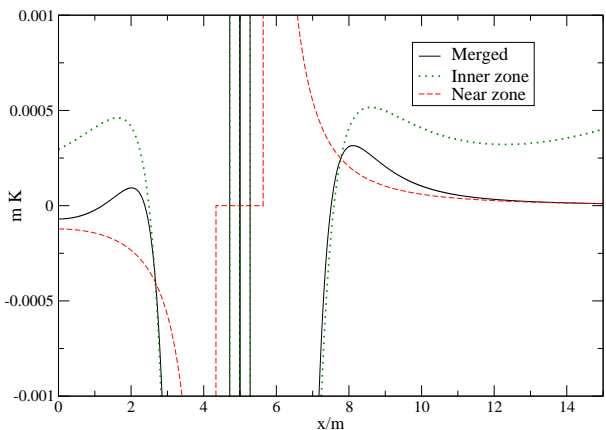}
\caption{\label{IZ-NZ-Ktrace} The trace of the extrinsic curvature of the inner zone, near
zone, and merged metrics near hole $1$ along the axis between the holes. As
always, these are all calculated for our equal-mass test system.
}
\end{figure}
\begin{figure}[htb]
\epsfig{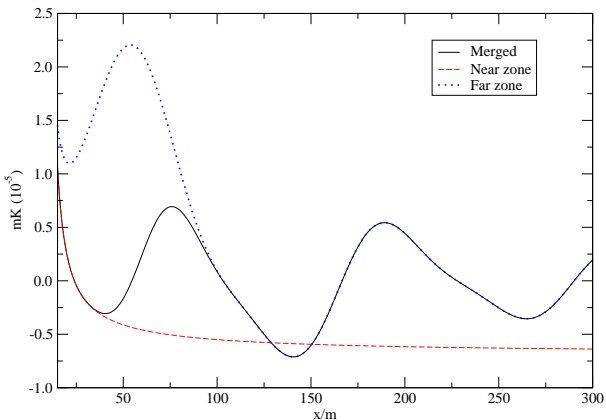} 
\caption{\label{NZ-FZ-Ktrace} The trace of the extrinsic curvature of the near zone, far
zone, and merged metrics to the right of hole $1$ along the axis between the holes. As
usual, these are all calculated for our equal-mass test system.
}
\end{figure}

To illustrate the matching, we have plotted the
volume elements of each zone's metric along with that of the merged metric,
all for our standard test system. (Recall that we use the full extended version of
the data in all plots unless otherwise noted.)
We did this for the inner-to-near zone transition in Fig.~\ref{IZ-NZ-vol-el-diffs}, plotting
the difference between the merged and inner zone metrics' volume elements, for
clarity. Similarly, the plot for the near-to-far transition was already given in
Fig.~\ref{NZ-FZ-merged}. 
We also need to consider the matching of the shift ($\beta^k$) separately,
since the volume element of the $4$-metric equals
the lapse times the volume element of the $3$-metric [see, e.g., Eq.~(E.2.25) in
Wald~\cite{Wald}]. We display its norm ($\sqrt{\beta^k\beta_k}$, with indices lowered by
the metric under consideration) in the near-to-far transition in
Fig.~\ref{NZ-FZ_shift_matching}.
The analogous plot in the inner-to-near transition looks similar enough to that of the
volume element (Fig.~\ref{IZ-NZ-vol-el-diffs}) that we do not
include it. For the same reason, we do not show the matching of the norm of the
extrinsic curvature
($\sqrt{K^{lp}K_{lp}}$, with indices raised by the metric under consideration) in either
transition region.
The trace of the extrinsic curvature, $K$, behaves differently enough that we
plot its matching in Figs.~\ref{IZ-NZ-Ktrace} and~\ref{NZ-FZ-Ktrace}.
Moreover, the trace of merged metric's extrinsic curvature is an intrinsically
interesting quantity: It tells us how much our data's slicing differs from maximal
slicing ($K = 0$).

\subsection{Constraint violations}
\label{CV}

\begin{figure}[htb]
\epsfig{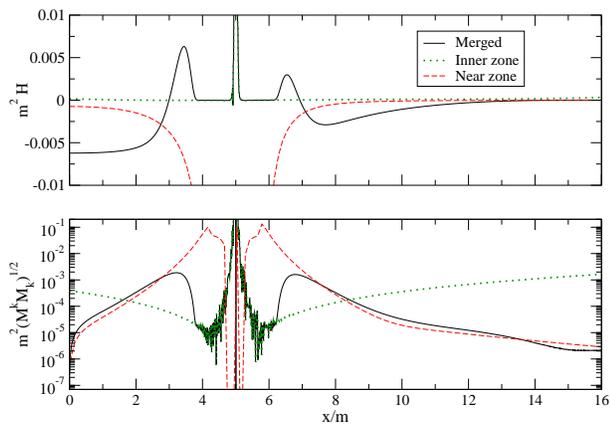} 
\caption{\label{IZ-NZ-constraint_comp} The Hamiltonian constraint violations
and norm of the momentum constraint violations of the merged, inner zone, and
near zone metrics. These are plotted along the $x$-axis around hole~$1$ for our standard
equal-mass test system.}
\end{figure}

\begin{figure}[htb]
\epsfig{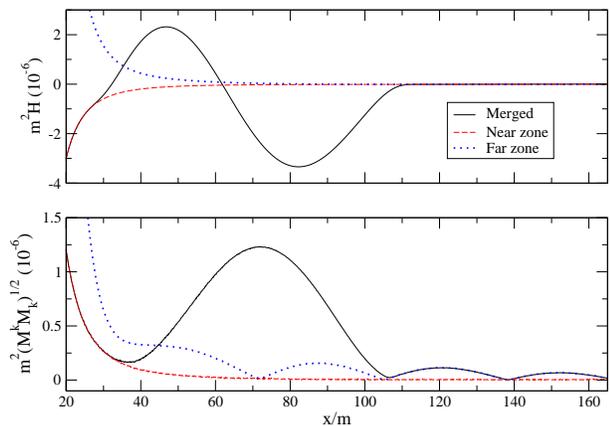} 
\caption{\label{NZ-FZ-constraint_comp} The Hamiltonian constraint violations
and norm of the momentum constraint violations of the merged, near zone, and
far zone metrics. These are plotted along the $x$-axis in a region including
the portion of the near-to-far transition region to the right of hole $1$.
As usual, this is done for our standard equal-mass binary test binary.}
\end{figure}

The constraint violations provide a much more sensitive check on how well the
transition functions are working than the previous subsection's plots, in
addition to giving a measure of the accuracy of
the entire initial data construction. We compare the constraint violations (computed
using {\sc{bam}}~\cite{BTJ, Bruegmannetal}) of the
individual metrics to those of the merged metric for our equal-mass test system
along the $x$-axis in Figs.~\ref{IZ-NZ-constraint_comp}
and~\ref{NZ-FZ-constraint_comp}. (We have
checked that the constraint violations behave roughly similarly---and are not
significantly worse---in the $y$- and $z$-directions.) For the norm of the momentum
constraint, $M_k$, we have chosen $\sqrt{M^kM_k}$, with indices raised by the (merged)
metric. [See, e.g., Eqs.~(14)--(15) in~\cite{CookLRR} for expressions for the constraint
equations.] N.B.: As one can see by looking at
the (norm of the) momentum constraint, the plot of the constraint violations in the
inner-to-near transition (Fig.~\ref{IZ-NZ-constraint_comp}) does not cover the entire
transition region.
However, the portion we do not show is not particularly interesting: The
momentum constraint violations of the near zone metric continue to decrease, and those of the merged metric
rapidly become indistinguishable from the near zone metric's momentum constraint 
violations.)

\begin{figure}[htb]
\epsfig{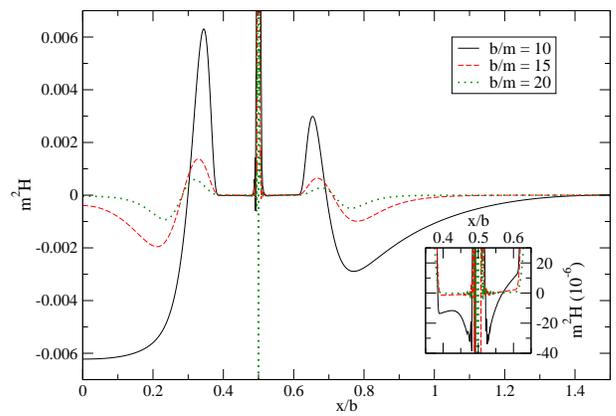} 
\caption{\label{ham_d_near} The Hamiltonian constraint violations around hole $1$
for an equal-mass binary with separations of $10m$, $15m$, and $20m$. For ease
of comparison, we have scaled the $x$-axis by $b$ so that (the point particle
associated with) hole~1 is always at the same position. In the inset, we zoom in
to show how the inner zone constraint violations vary with $b$.}
\end{figure}

\begin{figure}[htb]
\epsfig{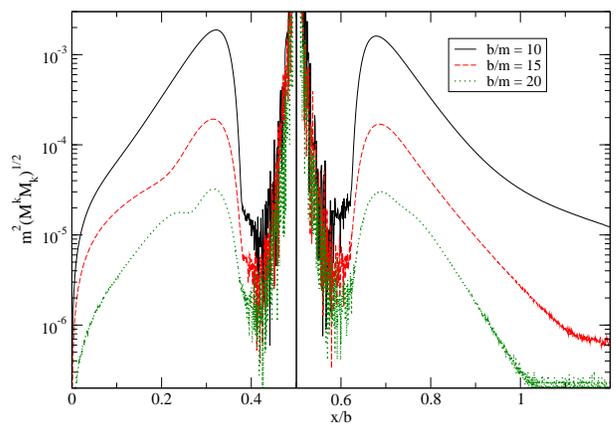} 
\caption{\label{mom_d_near} The norm of the momentum constraint violations
around hole $1$ for an equal-mass binary with separations of $10m$, $15m$, and $20m$. For
ease of comparison, we have scaled the $x$-axis by $b$ so that (the point particle
associated with) hole~1 is always at the same position.
}
\end{figure}

\begin{figure}[htb]
\epsfig{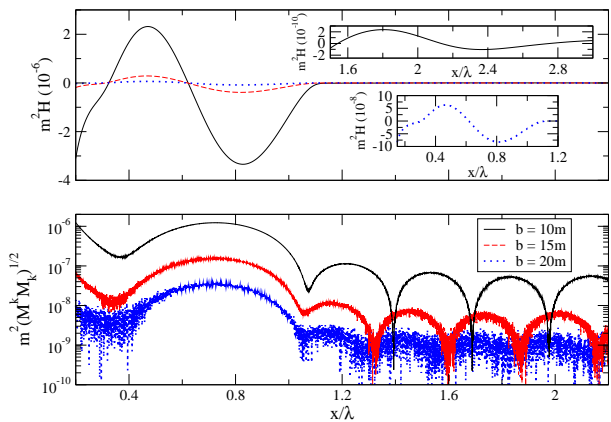} 
\caption{\label{constraint_d_far} The Hamiltonian constraint violations and norm of the
momentum constraint violations along the $x$-axis in the portion of the near-to-far
transition and far zone proper that lies to the right of hole~$1$. These are computed
for an equal-mass binary with separations of $10m$, $15m$, and $20m$. For ease
of comparison, we have scaled the $x$-axis by $\lambda = \pi\sqrt{b^3/m}$, the
Newtonian wavelength of the binary's gravitational radiation. In the two insets, we zoom in to better
illustrate the behavior of the Hamiltonian constraint violation in two situations: In the
lower inset, we consider the $b = 20m$ data in the transition region. In the upper inset,
we consider the $b = 10m$ data in the far zone proper. (The other data sets also display
similar oscillations in their Hamiltonian constraint violations in the far zone, though
the amplitude of these oscillations is too small to be visible on the scale we use to
display the oscillations of the $b = 10m$ data's Hamiltonian constraint.)
}
\end{figure}

The largest constraint violations in the merged metric outside of the horizons
occur in the inner-to-near transition regions. These are likely due in part to
the near zone metric's large constraint violations near the horizons due to
the coordinate singularity at the horizons. (Recall that the horizons are
approximately $.5m$ away from the positions of the point particles in our
test system---the effects of the new coordinates and tidal distortion are
small.) However, since the only knowledge we have about the size of the
uncontrolled remainders is their scaling with $v$ (and thus $b$), the
magnitude of the constraint violations does not really provide us with a check
that the uncontrolled remainders are in fact of the advertised orders. We can
obtain such a check by considering how the constraint violations vary as the
binary's separation ($b$) increases. As expected, they decrease at least as
rapidly as $b^{-5/2}$ (the expected scaling of the largest uncontrolled remainders) as the
binary's separation increases. This is illustrated in Figs.~\ref{ham_d_near}
and~\ref{mom_d_near} in the inner and near zones, and in Fig.~\ref{constraint_d_far} in the
far zone.


For a separation of $10m$, the relatively large constraint violations of the
merged metric compared to those of the individual metrics are an indication that
this separation is close to the minimum for which the hypotheses underlying the
data's construction are valid: For instance, as seen above, for this
separation (and an
equal-mass binary), the two inner-to-near transition regions overlap between
the holes, meaning that much of the inner-to-near transition is effected by
$f_\mathrm{near}$---see Table~\ref{Zone_boundaries} and the surrounding
discussion. Moreover, as we have seen previously, the $t = 0$ slice
of the near zone metric is not even spacelike at some points outside the
holes' horizons for $b = 10m$. However, this does not adversely affect the merged metric
with our choices of transition regions. If one tries closer separations, things
are significantly worse. For instance, for a separation of
$6m$, the maximum constraint violations are larger than those for $10m$ by a
factor of 10 or more. Moreover, with our choices for the transition regions,
the merged metric contains some of the portions of the near zone metric where
its $t = 0$ slice is not spacelike.

In the near-to-far transition (as shown in Fig.~\ref{NZ-FZ-constraint_comp}),
the relatively large constraint violations associated with the far zone metric
are not unexpected at the distances from the binary at which we have made our 
transition: We have uncontrolled remainders of, e.g., $O(mb^4/r^5)$ in the
multipolar expansion of the far zone metric, so we expect it to have
(dimensionless) constraint violations of $O(m^3b^3/r^7)$. For instance, at
$r = 30m$, $m^3b^3/r^7 \simeq 5 \times 10^{-7}$, reproducing the order of
magnitude of the constraint violations at that point.
We would have thus needed to
transition somewhat farther from the binary if we wanted
the near and far zone metrics' constraint violations be comparable in the
transition region. However,
we almost surely do not want to do this, since the near zone only accounts for
retardation perturbatively, and thus accumulates large phase errors beyond
$r \gtrsim \lambdabar$: It thus even has a considerable phase error in much of our current
transition region. As was (briefly) discussed in the previous subsection, a resummation of
the far zone metric might reduce its constraint violations closer to the
binary, though we did not pursue this here.

We compared the constraint violations of the new data with the old data (for our standard
test system) in Fig.~\ref{old_data_comp} and Table~\ref{constraint_norm_old_data} in the
introduction, though we deferred a more detailed discussion to here: First, it is important
to realize that the comparison is somewhat misleading, since each paper's data are in a
different coordinate system. (The data from Paper~I are in the same harmonic coordinate
system as this paper's data in the near zone. However, this is only true
perturbatively in the inner zone,
where the black hole background is in a coordinate system that is not horizon-penetrating.)

Second, while our data's Hamiltonian constraint violations are not appreciably better than
those of the data from Paper~II, even though we have matched to higher order, this is not
unexpected: Even though we used horizon-penetrating coordinates for the
black hole metrics, we used standard PN harmonic coordinates for the PN metric; these
coordinates are singular at the horizon. While the merged metric has no coordinate
singularities, the
PN metric's coordinate singularity increases the constraint violations in the transition
regions, making them comparable to those from Paper~II: Paper~II's data use a
PN metric with no
coordinate singularity as well as even further resummation of the black hole backgrounds
than we have employed here, leading to particularly small constraint violations.

Third, the increased momentum constraint violations
near the hole in the new data, compared with either of the old papers' data, come from
the $x$-component: The $y$-component of the new data's momentum constraint violations is
much smaller than
that of either of the two previous sets. However, the $x$-components of their momentum
constraints vanish. (The $x$-component of the momentum constraint of Paper~II's data only
vanishes before the transformation to horizon-penetrating coordinates; it develops two
spikes after that transformation.)

Finally, if one compares Fig.~\ref{old_data_comp} with,
e.g., Figs.~14 and~17 in Paper~II, one
notices differences in the behavior of the data from Papers~I and~II close to
the hole (and inside the horizon). This is because we
have generated the plot using higher-order finite
differencing (fourth order vs.\ second order) and a higher resolution ($0.002m$ vs.\
$0.008m$) in computing the constraint violations here than we did in first
computing them in Paper~II. It was necessary to do this to accurately resolve
the constraint violations in the inner zone, since the metric components diverge
rapidly there. Additionally, we have used the version of Paper~II's metric that
is in horizon-penetrating coordinates, while Figs.~14 and~17 in Paper~II were generated
using the version of the data without that additional coordinate transformation: The
transformation to horizon-penetrating coordinates introduces further structure in the
data's constraint violations in both the $x$- and $y$-components.

\begin{figure}[htb]
\epsfig{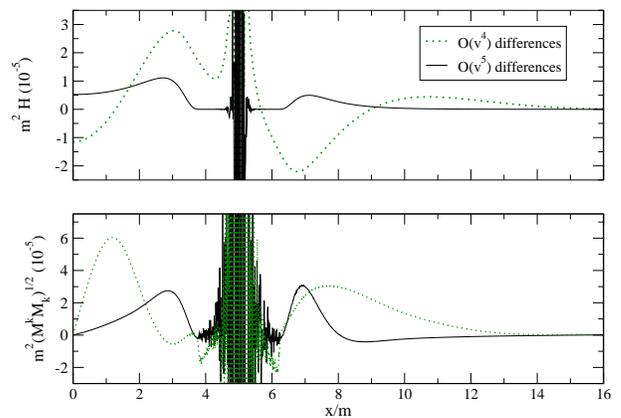} 
\caption{\label{IZ-NZ-version-constraint_comp} The constraint violations
of the $O(v^4)$ and $O(v^5)$ sets of data minus those of the full extended set of data.
These are computed along the $x$-axis near hole $1$ for our standard test binary.}
\end{figure}
\begin{figure}[htb]
\epsfig{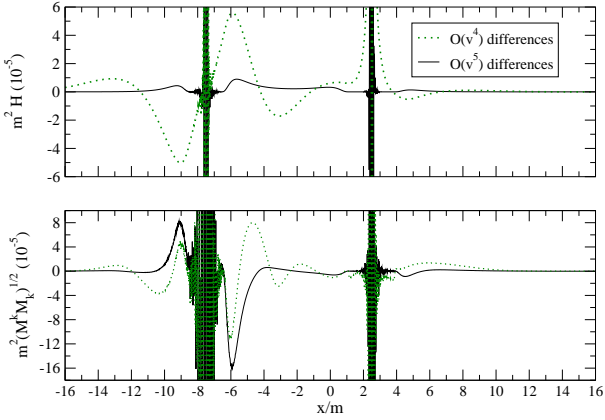} 
\caption{\label{IZ-NZ-version-constraint_comp_q3} The constraint violations
of the $O(v^4)$ and $O(v^5)$ sets of data minus those of the full extended set of data.
These are computed along the $x$-axis in the inner-to-near
transition region for a binary with $b = 10m$ and a mass ratio of $3:1$. (The more
massive hole is on the right.)}
\end{figure}
\begin{figure}[htb]
\epsfig{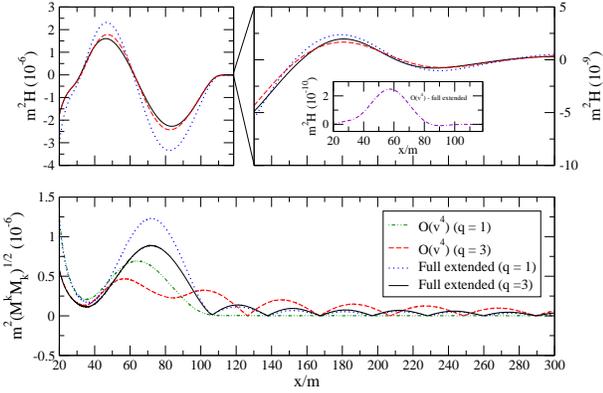} 
\caption{\label{NZ-FZ-version-constraint_comp} The constraint violations
of the $O(v^4)$ and full extended sets of data.
These are computed along the $x$-axis in the near-to-far
transition region (and far zone proper) to the right of hole $1$ for binaries with
$b = 10m$ and mass ratios
of $1:1$ and $3:1$. (The more massive hole is on the right.) In the right-hand panel, we
zoom in to show the differences in the oscillation of Hamiltonian constraint violations in
the far zone proper. In the inset, we 
show the difference between the Hamiltonian constraint violations of the $O(v^4)$ and full
extended versions of the data in the near-to-far transition for an equal-mass binary.}
\end{figure}
\begin{figure}[htb]
\epsfig{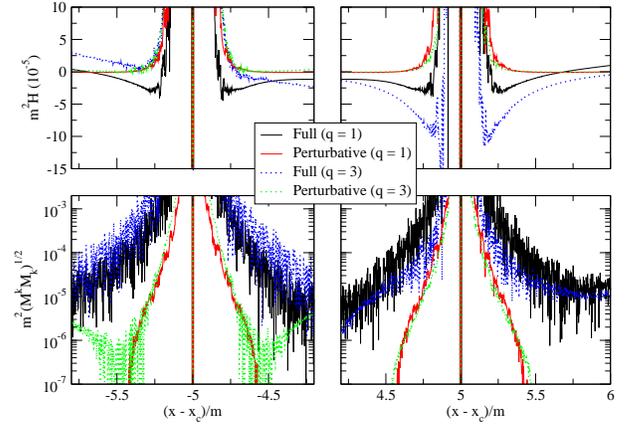} 
\caption{\label{IZ-comparison}  The Hamiltonian constraint violations and norm of the
momentum constraint violations of the versions of the inner zone metric with perturbative
and full time dependence for binaries with $b = 10m$ and mass ratios of $1:1$ and $3:1$.
We display these in the vicinity of both holes, which we shift in the $3:1$ case
so their associated point particles lie at
$x = \pm 5m$. [In the figure, $x_c := [1/(q+1) - 1/2]b$.]}
\end{figure}
\begin{figure}[htb]
\epsfig{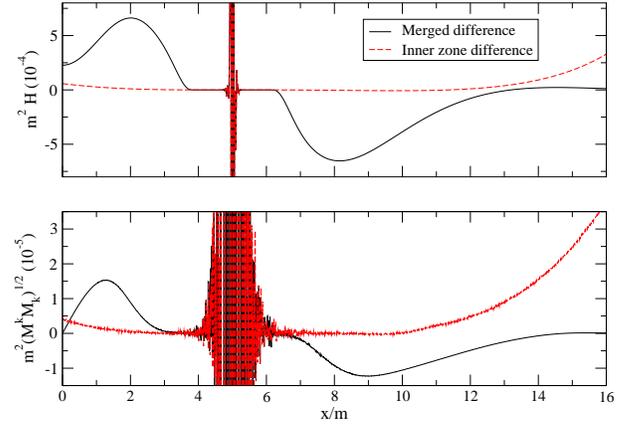} 
\caption{\label{IZ-NZ-oct_constraint_comp}  The differences between the
Hamiltonian constraint violations and norm of the momentum constraint violations of
the version of the $O(v^4)$ data without fourth-order octupole terms ({\tt{O4\_NoOct}}) in the inner
zone and the version including those terms ({\tt{O4}}).
These are computed along the $x$-axis near hole $1$ for our standard test binary.}
\end{figure}

We can also compare the constraint violations of the full extended data with those of
the $O(v^4)$ and $O(v^5)$ versions. We do this in the inner-to-near transition in
Figs.~\ref{IZ-NZ-version-constraint_comp} and \ref{IZ-NZ-version-constraint_comp_q3} for
an equal-mass and $3:1$ mass ratio binary, respectively, and in the near-to-far
transition, along with the far zone proper, for both
of those binaries in Fig.~\ref{NZ-FZ-version-constraint_comp}. [We consider a
unequal-mass binary to make the differences between the $O(v^4)$
and $O(v^5)$ versions more pronounced: Most of the $O(v^5)$ terms in the far zone
metric vanish for an equal-mass binary.]

In the latter plot (Fig.~\ref{NZ-FZ-version-constraint_comp}), we do not show the
differences between the $O(v^4)$ and full extended versions' Hamiltonian constraint in the
far zone proper, as they agree up to the level of numerical truncation error. We also do
not show the differences between the $O(v^5)$ and full
extended versions of the data. These two sets only differ substantially in the far
left-hand portion of the transition region, and even there the differences are
several orders of magnitude less than those between the $O(v^4)$ and full extended
versions. Additionally, we do not show the behavior of the constraint violations to the
left of the smaller hole (hole $2$): It
is qualitatively similar to their behavior to the right of the larger hole shown here,
except that for an equal-mass binary, the $O(v^4)$ version of the data has smaller
Hamiltonian constraint violations than
the full extended version in that region, and for a mass ratio of $3:1$, there is no
oscillation
in the transition region  in the $O(v^4)$ data's momentum constraint
violations.

We compare the constraint violations of the different inner zone versions in
Fig.~\ref{IZ-comparison} for binaries with mass ratios of $1:1$ and $3:1$. In this plot,
we do not include the version of the inner zone with no fourth order octupole pieces,
as the inclusion of those terms does not affect the constraint violations in the inner
zone proper at a level above numerical truncation error. However, these terms \emph{do}
have a noticable effect on the constraint violations in the inner-to-near transition, as
can be seen in Fig.~\ref{IZ-NZ-oct_constraint_comp}, which compares the $O(v^4)$ data with and without the
fourth-order octupole terms in the inner zone. (The fact that including the fourth
order octupole terms in the inner zone metric makes a much larger difference in the
merged metric than in the inner zone metric itself in the transition regions suggests that
the additional terms that have the most significant effect are those in the coordinate
transformation, not those in the tidal fields.) 

\begin{figure}[htb]
\epsfig{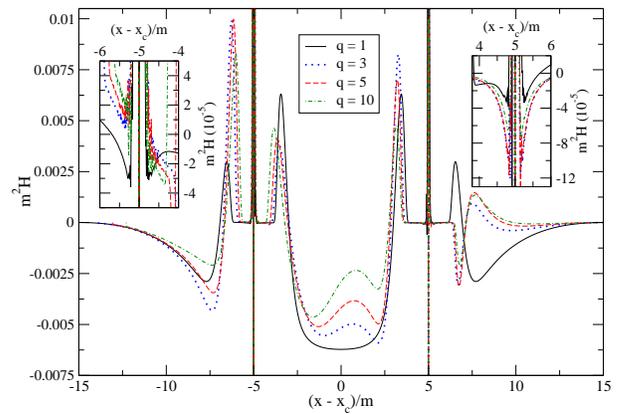} 
\caption{\label{ham_q} The Hamiltonian constraint violations along the $x$-axis
for a binary with a separation of $10m$ and mass ratios of $1:1$, $3:1$, $5:1$,
and $10:1$ ($q := m_1/m_2$). (The more massive hole---hole $1$---is on the right, and
the less massive hole---hole $2$---is on the left.) For ease of comparison, we have
shifted all the
data so that the point midway between the two particles is at $x = 0$. [In the
figure, $x_c := [1/(q+1) - 1/2]b$.] In the two insets, we zoom in to show how the 
inner zone metric's constraint violations vary with $q$, looking at the region around
each hole in the inset closest to it.}
\end{figure}

\begin{figure}[htb]
\epsfig{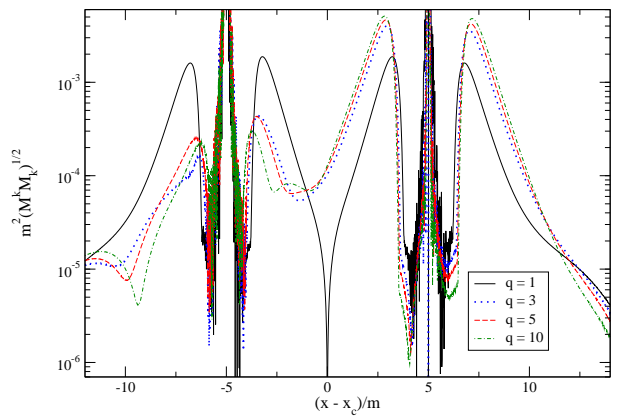} 
\caption{\label{mom_q} The norm of the momentum constraint violations along the
$x$-axis for a binary with a separation of $10m$ and mass ratios of $1:1$, $3:1$,
$5:1$, and $10:1$ ($q := m_1/m_2$). (The more massive hole---hole $1$---is on the right,
and the less massive hole---hole $2$---is on the left.) For ease of comparison, we have
shifted all the
data so that the point midway between the two particles is at $x = 0$. [In the
figure, $x_c := [1/(q+1) - 1/2]b$.]}
\end{figure}

It is also interesting to consider how accurate the data are for different mass ratios.
One finds that the constraint violations do not behave quite
as well as might be desired in the inner-to-near transition regions as one increases the
mass ratio.
This is shown in Figs.~\ref{ham_q} and~\ref{mom_q}, which plot the Hamiltonian
constraint and norm of the momentum constraint for mass ratios of $1:1$, $3:1$,
$5:1$, and $10:1$. The worst behavior is that of the
momentum constraint in the transition region near the more massive hole
(hole~$1$), which increases as the mass of that hole increases. The
Hamiltonian constraint also increases with mass ratio in the inner zones
around both holes. (This is due to the inclusion of the full time
dependence of the tidal fields---the inner zone constraint violations decrease with mass ratio
if one only uses the version of the inner zone metric with perturbative time dependence.)
The behavior of the other constraint violations is nonmonotonic.
When one looks at the near-to-far transition and far zone proper one finds much better
behavior: The constraint violations decrease with
increasing mass ratio in all of those regions, except for a slight increase in the
momentum constraint violations in the far zone proper for unequal mass ratios.
This is visible for a mass ratio of $3:1$
in Fig.~\ref{NZ-FZ-version-constraint_comp}; we do not display the results for higher mass
ratios, since they are not particularly interesting.

This behavior in the transition regions is primarily attributable to
the choices we have made for the transition functions. For instance, it is
possible to choose
parameters so that the momentum constraint violations \emph{decrease} around the more
massive hole as its mass increases. This can be accomplished by taking
$w_{\mathrm{inner},A} \propto r_A^\mathrm{T}$, as in Papers~I and~II. However, with this choice, the
decrease in momentum constraint violations around hole~$1$ occurs at the cost of the
aforementioned extreme increase in constraint violations around hole $2$ as its mass
goes to zero. It
should be possible to combine the two choices for $w_{\mathrm{inner},A}$ to obtain better
behavior for unequal mass ratios. However, we have chosen to leave such fine-tuning of
transition functions to future work, contenting ourselves with providing examples
of workable transition functions here.

\section{Conclusions}\label{Conclusions}

\subsection{Summary}

We have constructed approximate initial data for a nonspinning black hole binary in
a quasicircular orbit. This dSata set has uncontrolled remainders of $O(v^5)$ throughout
the timeslice (including the far zone), along with remainders of
$O(v^3[R/b]^4, v^5[R/b]^3)$ in the inner zone. We have verified the scaling of
the uncontrolled remainders with $v$ by checking
that the constraint violations decrease at least as rapidly as they should when
the binary's orbital separation is increased.
We constructed this data set by
asymptotically matching perturbed black hole metrics onto a PN metric and creating
transition functions to smoothly interpolate between the various metrics. The
resulting data do not assume conformal flatness and contain the binary's
outgoing radiation, in addition to the tidal deformations on the
holes. (We have included the quadrupole deformations through $1$PN along with
the lowest-order octupole deformations.)

The results of the
matching are given in Sec.~\ref{CT} and Eqs.~\eqref{TFs}: Sec.~\ref{CT} gives
directions for how to put together the coordinate transformation necessary to place
the inner zone metric in the same coordinate system as the near zone metric (to the
order we have matched). Eqs.~\eqref{TFs} give explicit expressions for the tidal
fields we
obtained. [We also found that the inner and near zone mass parameters---i.e., $M$ and
$m_1$---agree through at least $O(v^3)$.] These tidal fields are then inserted into
Detweiler's perturbed black
hole metric, given in Cook-Scheel coordinates in Eqs.~\eqref{hCS} and
\eqref{hCS_pieces}, to give the inner
zone metric. The near and far zone metrics are given in Eqs.~\eqref{gPN}
and~\eqref{FZ_metric}, respectively. We describe our
method of computing the binary's past phase evolution, needed for the far zone metric,
in Sec.~\ref{phasing}, and the specifics of how we put together the various zones'
metrics in Appendix~\ref{metric_comp}. Workable
(though surely not optimal) transition functions that smoothly interpolate
between the various zones' metrics are given in Sec.~\ref{Transitions}.

We have also constructed an extension of these data that is accurate through $O(v^5)$
in the near and far zones.
In addition, this extension includes other higher-order contributions to the
temporal components of the near and far zone metrics that were readily
available in the literature. We also calculated the fourth-order octupole
pieces of the tidal fields (and the associated piece of the coordinate
transformation) in Appendix~\ref{TFHO} and included them, as well. (The $1$PN
correction to the electric octupole tidal field is among the terms we
calculated and add here.) Additionally, we calculated the full time dependence
of the tidal fields (for times much less than the radiation reaction timescale)
and included it in the inner zone metric. This is discussed in
Appendices~\ref{TFHO} and~\ref{metric_comp}. See Appendix~\ref{ExtraTerms} for
a discussion of how we put together the extension to the near and far zone
metrics, and Eq.~\eqref{XP4OO} for the (polynomial part of the) coordinate
transformation that accompanies the fourth-order octupole pieces of the tidal
fields. See Table~\ref{Versions} (in Sec.~\ref{InclRad}) for an overview of the different versions of the
metrics we considered in the paper. N.B.: While we did not include this in the table,
for the sake of clarity, we considered all the versions of the near zone metrics with
and without background resummation, displaying the resulting differences in the volume
elements [for the $O(v^4)$ and full extended versions] in Figs.~\ref{IZ-NZ}
and~\ref{IZ-NZ-vol-el-diffs}. We found the full extended (\verb,all,) data set
(including background resummation in the near
zone) to be the best, overall, considering constraint violations as well as the
inclusion of terms that we expect to improve evolutions.

In the process of obtaining these data, we have developed a method of fixing the
matching parameters when matching a black hole onto a PN background that differs
from that presented by Taylor and Poisson~\cite{TP} and is more automatable.
We have also obtained the $1$PN corrections to
the magnetic quadrupole and electric octupole for a circular orbit using this
method, neither of which Taylor and Poisson computed.

The accurate description of
the tidal deformations on the holes contained in these data should substantially reduce
the high-frequency component of the initial spurious radiation; the use of a high-order
PN metric should do the same for the low-frequency component. 
(The combination of the high-order PN metric, including accurate expressions for
the trajectories, and the reduced junk radiation should also give a much
better quasicircular orbit---see, e.g.,~\cite{BSHH, Husaetal_ecc}.)
If these data do indeed reduce the initial spurious radiation, then they can
be used to directly quantify the effects of using the conformally flat
initial data currently employed, as opposed to data that include many more of the system's expected physical properties. In addition, the waveforms
generated using these data would be ideal for the construction of
hybrid numerical relativity/post-Newtonian waveforms (as in~\cite{Ajithetal}):
Since the initial data are directly connected to the PN approximation, the PN
parameters
and phasing that are input into the initial data should accurately describe
the subsequent evolution.

Of course,
one needs to evolve the data to see whether these putative improvements are indeed
realized. We have already coded this data set
into {\sc{Maple}} scripts, which were then converted to C code. Both the
scripts and codes are
freely available online at~\cite{Wolfgangs_website} to anyone who is interested
in evolving or otherwise studying the data. Any evolutions
of these data will need to use either excision~\cite{Shoemakeretal, Scheeletal_ex}, or the
turducken approach~\cite{Brownetal}: The
black hole perturbations are not valid all the way to the holes'
asymptotically flat ends, preventing the use of standard puncture methods.
Additionally, since the data only satisfy the
constraint equations approximately, one may want to project
them onto the constraint hypersurface before evolution. To do this, one
would need to use a code such as~\cite{Tichyetal, Pfeifferetal_CS, Tichy:2009BNS,Pretorius_solver}: The
standard solver for puncture
data~\cite{ABT} requires conformal flatness. However, evolutions of data that
only approximately solve the constraints are possible: See, e.g.,~\cite{Bodeetal,
Kellyetalev}.

\subsection{Possibilities for future initial data constructions with this method}

With this work, the asymptotic matching method for generating initial data for nonspinning binary black holes in a quasicircular orbit
first essayed by Alvi~\cite{Alvi} and further developed in Papers~I and~II has been taken to the highest order possible
without further development in its constituent parts: Including
higher multipole orders in the inner zone would require the input of nonlinear black hole perturbation theory. Iterating
to higher orders in $v$ would not only require the higher multipoles, but also an explicit expression for the $O(v^6)$ pieces of the purely spatial
components of the near
zone metric. Determining the higher-order-in-$v$ pieces of the far zone metric
would either require calculation of further contributions from the outer integrals
in the DIRE approach~\cite{PW1},
or obtaining the far zone metric via matching, following the post-Minkowskian
approach of~\cite{PB}.

Nevertheless, it might still be possible to improve the initial data at the
(formal) order presented here, as was done in Paper~II for the data in Paper~I.
For instance, one could contemplate converting the near zone metric to
Cook-Scheel coordinates (or some other horizon-penetrating coordinate system)
in a neighborhood of each black hole. This would regain the complete agreement
between background coordinates, with no coordinate singularity at the horizon
in the near zone metric, that was found to improve the numerical agreement of the metrics
in Paper~II. One could also further tweak the transition functions, though this
is not likely to produce any dramatic improvements in constraint violations.
However, it is possible that different choices for the transition
functions could improve the data's properties in evolutions. For instance, in
the current near-to-far transition, the near zone metric is used
(blended with the far zone metric) for $r \gg \lambdabar$, where its
perturbative treatment of retardation leads to large phasing errors.
It might thus be preferable to transition closer to the binary, even at the
cost of greater constraint violations. (Resumming the far zone metric is a 
possibility for reducing these constraint violations, as was discussed
briefly in Sec.~\ref{BR}.)

The prospects for generalizing this construction to include eccentricity or spin are good:
The ingredients are nearly all readily available in the literature. Such
generalization---particularly the inclusion of spin---would be an obvious next step if
this initial data set indeed significantly reduces the spurious radiation. Including
eccentricity would be straightforward, though algebraically involved, and
can be carried out with the ingredients we have used here, perhaps supplemented with the
results from~\cite{TP}. (The evolution of the binary's phase and separation
needed in the far zone metric can be obtained through $3.5$PN order using the
results of~\cite{KG}.)

However, it would only be possible to obtain the generalization
of these data for a spinning binary to $O(v^2)$, formally, while still including all
the formal quadrupole pieces in the inner zone. This is true even though the
generalization of Blanchet, Faye, and Ponsot's metric to include spin (and spin-orbit
coupling) is available~\cite{TOO},\footnote{Errata for the potentials and equations of
motion in~\cite{TOO} are given in footnotes 6 and 10 of~\cite{FBB},
respectively.} as are the expressions for the source multipoles necessary to
obtain the matching far zone metric~\cite{Will}. (The binary's evolution under
radiation reaction is also known through $2.5$PN~\cite{FBB}.)
The bottleneck is the available tidally perturbed Kerr metric~\cite{YG}, which
only includes the quadrupole perturbations. We have seen that knowledge of the octupole
perturbations is necessary to carry out the matching of all the formal quadrupole pieces
at $O(v^4)$ [which one would need to do to obtain data that are formally valid through
$O(v^3)$]. However, one could use the same
philosophy we did when computing our extension and add the higher-order terms in the
near and far zones without computing the matching inner zone terms.

\acknowledgments

We would like to thank Luc Blanchet, Steven Detweiler, Lee Lindblom, Eric Poisson, Frans
Pretorius, and Clifford Will for useful discussions and clarifications.
This work was supported by the Penn State Center for Gravitational Wave
Physics under NSF cooperative agreement PHY-0114375, as well as by NSF
grants PHY-0555628 to Penn State, PHY-0652874 and PHY-0855315 to FAU,
and PHY-0745779 to Princeton.

\appendix

\section{Comparing Cook-Scheel and PN harmonic coordinates}
\label{Coord_comp}

In the quasi-Cartesian form of Cook-Scheel coordinates~\cite{CS}, the
Schwarzschild metric is
\<
\begin{split}
g^\mathrm{CS}_{00} &= - \frac{R - M}{R+M},\\
g^\mathrm{CS}_{0k} &= \frac{4 M^{2}}{(R + M)^{2}} \frac{X_{k}}{R},\\
g^\mathrm{CS}_{kl} &= \left(1 + \frac{M}{R} \right)^{2} \delta_{kl}\\
&\quad + \frac{M^{2}}{R^2}\frac{R - M}{R + M}\left[1 + \frac{4MR}{(R + M)^2}\right] \frac{X_{k} X_{l}}{R^2}.
\end{split}
\?
[The transformation from Schwarzschild to Cook-Scheel coordinates is given in
Eq.~\eqref{CS_trans}.] For comparison,
the Schwarzschild metric in PN harmonic coordinates is
\<
\begin{split}
g_{00}^\mathrm{har} &= - \frac{R - M}{R + M},\\
g_{kl}^\mathrm{har} &= \left(1 + \frac{M}{R}\right)^{2} \left(\delta_{kl} - \frac{X_{k} X_{l}}{R^{2}} \right) + \frac{R + M}{R - M} \frac{X_{k} X_{l}}{R^{2}},
\end{split}
\?
where $g_{0k}^\mathrm{har} = 0$. (One obtains PN harmonic coordinates from
Schwarzschild coordinates by $R = \sP - M$, where $\sP$ is the
Schwarzschild radial coordinate. This is just the spatial
part of the Schwarzschild-to-Cook-Scheel transformation.) Obviously, these
coordinates have preserved the
coordinate singularity of the Schwarzschild metric in Schwarzschild coordinates.

Note that the purely temporal component of the Cook-Scheel version has
the same form
as in PN harmonic coordinates, but all the other
components are
different. In particular, the Cook-Scheel version has
a nonzero shift component as well as a slightly more involved nondiagonal
piece of the spatial metric. Explicitly, the differences between the PN
harmonic and Cook-Scheel versions
of the Schwarzschild metric components are
\<
\begin{split}
g_{00}^\mathrm{CS} - g_{00}^\mathrm{har} &= 0,\\
g_{0k}^\mathrm{CS} - g_{0k}^\mathrm{har} &= \frac{4 M^{2}}{(R + M)^{2}} \frac{X_{k}}{R},\\
g_{kl}^\mathrm{CS} - g_{kl}^\mathrm{har} &= -\frac{16 M^{4}}{\left(R - M\right) \left(R + M\right)^{3}} \frac{X_{k} X_{l}}{R^2}.
\end{split}
\?
(Of course, this must be interpreted purely algebraically, since we are
subtracting components in different coordinate systems.)
The difference of the purely spatial components scales as
$O([M/R]^{4})$ as $M/R \to 0$, which is clearly higher order in the near zone.
However, the difference
of the spatiotemporal components is $O([M/R]^{2})$, which is of the same order as the
terms we do keep in the spatial metric. And, indeed, we do see the first term in this
 expansion appearing in our coordinate transformation---see Sec.~\ref{CT}. This
shows that we have given up on the exact agreement of background metrics---which was
seen to significantly improve the quality of the merged metric in Paper~II---in
order to have horizon-penetrating coordinates.

\section{Tidal fields}\label{TidalFields}

\subsection{Comparison with Taylor and Poisson's results}\label{TFTP}

To facilitate the comparison of our expressions for the tidal fields with
those obtained
by Taylor and Poisson~\cite{TP}, we collect the results of our matching here.
These include the results of the fourth order octupole matching from the next
subsection, and are all put together using Eqs.~\eqref{TFexpansion} to give
explicit expressions for the tidal fields about hole~1:
\begin{widetext}
\begin{subequations}\label{TFs}
\begin{align}
\sE_{kl}(t) &= \frac{m_2}{b^3}\left\{\left[1 - \frac{1}{2}\frac{m_2}{b}\right][\delta_{kl} - 3\hat{x}_k\hat{x}_l] + \frac{1}{2}\frac{m}{b}[4\hat{x}_k\hat{x}_l -
5\hat{y}_k\hat{y}_l + \hat{z}_k\hat{z}_l] - 6\sqrt{\frac{m}{b}}\frac{t}{b}\hat{x}_{(k}\hat{y}_{l)} + O\left(\left[\frac{m_2}{b}\right]^2,\frac{t^2}{b^2}\right)\right\},\\
\begin{split}
\sB_{kl}(t) &= \frac{m_2}{b^3}\sqrt{\frac{m_2}{b}}\biggl\{\biggl[-6\sqrt{\frac{m}{m_2}} +
\frac{m_2}{b}\biggl\{5\left(\frac{m}{m_2}\right)^{3/2} + 7\sqrt{\frac{m}{m_2}} - 3\sqrt{\frac{m_2}{m}}\biggr\}\biggr]\hat{x}_{(k}\hat{z}_{l)} - 6\frac{m}{m_2}\sqrt{\frac{m_2}{b}}\frac{t}{b}\hat{y}_{(k}\hat{z}_{l)}\\
&\quad + O\left(\left[\frac{m_2}{b}\right]^{3/2},\frac{t^2}{b^2}\right)\biggr\},
\end{split}\\
\sE_{klp}(t) &= \frac{m_2}{b^4}\left\{\left[1 - 3\frac{m_2}{b}\right][15\hat{x}_k\hat{x}_{\vphantom{(}l}\hat{x}_p - 9\delta_{(kl}\hat{x}_{p)}] -
3\frac{m}{b}[\hat{x}_k\hat{x}_l\hat{x}_p - 4\hat{y}_{(k}\hat{y}_l\hat{x}_{p)} + \hat{z}_{(k}\hat{z}_l\hat{x}_{p)}] + O\left(\left[\frac{m_2}{b}\right]^2,\frac{t}{b}\right)\right\},\\
\sB_{klp}(t) &= \frac{9}{2}\frac{m_2}{b^4}\sqrt{\frac{m}{b}}\left\{5\hat{x}_{(k}\hat{x}_l\hat{z}_{p)} - \delta_{(kl}\hat{z}_{p)} + O\left(\left[\frac{m_2}{b}\right]^{1/2},\frac{t}{b}\right)\right\}.
\end{align}
\end{subequations}
\end{widetext}
Here, for the purposes of comparison with Taylor and Poisson's results, we have included the time dependent pieces we know in the quadrupole fields (the time dependence
falls into the uncontrolled remainders in the octupole fields). These are ordinarily contained in $\dot{\sE}_{kl}$ and $\dot{\sB}_{kl}$, since we usually treat them as 
independent tidal fields. 

Taylor and Poisson give explicit expressions for the quadrupole tidal fields for a binary in a circular orbit in their Eqs.~(1.10)--(1.14) [and with alternate notation in
Eqs.~(7.25)--(7.29)]. The parts of the quadrupole fields that Taylor and Poisson
and we have both computed agree: These consist of the electric quadrupole,
including its $1$PN corrections; the time derivative of the electric
quadrupole with no corrections; and the magnetic quadrupole with no
corrections, all evaluated at $t = 0$. In fact, we can recover all of Taylor
and Poisson's expressions for the tidal fields, including the full time
dependence, if we evaluate our expressions for the tidal fields at $t = 0$ and
then make the substitutions $\hat{x}_k \to \hat{x}_k \cos \omega t  +
\hat{y}_k \sin \omega t$ and $\hat{y}_k \to -\hat{x}_k \sin \omega t +
\hat{y}_k \cos \omega t$. (This is, of course, only accurate for times much less than
the radiation reaction time scale.) Thus, even though we are not given the full
time dependence directly from the matching, we can obtain it from our results,
since they are true for any point in the orbit.

While we have computed certain higher-order contributions to the
tidal fields that Taylor and Poisson did not, we cannot improve upon the formal
accuracy of their result for the tidal heating [given through $1$PN in
their Eq.~(9.4)].
This is the case even though Poisson gives the contribution of the octupole fields to
the tidal heating in Eq.~(12) of~\cite{PoissonPRL}: The $2$PN correction to the
expression for the tidal heating involves the unknown
$2$PN correction to the electric quadrupole (along with the known $1$PN correction to
the magnetic quadrupole and the Newtonian piece of the electric octupole).

We can also check the lowest-order pieces of all the tidal fields we found are the
expected Newtonian ones: The Newtonian pieces of the tidal fields can be
computed independently using Eqs.~(5.45), (5.50), and (5.56) in~\cite{BHTMPV}, along
with the obvious generalization of that reference's Eqs.~(5.45) and (5.50) for the
magnetic octupole,
viz., $\sB_{klp}^\mathrm{Newt} = -(3/8)\epsilon^{su}{}_{(k}\partial_{lp)s}\beta_u$, where
$\beta^k$ is given by Eq.~(5.56b) of~\cite{BHTMPV}. (To reproduce our results
exactly, one needs to evaluate all of these expressions for the Newtonian parts of the
tidal fields at $r_1 = 0$. Also, since these expressions for the Newtonian
parts of the tidal fields are valid in the rest frame of
hole~1, we need to use the relative velocity of the holes in calculating
$\beta^k$.) Our expression for the $1$PN correction to the magnetic
quadrupole can be checked in the extreme mass ratio limit against that computed by Poisson
in~\cite{Poisson2004}: This result is given in an appropriate form for comparison in an
unnumbered equation in Sec.~VII~E of Taylor and Poisson~\cite{TP} and agrees with our computation.

\subsection{$1$PN corrections to the fourth order octupole fields}\label{TFHO}

Even though we cannot obtain all the inner zone octupole pieces at fourth and fifth
orders without including the hexadecapole fields (since these will give octupolar
contributions to the nonpolynomial part at these orders), it \emph{is} possible to
match only the polynomial parts, and, in doing so, read off the $1$PN corrections to
the octupolar tidal fields. As an illustration, we shall read off the $1$PN correction
to the electric octupole (along with the lowest-order piece of the time derivative of
the magnetic quadrupole) by matching the octupole parts of the polynomial pieces at
fourth order.

Except for the added algebraic complication of keeping higher-order multipole
terms, this calculation proceeds precisely analogously to the fourth order
calculation involving the
quadrupole-and-lower multipoles of the polynomial part in Sec.~\ref{4O5O}. The only subtlety that we should mention is one that was already
present in our original fourth order calculation. However, it was not a
potential source of confusion there because
we were computing both the polynomial and nonpolynomial parts at once.
Now that we want to compute the polynomial part by itself, we need to bear in
mind that two nonpolynomial pieces can produce a polynomial piece when
multiplied together (e.g., $\tr$ is a nonpolynomial piece, but $\tr^2$ is a
polynomial piece). Therefore, contributions to the near zone metric involving
terms such as $n_1^{(k}n_{12}^{l)}/S^2$ will contain polynomial pieces, since
$n_1^k$ contains a factor of $1/\tr$, and $1/S^2$ contributes a factor of
$\tr$ [see Eq.~\eqref{S^n}]. The final results are
\begin{widetext}
\begin{subequations}
\begin{align}
\ods{\dot{\bE}_{kl}}{2} &= 0,& \od{\bE_{klp}}{2} &= 9[3\delta_{(kl}\hat{x}_{p)} - 5\hat{x}_k\hat{x}_l\hat{x}_p] - \frac{m}{m_2}[3\hat{x}_k\hat{x}_l\hat{x}_p -
12\hat{y}_{(k}\hat{y}_{\vphantom{(}l}\hat{x}_{p)} + 3\hat{z}_{(k}\hat{z}_{\vphantom{(}l}\hat{x}_{p)}],\\
\qquad \ods{\dot{\bB}_{kl}}{1} &= -6\frac{m}{m_2}\frac{1}{b}\hat{y}_{(k}\hat{z}_{l)},& \od{\bB_{klp}}{1} &= 0,
\end{align}
\end{subequations}
with an accompanying polynomial piece of the coordinate transformation of
\<
\begin{split}\label{XP4OO}
\od{X^\rP_\alpha}{4,3} &= \frac{\tx t}{b^3}\biggl\{\biggl[\frac{3}{2}\frac{m}{m_2} - 2\biggr]\tx^2 - \frac{m}{m_2}\frac{t^2}{6} + \biggl(\frac{3}{2} -
\frac{9}{2}\frac{m}{m_2}\biggr)y^2 + \frac{1}{2}\biggl(1 - \frac{m}{m_2}\biggr)z^2\biggr\}\hat{t}_\alpha\\
&\quad + \frac{1}{b^3}\biggl\{\biggl[\frac{m}{m_2}\biggl(\frac{t^2}{4} + \frac{\tx^2}{8} + y^2 - \frac{3}{2}z^2\biggr) + y^2 + z^2\biggr]\tx^2 - \frac{m}{m_2}\biggl[\frac{t^4}{24} + \frac{y^4 - z^4}{8} + \frac{(y^2 + z^2)t^2}{4}\biggr] - \frac{(y^2 + z^2)^2}{4}\biggr\}\hat{x}_\alpha\\
&\quad + \frac{\tx y}{b^3}\biggl\{3\tx^2 - \frac{9}{4}(y^2 + z^2) + \frac{m}{m_2}\biggl[\frac{t^2}{2} - \frac{5}{2}\tx^2 + \frac{9}{4}y^2 +
\frac{3}{4}z^2\biggr]\biggr\}\hat{y}_\alpha\\
&\quad + \frac{\tx z}{b^3}\biggl\{3\tx^2 - \frac{9}{4}(y^2 + z^2) + \frac{m}{m_2}\biggl[\frac{t^2}{2} - \frac{\tx^2}{2} + \frac{7}{4}y^2 +
\frac{z^2}{4}\biggr]\biggr\}\hat{z}_\alpha.
\end{split}
\?
\end{widetext}

Continuing on to fifth order to obtain the $1$PN corrections to the magnetic octupole
and time derivative of the electric quadrupole would be algebraically more
complicated, but would otherwise proceed as above. At sixth order we are not so
fortunate. If we tried to carry out even the quadrupolar part of the sixth order
polynomial matching, so
as, e.g., to read off the $2$PN correction to the electric quadrupole, we would be stymied by our lack of knowledge of $\od{X^\rNP_\alpha}{4,3}$: The
$\tr$-times-a-polynomial-in-$\tx^\alpha$ pieces of $\od{X^\rNP_\alpha}{4,3}$ (which we expect to be present, as there have been such terms at all lower
multipole orders) will contribute to $\ods{h^\rP_{\alpha\beta}}{6}$ via the $b\od{1/R}{4}$ term.

\section{Higher-order terms}
\label{ExtraTerms}

\subsection{Discussion}

Let us first catalogue all the higher-order near and far zone pieces that are readily
available in the literature (besides those already discussed in
Sec.~\ref{phasing}): Blanchet, Faye, and Ponsot~\cite{BFP} give an explicit
expression for the near zone $2.5$PN metric with the standard PN order counting---i.e.,
with remainders of $O(v^8)$ in $g_{00}$, $O(v^7)$ in $g_{0k}$, and $O(v^6)$ in $g_{kl}$.
The expression Pati and Will give for the far zone metric perturbation in terms of
derivatives of the multipoles [reproduced here in Eq.~\eqref{metric-perturbation}] has
remainders of at least $O(v^6)$ in all components. Additionally, Pati and Will express the
metric in terms of the metric perturbation through $3.5$PN order (with the standard PN
order counting) in, e.g., Eqs.~(4.2) of~\cite{PW1}. 
Further contributions to the far zone metric perturbation can be obtained from the full
expression for the field multipole expansion of the near zone contributions to the far
zone [Eq.~(2.13) in~\cite{PW1}], combined with the expressions for the
field multipoles, $M^Q$, in terms of the source multipoles, $\sI^Q$ and $\sJ^Q$, in
Eqs.~(4.7) of~\cite{PW1}.

The lowest-order (Newtonian) pieces of all the source multipoles are known---see
Eq.~(D1) in~\cite{PW2}. Additionally, further PN corrections to the source multipole
moments than we used in Sec.~\ref{FZ} can be found in various places: Higher-order
corrections to $\sI$ can be obtained from the system's binding energy,
which Blanchet gives through $3$PN in Eq.~(170) of~\cite{BlanchetLRR}.  Similarly, one can get higher-order corrections to $\sJ^k$ from
the expression Kidder, Will, and
Wiseman give for the system's angular momentum (through $2$PN) in Eq.~(2.13b)
of~\cite{KWW}. We gave the $1$PN correction to
the mass quadrupole in Eq.~\eqref{Ikl}, even though we did not need it to construct our
$O(v^4)$ data. The $1$PN correction to the mass octupole can be obtained (up to the
caveats mentioned below) from the expression for the three-index Epstein-Wagoner (EW)
moment given in Eq.~(6.6b) in Will and Wiseman~\cite{WW}. This EW moment also yields
the $1$PN correction to the current quadrupole. Even further corrections to
various source multipoles can be obtained from the expressions for the two- and
four-index EW moments that Will and Wiseman give. 

We can use Blanchet, Faye, and Ponsot's results to obtain initial data in the
near zone that are formally valid through $O(v^5)$. However, to obtain $O(v^5)$
initial data in the far zone, one needs the $O(v^6)$ pieces of the
spatiotemporal components of the metric. As mentioned in Sec.~\ref{InclRad},
including these is slightly problematic: There is an outer integral term in
$h^{0k}$ that is $O(v^6)$ with the order counting used in Sec.~\ref{FZ} and is not known at present. This term looks, schematically, like $\sI\sJ^k/r^3$
and comes from the first term in the general expression Pati and Will give for $\Lambda^{0k}$ in Eq.~(4.4b)
of~\cite{PW1}. [Here $\Lambda^{\alpha\beta}$ gives the contributions of the gravitational field to $\tau^{\alpha\beta}$, the ``effective'' stress-energy pseudotensor. See Eqs.~(2.5)--(2.7)
in~\cite{PW1} for the explicit definitions.] However, it is \emph{not} present in the explicit expression for $\Lambda^{0k}$ in the
far zone that Pati and Will give in Eq.~(6.5b) of~\cite{PW1}. This is not a problem, since
Pati and Will derive this expression specifically for computing the contributions to the near zone
field from the outer integrals. They have thus used the ``quick and dirty'' rule from their
Eq.~(4.9) to eliminate any pieces that do not contribute there to the order considered. This
piece is one of those eliminated, as it will contribute an $\sR$-independent
term only through a time derivative:\footnote{Here $\sR$ is the artificial radius of separation between the near and
far zones in the DIRE approach, which, as Pati and Will demonstrate in Sec.~II~I
of~\cite{PW1}, cannot appear in any final results.} Since both $\sI$ and $\sJ^k$ are constants of the motion, up to
radiative losses, the resulting contribution would be of considerably higher order than
Pati and Will
are keeping. However, the ``quick and dirty'' rule is not applicable to the far zone, and we indeed have to
consider this term.

While it would be, in principle, reasonably straightforward to
compute the unknown term, following the procedure given by Pati and Will in~\cite{PW1}, the
calculation would be involved enough that we do not attempt it here. In fact, it is possible
to argue that we can ignore this term entirely: Recalling that, in practice, extra factors of
$1/r$ always make a term smaller by at least a factor of $v$ in the far zone [see
Eq.~\eqref{ordering} and the surrounding discussion], we can choose to count \emph{all} powers of
$1/r$ after the first as $O(v)$, disregarding  post-Minkowskian considerations. This
``practical'' order counting is justified numerically, since the factors of $1/r$ will make
the highest-order near zone terms (whose far zone analogues will not be present) similarly small when evaluated in the buffer zone. Since this is where
we stitch the near and far zone metrics together numerically, it is thus the only place where
we are concerned with their agreement, in practice.

We therefore disregard post-Minkowskian powers of $G$ in our counting, giving, e.g., 
$(\sI/r)^2 = O(v^5)$, as opposed to $O(v^4)$, as it was before. Nevertheless, this new order
counting still allows us to keep all the outer integral contributions that we know
(and, indeed, all the terms we calculated in Sec.~\ref{FZ}), since we
are now keeping terms to one order in $v$ higher than before. It also gives
$\sI\sJ^k/r^3 = O(v^7)$, so we can safely ignore this unknown term. Thus, all we need
to consistently calculate initial data through $O(v^5)$ in the far zone are the $1$PN
corrections to the current dipole and the contributions to $h^{0k}$ from the mass
hexadecapole and current quadrupole. As mentioned above, all of these are easily
obtainable.

Now, even though we can obtain initial data that are formally accurate to $O(v^5)$ in
the near and far zones (with the new far zone order counting), this does not utilize
the $O(v^6)$ and $O(v^7)$ terms in the purely temporal component of the metric that
Blanchet, Faye, and Ponsot give us. While including these terms does not increase the
formal accuracy of the data, even just in the near and far zones, it is nevertheless
possible that adding such terms will improve the data's quality in practice. Of
course, if we include terms through $O(v^7)$ in the
purely temporal component of the near zone metric, it seems desirable to also
include them in the
purely temporal component of the far zone metric,
and it is possible to do so, up to outer integral terms. Here we shall simply neglect
the outer integral terms we do not know, since there appear to be some in $h^{00}$
that are $O(v^7)$ even with the new ``practical'' order counting, e.g., ones that look
like $\sI\ddot{\sI}^{kl}\hat{n}^{<kl>}/r^2$. [See Eq.~(6.5a)
in~\cite{PW1}.\footnote{This
equation only contains the parts of $\Lambda^{00}$ that contribute to the near zone
metric. But one can check, starting from the general expression for $\Lambda^{00}$
Pati and Will give in Eq.~(4.4a) of~\cite{PW1}, that there are not any lower-order
contributions that only appear in the far zone.}] Since we are already
adding pieces without regard to formal accuracy, it does not make much sense to go to
the trouble of calculating these outer integral terms here. This is
particularly true since we
are not even sure that we have all the terms we need in the $1$PN correction to the
mass octupole. (We \emph{do} know some even higher-order contributions to the
far zone metric, but do not include those, since we do not know the matching
near zone contributions.)

This uncertainty arises because the EW moments presented by Will and Wiseman are missing
any pieces that are pure traces in the first two indices: Will and
Wiseman were only interested in using their results to compute the gravitational
waveform [via their Eq.~(2.18)], which is transverse and tracefree. For instance, the
$(7/4)m_1m_2b\delta^{kl}$ term present in the $1$PN correction to $\sI^{kl}$ in
Eq.~\eqref{Ikl} is not present in the expression for $I^{kl}_\rEW$ in Eq.~(4.17)
of~\cite{WW}. If not for the missing trace term, these two expressions would be
identical, up to a surface term, as can be seen from their definitions: $I^{kl}_\rEW$
and $\sI^{kl}$ are defined in Eq.~(2.19a) of~\cite{WW} and Eq.~(4.5b) of~\cite{PW1}, respectively. Unlike for $\sI^{kl}$, we do not have an independent
calculation of the $1$PN corrections to $\sI^{klp}$. It is thus possible that we are
missing terms such as $m_1^2b\delta^{kl}x_1^{p} + (1 \leftrightarrow 2)$ in
$I^{klp}_\rEW$, which (as demonstrated in the next subsection) would
contribute to the $1$PN correction to $\sI^{klp}$ and thus to the $O(v^7)$ pieces of
$g_{00}$.

\subsection{Outline of the calculation}

For the far zone metric, we first calculate the higher-order multipole
contributions that need to be added to the expression for $h^{\alpha\beta}$ given in
Eq.~\eqref{metric-perturbation}. The $O(v^6)$ and $O(v^7)$ pieces of $h^{00}$ are
\<
\frac{1}{6}\partial_{klps}\left[\frac{\sI^{klps}(u)}{r}\right] -
\frac{1}{30}\partial_{klpsv}\left[\frac{\sI^{klpsv}(u)}{r}\right],
\?
and the $O(v^6)$ pieces of $h^{0k}$ are
\<
-\frac{1}{6}\partial_{lps}\left[\frac{\dot{\sI}^{klps}(u)}{r}\right] +
\frac{1}{2}\epsilon^{lkp}\partial_{psv}\left[\frac{\sJ^{lsv}(u)}{r}\right].
\?
These were obtained by substituting the expressions for $M^Q$ (in terms of
$\sI^Q$ and $\sJ^Q$) from Pati and Will's Eqs.~(4.7) into their Eq.~(2.13) (both
from~\cite{PW1}).
One also needs higher-order contributions to the expression for $g_{00}$ in terms of
$h^{\alpha\beta}$. We do not need to add any other new terms to the expression for
$g_{\alpha\beta}$, since the expressions required to obtain the $O(v^6)$ pieces of
$g_{0k}$ and $O(v^5)$ pieces of $g_{kl}$ are the same as those given previously in
Eq.~\eqref{metric}. This follows because we already included the $-(1/2)h^{00}$ term
in $g_{0k}$ for formal consistency, even
though it only gives $O(v^6)$ terms in actuality with our order counting. The
expression we need to obtain $g_{00}$ consistently through $O(v^7)$ is
\<
\begin{split}
g_{00} &= -\left[1 - \frac{1}{2}h^{00} + \frac{3}{8}\left(h^{00}\right)^2 - \frac{5}{16}\left(h^{00}\right)^3\right]\\
&\quad + \frac{1}{2}\left[1 - \frac{1}{2}h^{00}\right]h^{kk},
\end{split}
\?
taken from Eq.~(4.2a) in~\cite{PW1}.

We also need the $2$PN corrections to the mass monopole, along with the
$1$PN corrections to the mass quadrupole, mass octupole, and current dipole.
As discussed previously, all of these except the $1$PN correction to the mass
octupole are given directly in the literature: The corrections to the mass
monopole come from Blanchet's Eq.~(170)~\cite{BlanchetLRR} and those to the
current dipole from Eq.~(2.13b) in~\cite{KWW}; the corrections to the mass
quadrupole are given in our Eq.~\eqref{Ikl}. 
However, it is possible to obtain a very simple expression for its time derivative (up to the caveats mentioned above) in
terms of the three-index EW moment $I_\rEW^{klp}$ [which is given
in, e.g., Eq.~(6.6b) in~\cite{WW}], viz.,
$\dot{\sI}^{klp} = 3I_\rEW^{(klp)}$. Here the equality holds up to surface terms
(i.e., ones involving $\sR$), which we can neglect here. To obtain this equality,
take a time derivative of the definition of the mass octupole, use the conservation law
$\partial_\beta\tau^{\alpha\beta} = 0$ to write $\partial_0\tau^{\alpha 0} =
-\partial_k\tau^{\alpha k}$, and integrate the result by parts, giving an
expression that equals $3I_\rEW^{(klp)}$ up to surface terms.

We can now antidifferentiate the resulting expression for $\dot{\sI}^{klp}$ to obtain
$\sI^{klp}$ (up to the caveats mentioned in the previous subsection). If we do this for
an arbitrary orbit, we do not need to worry
about missing terms due to the constant of integration: As we shall
demonstrate below, any contributions to
$\sI_{klp}$ that are constant at Newtonian order vanish for a circular orbit.
(It also turns out that any terms that are time-independent for a circular
orbit also vanish.)
For a generic orbit, it is easiest to start from the explicit $2$-body
reduction Will and Wiseman give in Eq.~(6.6b). The resulting expression is
exactly what one would expect from the forms of the $1$PN corrections to the
mass monopole and quadrupole, viz.,
$\sI^{klp} = m_1x_1^{klp}(1 + v_1^2/2 - m_2/2b) + (1 \leftrightarrow
2) + O(v^4)$.

To prove our claim that any terms in the $1$PN correction to $\sI^{klp}$ that are
constant at Newtonian order for an arbitrary orbit vanish for a circular orbit, we
construct all such possible terms. To do so, we note that the the binary's only
(nonzero) vectorial Newtonian constants of the motion are its Newtonian angular
momentum $\vec{L}$ and Laplace-Runge-Lenz vector $\vec{A} := \vec{p} \times
\vec{L} - \mu^2m\hat{b}$. (Here $\vec{p}$ is the momentum
of the reduced mass $\mu := m_1m_2/m$.) Noting that these are
$O(c^{-1})$ and $O(c^{-2})$, respectively, we can write down
the only two possible $O(c^{-2})$ symmetric $3$-index (Cartesian) tensors
involving only those two vectors (and constants). These will then be the
only possible $1$PN contributions to $\sI^{klp}$ that are constant at
Newtonian order. They are, up to numerical factors (which could include 
contributions of $\|\vec{A}\|/\|\vec{L}\|^2$),
$A^{(k}\delta^{lp)}$ and $\|\vec{L}\|L^{(k}\delta^{lp)}/m$. The first of these vanishes for a circular
orbit, since $\vec{A}$ does. The second is odd under time reversal and is thus
inadmissable, since $\sI^{klp}$ should be even, from its definition, given in
Eq.~(4.5b) in~\cite{PW1}.

With all these ingredients, we can put together
the far zone metric in the same way as we did in Sec.~\ref{FZ}.
 We thus obtain its various components to the same order as we are keeping the near
zone metric (with the above caveats about missing terms), viz., with uncontrolled
remainders of $O(v^8)$ in the purely temporal component, $O(v^7)$ in the
spatiotemporal components, and $O(v^6)$ in the purely spatial components.

\section{Computation of the metrics}\label{metric_comp}

Here we detail exactly how the metrics are computed in the
{\sc{Maple}} scripts that were used (along with {\sc{BAM}}) to compute the constraint
violations and
create the plots. (The scripts themselves, and the resulting C code that {\sc{Maple}}
outputs are available online at~\cite{Wolfgangs_website}).

\subsection{Inner zone}\label{IZ_comp}

To compute the inner zone metric around hole~$1$, we substitute the tidal
fields given in Eq.~\eqref{TFs} into the expression for Detweiler's perturbed
Schwarzschild metric in Cook-Scheel coordinates that we give in
Eqs.~\eqref{hCS} and~\eqref{hCS_pieces}, taking $M = m_1$. We then transform
using the coordinate transformation given in Sec.~\ref{CT}. The resulting
metric thus includes the $O(v^5)$ terms in the purely temporal and spatial
components, though these do not increase the formal accuracy of the initial
data. We also do not perform any expansions after substituting the tidal fields
and performing the coordinate transformation, so the final, transformed metric
also contains various other higher-order-in-$v$ terms. The inner zone metric
around hole~$2$ is obtained by the same procedure, along with the
transformations detailed at the beginning of Sec.~\ref{Matching1}.

We have considered three versions of the inner zone metric: The first version
(contained in \verb,O4_NoOct,)
comes directly from the matching performed in Sec.~\ref{Matching} and only
contains the pieces that we were able to match onto the near zone metric while
including all of the multipolar contributions at a given order [i.e., up to
octupolar order through $O(v^3)$ and then only up to quadrupolar order
through $O(v^5)$]. The second version (contained in \verb,O4,) also incorporates the
results of the fourth-order octupole matching carried out in
Appendix~\ref{TFHO}---this includes the $1$PN
correction to the electric octupole, but only the polynomial part of the accompanying
coordinate transformation. The third version (contained in \verb,O5, and \verb,all,)
adds on the time dependence of all the
tidal fields (for a circular orbit), obtained in the manner described in
Appendix~\ref{TFHO}, though it still uses the same coordinate transformation as before. (We use
the $1$PN expression for $\omega$ when substituting for the
unit vectors in obtaining the full time dependence. We leave off the
known higher-order corrections to $\omega$ here since the expressions for the tidal
fields we obtained by matching came from using the $1$PN version of $\omega$.)

We calculate the third version by substituting $T\dot{\sE}_{kl} \to
(T - t)\dot{\sE}_{kl}(0)$
and similarly for $\dot{\sC}_{klp}$ in Eqs.~\eqref{hCS_pieces} before
substituting in the tidal fields (with full time dependence). 
These substitutions are necessary because the $T\dot{\sE}_{kl}$ and
$T\dot{\sC}_{klp}$ terms in Eqs.~\eqref{hCS_pieces} come from the expansions of
$\sE_{kl}$ and $\sC_{klp}$ (with full dependence on $V$, the ingoing
Eddington-Finkelstein coordinate) about $V = 0$. We, however, are only
including the full time dependence on $t$, the near zone time coordinate, in
the tidal fields, due to our method of obtaining this dependence. These
expressions will thus will only contain the $t\dot{\sE}_{kl}$ and
$t\dot{\sC}_{klp}$ pieces (when expanded about $t = 0$), so we make the above
substitutions to retain the linear $T$-dependence given by the matching while
not including the linear $t$-dependence twice. (We experimented with including the full
time dependence of the tidal fields using $T$ instead of $t$ and found that the
constraint violations increased.)

This method of computing the metric deliberately does \emph{not} include the effects of the
full time dependence on the spatial variation of the tidal fields'
contributions to the metric [due to their dependence on $V$; see the
discussion following Eq.~\eqref{hCS_pieces}], since these would
enter at the same order as the unknown time derivatives of the tidal fields
(second derivatives of the quadrupole fields, and first derivatives of the
octupole fields). Of course, the terms we are keeping are higher-order as well,
but since they would enter with explicit factors of $t$ in the multipole
expansion, they would not be entangled with the explicit appearances of
unknown time derivatives. (This follows because the Schwarzschild metric is
time-independent in the coordinates we use.) What we have done is equivalent
to repeating the matching we have
performed at each value of (near zone time) $t$, up to orbital shrinkage
effects, which are higher-order than the terms we are considering here. (We
also have not attempted to include the full time dependence of the coordinate
transformation for the reasons discussed in Sec.~\ref{InclRad}.)

\subsection{Near zone}

We compute the near zone metric by substituting the trajectories for the point
particles [obtained in the manner discussed in Sec.~\ref{phasing}]
into Blanchet, Faye, and Ponsot's metric [given in Eqs.~(7.2) of~\cite{BFP}],
including the background resummation given in Sec.~\ref{BR}. Here there are,
again, three versions of the metric, one giving $O(v^4)$ data, one $O(v^5)$
data, and one containing the complete $2.5$PN metric.
Recall that one needs all the components through $O(v^4)$ [resp.\ $O(v^5)$] in addition to the $O(v^5)$ [resp.\ $O(v^6)$]
terms in the spatiotemporal components in order to obtain $O(v^4)$ [resp.\ $O(v^5)$] data.
N.B.: In order to perform background resummation
on the purely temporal component of the complete $2.5$PN metric, one needs
to also subtract the $O([m_1/r_1]^3)$ portion of the expansion of the background, viz.,
$2m_1^3/r_1^3$, in Eq.~\eqref{g00resum}. All these versions are constructed
by truncating the metric components to the desired order before substituting in the
trajectories. No expansions are performed after that substitution, since we do not want to
drop the higher-order terms that we are keeping in the trajectories (discussed in
Sec.~\ref{phasing}) for conformity with the far zone metric. Due to an oversight, we
did not include the $3$PN corrections to the relative-to-COM relation in the $O(v^4)$
version of the near zone metric.

\subsection{Far zone}

The calculation of the far zone metric follows the DIRE approach, detailed in Sec.~\ref{FZ},
differentiating multipole moments to obtain the metric. We have the same
three versions of the far zone metric as for the near zone
metric, and obtain them in the same manner: We expand to the desired
order after performing all the substitutions except for $\phi$, $\omega$, and $b$
(i.e., the contributions that vary
due to secular radiation reaction effects and are discussed in Sec.~\ref{phasing}). We also
do not include any of the terms due to derivatives acting on $b$: These would
give nonzero contributions, due to the separation's retarded time dependence from orbital
shrinkage, but these terms are quite small, both formally and practically---unlike those
due to the radiation reaction effects in the phase or undifferentiated
separation---so we neglect them. [For instance, the lowest-order contribution
due to the nonzero time derivative of $b$ in the
$\partial_{kl}[\sI^{kl}(u)/r]$ contribution to $g_{00}$ is formally $O(v^9)$. It is also
numerically small: The largest contribution has a magnitude of $\sim 10^{-7}$ when
evaluated at the intersection of the $x$-axis
and the inner boundary of the near-to-far transition for our equal-mass test
binary, viz., $x \simeq 20m$---see Sec.~\ref{Transitions}. For comparison, the
contribution of the uncorrected $(m_1/r)v_1^2 + (1 \leftrightarrow 2)$ term in that
situation is $\sim 10^{-4}$.]
The $O(v^4)$ version uses a slightly different order counting than
the $O(v^5)$ and full extended versions: As
discussed in Sec.~\ref{FZ}, we choose to keep the outer integral terms---here these are the
terms that look like $(m/r)^2$---in the $O(v^4)$ data due to
post-Minkowskian considerations, even though those terms are
$O(v^5)$ if one interprets the Pati-Will order counting strictly. However,
as mentioned in Appendix~\ref{ExtraTerms}, we
do not know any of the higher-order outer integral terms, so we simply drop them
in the $O(v^5)$ and full extended data. 

\bibliography{paper}

\end{document}